\documentclass[12pt]{article}
\usepackage{amsmath, amssymb}

\usepackage{epsfig}

\newtheorem{theorem}{Theorem}[section]
\newtheorem{proposition}[theorem]{Proposition}

\newtheorem{lemma}[theorem]{Lemma}
\newtheorem{definition}[theorem]{Definition}

\newcommand{\cs}{\clubsuit}
\newcommand{\ds}{\diamondsuit}
\newcommand{\hs}{\heartsuit}

\newcommand{\rd}{{\rm d}}
\newcommand{\be}{\begin{equation}}
\newcommand{\ee}{\end{equation}}
\newcommand{\bey}{\begin{eqnarray}}
\newcommand{\eey}{\end{eqnarray}}

\newcommand{\bef}{\begin{figure}}
\newcommand{\eef}{\end{figure}}
\newcommand{\bec}{\begin{center}}
\newcommand{\eec}{\end{center}}

\newcommand{\tri}{| \! |\!|}

\newcommand{\eqn}{\begin{eqnarray}}
\newcommand{\eeqn}{\end{eqnarray}}

\newcommand{\bP}{{\bf P}}

\newcommand{\bD}{{\bf D}}

\newcommand{\bA}{{\bf A}}

\newcommand{\bw}{{\bf w}}
\newcommand{\bp}{{\bf p}}

\newcommand{\bu}{{\bf u}}

\newcommand{\tp}{{\tilde p}}

\newcommand{\tu}{{\tilde u}}

\newcommand{\tgamma}{{\tilde\gamma}}
\newcommand{\tGamma}{{\tilde\Gamma}}

\newcommand{\tbp}{{\tilde{\bf p}}}

\renewcommand{\a}{\alpha}

\newcommand{\e}{\varepsilon}
\newcommand{\s}{\sigma}

\newcommand{\om}{{\omega}}

\renewcommand{\t}{\tau}

\newcommand{\bE}{{\bf E}}

\newcommand{\cM}{{\cal M}}

\renewcommand{\iint}{\int \!\! \int}
\newcommand{\bR}{{\mathbb R}}
\newcommand{\bC}{{\bf C}}
\newcommand{\bN}{{\bf N}}

\newcommand{\wt}{\widetilde}
\newcommand{\wh}{\widehat}
\newcommand{\ov}{\overline}
\newcommand{\balpha}{\mbox{\boldmath $\alpha$}}

\newcommand{\cI}{{\cal I}}
\newcommand{\cR}{{\cal R}}
\newcommand{\cG}{{\cal G}}

\newcommand{\cF}{{\cal F}}
\newcommand{\cA}{{\cal A}}

\newcommand{\cE}{{\cal E}}
\newcommand{\cP}{{\cal P}}

\newcommand{\cV}{{\cal V}}

\newcommand{\cL}{{\cal L}}
\newcommand{\cN}{{\cal N}}
\newcommand{\cO}{{\cal O}}

\input epsf

\def\fS{{\frak S}}

\date{Dec 4, 2005  (Revised on May 15, 2006)}

\begin{document}

\title{Quantum diffusion of the random Schr\"odinger evolution 
\\
in the scaling limit II. The recollision diagrams.}
\author{L\'aszl\'o Erd\H os${}^1$\thanks{Partially
supported by NSF grant DMS-0200235 and EU-IHP Network 
``Analysis and Quantum'' HPRN-CT-2002-0027.}
\\ Manfred  Salmhofer${}^2$
\\ Horng-Tzer Yau${}^3$\thanks{Partially supported by NSF grant
DMS-0602038.} \\
\\
\small
${}^1\;$Institute of Mathematics, University of Munich, 
\\
\small
Theresienstr. 39, D-80333 Munich, Germany
\\
\small
${}^2\;$Max--Planck Institute for Mathematics, Inselstr.\  22, 04103 Leipzig,
and
\\ 
\small
Theoretical Physics, University of Leipzig, Postfach 100920, 04009 Leipzig, 
Germany
\\
\small
${}^3\;$Department of Mathematics, Harvard University, Cambridge
MA-02138, USA \\ }

\maketitle

\abstract{  We consider  random Schr\"odinger equations
on $\bR^d$ for $d\ge 3$
with a homogeneous Anderson-Poisson type
random potential.
Denote by $\lambda$ the coupling constant  and $\psi_t$ the solution
with initial data $\psi_0$. The space and time variables
scale as $x\sim \lambda^{-2 -\kappa/2}, t \sim \lambda^{-2 -\kappa}$
with $0< \kappa < \kappa_0(d)$. 
We prove that, in the limit $\lambda \to 0$,
the expectation of the Wigner distribution of $\psi_t$
converges weakly to the
solution of a heat equation
in the space variable $x$ for arbitrary $L^2$  initial data.
The proof is based on a rigorous analysis of Feynman diagrams.
In the companion paper \cite{ESYI}
 the analysis of the non-repetition diagrams was presented.
In this  paper we complete the proof by
estimating the recollision diagrams and
showing that the main terms, i.e. the ladder diagrams
with renormalized propagator, converge to the heat equation.}

\bigskip\noindent
{\bf AMS 2000 Subject Classification:} 60J65, 81T18, 82C10, 82C44


\section{Introduction}\label{sec:intro}

The Schr\"odinger equation  is time  reversible and has no dissipation.
The long time dynamics of a quantum particle in a small random  
environment
nevertheless exhibits a stochastic behavior that can be described by
a diffusion equation.  Our goal is to establish
this fact rigorously  in the weak coupling regime.
We have announced in \cite{ESYI} that under a scaling of space
and time with inverse powers of the coupling constant $\lambda$,
the Wigner distribution of the solution to the random
Schr\" odinger equation converges to the solution of
a heat equation as $\lambda \to 0$.
Our approach is  
based on graphical expansion
methods coupled with a certain truncation scheme. The first part of the  
proof  was given in \cite{ESYI};
the current  paper contains the second and final part.
To help the orientation of the reader, we now summarize the
important notations and the main result below.

The quantum dynamics of a single particle in a random potential
is given by the Schr\"odinger equation
\be 
  i\partial_t\psi_t = H\psi_t, \qquad \psi_t\in L^2(\bR^d), \;\;
  t\in \bR \; .
\label{sch}
\ee
The Hamiltonian is a  Schr\"odinger operator,
\be\label{H}
        H := -\frac{1}{2}\Delta + \lambda V\;,
\ee
acting on $L^2(\bR^d)$
with a random potential $V=V_\om(x)$ and 
 a small positive coupling constant $\lambda$.
The potential is given by
\be
   V_\om (x): = \int_{\bR^d} B(x-y)\rd \mu_\om(y)\; ,
\label{Vom}
\ee
where $B$ is a single site potential profile and
$\mu_\om$ is a Poisson point process on $\bR^d$ with
homogeneous unit density and with independent, identically
distributed random coupling constants.
 More precisely, for
almost all realizations $\om$ consists of a countable, locally 
finite collection of points, $\{ y_\gamma(\om) \; : \; 
\gamma=1,2, \ldots \}$,
and random charges $\{ v_\gamma(\om) \; : \; \gamma=1,2,\ldots \}$ such that
the random measure is given by
\be
      \mu_\om = \sum_{\gamma=1}^\infty v_\gamma(\om) \delta_{y_\gamma(\om)}
\label{def:muom}
\ee
where $\delta_y$ denotes the Dirac mass at $y\in\bR^d$.
The Poisson process $\{ y_\gamma(\om)\}$ is independent of
the charges $\{ v_\gamma(\om)\}$.  The charges are real
i.i.d. random variables with distribution $\bP_v$
and with moments   $m_k:=\bE_v \, v_\gamma^{k}$ satisfying
\be
m_2=1,\;\;  m_{2d}<\infty,\;\; \;\;
m_1=m_3=m_5=0\, .
\label{momm}
\ee
The expectation with respect to the random process $\{ y_\gamma, v_\gamma\}$
is denoted by $\bE$.
For the single-site potential, we assume that $B$
 is a spherically symmetric Schwarz function
with $0$ in the support of its Fourier transform.

A quantum wave $\psi\in L^2(\bR^d)$  function can be represented on the phase
space by its Wigner transform
\be
        W_\psi(x,v): = \int
        e^{2\pi iv\cdot \eta} \overline{\psi(x+\frac{\eta}{2})}
  \psi(x-\frac{\eta}{2}) \rd \eta\; ,
\label{WP}
\ee
with the convention that integrals without explicit domains
denote integration over $\bR^d$ with respect to the Lebesgue measure.
Define the rescaled Wigner distribution as
\be
        W^\e_\psi (X, V) : = \e^{-d}W_\psi\Big( \frac{X}{\e}, V\Big)\; .
\label{wig}
\ee
The Fourier transform of the kinetic energy (dispersion relation)
is $e(p):=\frac{1}{2}p^2$, the velocity is given by $\frac{1}{2\pi}
\nabla e(p) =\frac{1}{2\pi} p$.

For any function $h:\bR^d\to\bC$ and energy value $e\ge0$ we introduce the
notation
\be
   [h](e):= \int h(v) \delta(e-e(v))\rd v:=
  \int_{\Sigma_e} h(q) \;\frac{\rd\nu(q)}{|\nabla e(q)|} \; ,
\label{coarea}
\ee
where $\rd\nu(q)$ is the restriction of the Lebesgue measure
onto the energy surface $\Sigma_e:=\{ q \; : \; e(q)=e\}$
that is the ball of radius $\sqrt{2e}$. 
The normalization of the measure $[\cdot ]_e$ is given by
\be
[1](e):= c_{d-1} (2e)^{\frac{d}{2}-1}\; ,
\label{1edef}
\ee
where $c_{d-1}$ is the volume of the unit sphere $S^{d-1}$.

We consider the scaling
\be\label{scale}
t=\lambda^{-\kappa}\big( \lambda^{-2}T\big), \quad 
x= \lambda^{-\kappa/2}\big(\lambda^{-2}X\big) = X/\e, 
\quad \e = \lambda^{\kappa/2+2} \; 
\ee
with some $\kappa \ge 0$. On the kinetic scale, $\kappa=0$,
the limiting equation is
the linear Boltzmann equation with a collision kernel
\be
\sigma(u,v):=2\pi|\wh B(u-v)|^2\delta( e(u)-e(v)) \; .
\label{def:sigma}
\ee
The generator of the corresponding momentum jump process $v(t)$
on the energy surface $\Sigma_e$, $e>0$, is
\be
   L_e f(v): = \int \rd u \; \sigma(u, v) [ f(u)-f(v)], \qquad e(v)=e\; .
\label{Lgen}
\ee
The choice $\kappa>0$ corresponds 
to the long time limit of the
 Boltzmann equation with diffusive scaling.  The
Markov process $\{ v(t)\}_{t\ge0}$ with generator $L_e$
is exponentially mixing (see Lemma \ref{lemma:erg} in the Appendix).
Let
$$
   D_{ij}(e): = \frac{1}{(2\pi)^2}
\int_0^\infty \cE_e \big[ v^{(i)}(t) v^{(j)}(0)\big] 
\rd t\; ,\qquad v= (v^{(1)}, \ldots , v^{(d)}), \quad
 i,j=1,2,\ldots d,
$$
be the  velocity autocorrelation matrix, where $\cE_e$
denotes the expectation with respect to this Markov process
in equilibrium.
By the spherical symmetry of $\wh B$ and $e(U)$, 
the autocorrelation matrix is constant times the identity:
\be
     D_{ij}(e) = D_e \; \delta_{ij}, \qquad  D_e: = \frac{1}{(2\pi)^2d} \;
    \int_0^\infty \cE_e \big[ v(t)\cdot v(0) \big] \rd t \; .
\label{diffconst}
\ee
The main result is the following theorem.

\begin{theorem}\label{main}
Let $d\ge3$ and $\psi_0 \in L^2(\bR^d)$ be a normalized initial wave function.
Let $\psi(t):= \exp(-itH)\psi_0$ solve the Schr\"odinger equation
(\ref{sch}). Let $\cO(x, v)$ be a Schwarz function on $\bR^d\times \bR^d$.
For any $e>0$, let $f$ be the solution to the heat equation
\be
\partial_T f(T, X, e) = \; D_e \;
\Delta_X f(T, X, e) \label{eq:heat}
\ee
with the initial condition
$$
     f(0, X, e): = \delta(X) \Big[ |\wh \psi_0(v)|^2 \Big](e)\; .
$$
Then there exist $0<\kappa_0(d)\leq 2$ such that
for  $0<\kappa<\kappa_0(d)$ and for $\e$ and $\lambda$ 
related by \eqref{scale},
the rescaled  Wigner distribution  satisfies
\be
\lim_{\lambda \to 0} \int\rd X \! \int \! \rd v \;  \cO(X, v)  \bE
 W^\e_{\psi(\lambda^{-\kappa-2} T)} (X, v)
= \int\rd X \int \!\! \rd v \; \cO(X, v) f(T, X, e(v)) \; ,
\label{fint}
\ee
and the limit is uniform on $ T\in [0, T_0]$ with any fixed $T_0$.
In $d=3$ one can choose $\kappa_0(3) = 1/500$.
\end{theorem}

Our result shows that the quantum dynamics on the time scale
$t\sim \lambda^{-2-\kappa}$ is given by a heat equation
and the diffusion coefficient can be computed from
the long time behavior of the underlying Boltzmann dynamics.
Heuristically, this statement can be understood from two facts.
First,  the Boltzmann equation correctly describes the limit of
the quantum evolution 
on the kinetic time scale $t\sim \lambda^{-2}$, $\lambda\to0$
(see \cite{Sp1}, \cite{EY}, \cite{Ch} and \cite{LS}).
Second, the long time
limit of the linear Boltzmann evolution is a diffusion. 
This two step limiting argument is, however,  misleading
(e.g.\ in $d=2$, localization is expected to occur at all
values of $\lambda$, so that no diffusion occurs).
In higher dimensions, 
quantum correlations that are small on the kinetic scale and are
neglected in the first limit may become 
important on the longer time scale, too.
To prove Theorem \ref{main}, we need to control the full quantum dynamics up
to the time scale $t\sim \lambda^{-2-\kappa}$
and prove that the quantum correlations
are not sufficiently strong to destroy the heuristic picture.

\bigskip

The approach of this paper applies also to lattice models
and yields a derivation of  Brownian motion from the Anderson model 
\cite{ESY1, ESY2}.
The dynamics of the Anderson model was postulated by Anderson \cite{A}
to be localized  for large coupling constant
$\lambda$ and  extended for small coupling constant
(away from the band edges
and in dimension $d \ge 3$).
The localization conjecture 
was first established rigorously by
Goldsheid, Molchanov and Pastur \cite{GMP} in
one dimension, by Fr\"ohlich-Spencer \cite{FS},
and later by Aizenman-Molchanov \cite{AM} in several
dimensions, and many other works have since contributed to this field.

The progress for the extended state regime, however, has  been limited. 
It was proved by Klein \cite{Kl}  that
all eigenfuctions are extended on the Bethe lattice (see also 
\cite{ASW, FHS}).  In Euclidean space,
Schlag, Shubin  and Wolff \cite{SSW}
proved that the eigenfunctions cannot be localized in a region smaller
than $\lambda^{-2+ \delta}$ for some $\delta > 0$ in $d=2$. Chen \cite{Ch},
extended this result to all
dimensions $d\ge 2$ and $\delta=0$ (with logarithmic corrections).
 Extended states 
for Schr\"odinger equation with a sufficiently decaying random potential were 
proven by Rodnianski and Schlag \cite{RS} and Bourgain \cite{B}  
(see also  \cite{D}).
However, all known results 
for Anderson-type models in Euclidean space
are  in regions where the dynamics
have typically finitely many effective collisions.
Under the  diffusive scaling of this paper, see (\ref{scale}), 
the number of effective scatterings is a negative fractional power of the
scaling parameter.  In particular, it goes to infinity in the scaling limit,
 as it should be the case if we aim to obtain a Brownian motion.

\medskip

{\it Acknowledgement.} The authors are grateful
for the financial support and hospitality of
 the Erwin Schr\"odinger Institut, Vienna,  Max Planck Institut, Leipzig,
Stanford University and Harvard University, where part of this
work has been done.

\section{Summary of Part I}
\setcounter{equation}{0}

\subsection{Notations}

We introduce a few notations.
The letters   $x,y,z$ will  denote configuration space variables, while
$p, q, r, u, v, w$ will be used  for $d$-dimensional momentum
variables. The norm  $\| \cdot \|$
 denotes the standard $L^2(\bR^d)$ norm and we set
$$
   \| f \|_{m,n} := \sum_{|\alpha|\leq n}
\| \langle x \rangle^m  \partial^\alpha f(x)\|_\infty
$$
with $\langle x \rangle : = (2+x^2)^{1/2}$ (here $\alpha$ is a 
multiindex).  The bracket
$(\cdot \, , \cdot)$ denotes the standard scalar product on $L^2(\bR^d)$
and $\langle \cdot \, , \cdot \rangle$ will denote the pairing
between the Schwarz space and its dual 
 on the phase space $\bR^d\times \bR^d$. The Fourier transform
is denoted by hat. For functions on the phase space, $f(x,v)$,
$x,v\in \bR^d$, Fourier transform will always be understood
in the first variable only:
$$
     \wh f(\xi, v): = \int e^{-2\pi i \xi\cdot x} f(x, v)\rd v \; .
$$
By the regularization argument in
Section~3.2 of \cite{ESYI}, we can cutoff
the high momentum regime of $\psi_0$ and
we can  multiply $B$ by a $\lambda$-dependent
cutoff function in momentum space at a negligible error.
 When dealing with
estimates for any fixed $\lambda>0$, we can thus
assume that 
\be
\mbox{supp} \, \wh\psi_0 \;\; \mbox{is compact}, \qquad
\mbox{supp} \, \wh B(p)\subset
\{ |p|\leq \lambda^{-\delta}\}
\label{suppM}
\ee
 for any fixed $\delta>0$.
Universal constants and constants that depend only on the
dimension $d$, on the final time $T_0$ and on $\psi_0$ and $B$
will be denoted by $C$.
 The same applies to the
hidden constants in the $O(\cdot)$ and $o(\cdot)$ notations.

\medskip

In \cite{ESYI}
the self-energy operator was defined as the multiplication operator
in momentum space
\be
 \theta(p) := \Theta(e(p)), \qquad    \Theta (\alpha): = \lim_{\e \to 0+ }
 \Theta_\e (\alpha) \; , \qquad \Theta_\e(\alpha) : = \Theta_\e
(\alpha, r)
\label{eq:thetalim}
\ee
for any $r$ with $e(r)=\alpha$, 
where
\be\label{theta}
    \Theta_\e (\alpha, r): =  \int \frac  { |\wh B(q-r)|^2
\rd q} { \alpha- e(q) + i \e} \; .  
\ee
The function $\Theta(\alpha)$ is H\"older continuous with exponent 
$\frac{1}{2}$, 
and  it 
decays as $\langle \a\rangle^{-1/2}$ (Lemma 3.1 and 3.2 in \cite{ESYI}).
If we write $\Theta(\a)=  {\cal R} (\a)- i{\cal I}(\a)$, 
where ${\cal R} (\a)$ and
${\cal I}(\a)$ are real functions,
and recall  $ Im (x+  i 0)^{-1} = - \pi \delta(x)$,  we have
\be
   {\cal I}(\a)=- \mbox{Im} \,\Theta (\a) = 
\pi\int \delta(e(q)-\a) |\wh B(q-r)|^2 \rd q
\label{eq:opt} \ee 
for any $r$ satisfying $\a=e(r)$.

The dispersion relation was renormalized by adding the self-energy term:
$$
      \om (p): = e(p) + \lambda^2\theta(p) \; ,
$$
and the Hamiltonian was rewritten as
\be\label{renH} H= H_0 + \wt V, \qquad
  H_0:=\omega(p), \qquad \wt V := \lambda
    V -\lambda^2 \theta(p)\; .
\ee
The renormalization compensates for the immediate recollisions
in the Duhamel expansion
(see Section \ref{sec:stopping}). The rate of the immediate
recollisions is of order $\lambda^2$, thus their total effect
is $\lambda^2t \gg1$. The renormalization removes this
instability. We note that  $\omega(p)$ is
not the self-consistently renormalized dispersion relation,
but only its approximation up to $O(\lambda^4)$.  
Since this error is negligible on
our time scale due to $\lambda^4 t\ll 1$, we use $\omega$ to simplify the
technicalities associated with the analysis of the self-consistent 
dispersion relation.

We define the renormalized propagator (with $\eta$-regularization):
$$
        R_\eta(\a, v):= \frac{1}{\alpha - \om(v) + i\eta} \; .
$$
In Appendix \ref{sec:propag} we prove the following estimates
 on the renormalized  propagator.

\begin{lemma}\label{le:opt}
Suppose that $\lambda^2 \ge \eta \ge \lambda^{2+ 4 \kappa}$ with
$\kappa \le 1/12$. Then we have,
\be
    \int \frac{|h(p-q)| \rd p}{|\alpha - \om(p)+ i\eta|}
    \leq \frac{C \| h\|_{2d,0}\, |\log\lambda| \; 
\log \langle \alpha\rangle}{\langle \alpha\rangle^{1/2}
 \langle |q| -\sqrt{2|\alpha|}\rangle } \; ,
\label{eq:logest}
\ee
and for $0\le a<1$
\begin{align}
     \int \frac{|h(p-q)| \rd p}{|\a -\om(p) + i\eta|^{2-a}}
 \leq & \frac{C_a \| h \|_{2d, 0}\,\lambda^{-2(1-a)}}{\langle \alpha\rangle^{a/2}
\langle |q| -\sqrt{2|\alpha|}\rangle } \; ,
\label{eq:2aint}
\\
  \int \frac{|h(p-q)| \rd p}{|\a -e(p) + i\eta|^{2-a}}
 \leq & \frac{C_a \| h \|_{2d, 0}\,\eta^{-2(1-a)}}{\langle \alpha\rangle^{a/2}
\langle |q| -\sqrt{2|\alpha|}\rangle } \; .
\label{eq:3aint}
\end{align}
For $a=0$ and with $h:= \wh B$, the following more precise estimate holds.
There exists a  constant $C_0$, depending only
on finitely many $\| B\|_{k,k}$ norms,  such that
\be
  \int \frac{ \lambda^{2}|\wh B(p-q)|^2 \; \rd p 
 }{|\alpha-\ov\om(p)-i\eta|^{2}}  \leq
    1+ C_0\lambda^{-12\kappa}\big[\lambda + |\a- \om(q)|^{1/2}\big] \, .
\label{eq:ladderint}
\ee
\end{lemma}

\subsection{The expansion}\label{sec:stoprule}

We expand the unitary kernel of $H= H_0 + \wt V$ (see \eqref{renH})
by the Duhamel formula and after taking the expectation,
we organize the expansion into sums of Feynman diagrams.
In order to avoid the infinite summations \eqref{def:muom}
in the expansion, we have
reduced the problem to a large
finite box, $\Lambda_L=[-L/2, L/2]^d\subset \bR^d$
with periodic boundary conditions (see  Section 3.3 of \cite{ESYI}).
The infinite volume Poisson process $\mu_\om$ was replaced with
$$
  \mu_\om' = \sum_{\gamma=1}^M v'_\gamma \delta_{y_\gamma'}
$$
where $M$ is a Poisson random variable with mean $|\Lambda_L|$,
the points $\{ y_\gamma'\}_{\gamma=1}^M$ are uniformly distributed
on $\Lambda_L$ and the real charges $v_\gamma'$  have distribution $\bP_v$.
All random variables are independent. 
 Lemma 3.4 of \cite{ESYI} guarantees that these modifications
 have no effect on the final
result if  $L\to\infty$ is taken before
any other limit. 

After the Duhamel expansion, taking the expectation and letting
 $L\to\infty$, we regain the infinite volume formulas for
the Feynman graphs. In \cite{ESYI} we used primes
to denote the restricted quantities, but to avoid the heavy
notation here we will omit them, except when stating
the theorems. All quantities throughout this section are
understood on $\Lambda_L$ with $M$ random points.

We recall the Duhamel formula  from Section 4 of \cite{ESYI}.
For any fixed integer $N\ge 1$
\begin{equation}\label{duh}
        \psi_t : = e^{-itH}\psi_0 = \sum_{n=0}^{N-1} \psi_n (t)
     + \Psi_{N}(t) \; ,
\end{equation}
with
\begin{align}
        \psi_n(t) : = & (-i)^n\int_0^t [\rd s_j]_1^{n+1} \; \;
    e^{-is_{n+1}H_0}\wt V e^{-is_nH_0}\wt V\ldots
       \wt V e^{-is_1 H_0}\psi_0
\label{eq:psin}
\\
         \Psi_{N} (t): = & (-i) \int_0^t \rd s \, e^{-i(t-s)H}
   \wt V \psi_{N-1}(s)
\label{eq:PsiN}
\end{align}
with the notation
$$
    \int_0^t [\rd s_j]_1^n : = \int_0^t\ldots \int_0^t
 \Big(\prod_{j=1}^n \rd s_j\Big)
        \delta\Big( t- \sum_{j=1}^n s_j\Big) \; .
$$
Substituting  $\wt V=
-\lambda^2\theta(p)+\sum_{\gamma=1}^M \lambda
V_\gamma$ with $V_\gamma(x):=v_\gamma
B(x-y_\gamma)$, 
  the terms in (\ref{eq:psin}) and (\ref{eq:PsiN})
are summations over 
 collision history.
Denote by
$\tGamma_n$, $n\le\infty$,  the set of sequences
\be\label{Gamman}
\tgamma = (\tgamma_1, \tgamma_2, \ldots , \tgamma_n), \qquad
\tgamma_j\in \{ 1, 2, \ldots , M\}\cup \{ \vartheta\}
\ee
and by $ W_\tgamma$ the associated  potential
$$
        W_\tgamma :=  \left\{ \begin{array}{cll} \lambda V_\tgamma & \qquad
\mbox{if} \quad & \tgamma\in \{ 1, \ldots, M\} \\
     - \lambda^2 \theta(p)  & \qquad \mbox{if} \quad & \tgamma =\vartheta
 \; . \end{array} \right.
$$
The tilde refers to the fact that the additional $\{ \vartheta\}$
symbol is also allowed. An element $\tgamma_j$ of the
sequence $\tgamma\in \tGamma$ is called {\it potential label}
and $j$ is called {\it potential index} if $\tgamma_j\in \{1, 2, \ldots M\}$,
otherwise they are called  {\it $\vartheta$-label}
and {\it $\vartheta$-index}, respectively.
 A potential label carries a factor $\lambda$,
a $\vartheta$-label carries $\lambda^2$.

For any $\tgamma\in\tGamma_n$ we define the
following fully expanded wave function with truncation
\be
    \psi_{*t, \tgamma}: = (-i)^{n-1}\int_0^t [\rd s_j]_1^{n}
 \; \; W_{\tgamma_n}
    e^{-is_nH_0} W_{\tgamma_{n-1}} \ldots
    e^{-is_2H_0} W_{\tgamma_1} e^{-is_1H_0} \psi_0
\label{eq:trunc}
\ee
and without truncation
\be
    \psi_{t, \tgamma}: =  (-i)^{n}\int_0^t [\rd s_j]_1^{n+1} \; \;
    e^{-is_{n+1}H_0} W_{\tgamma_n}
    e^{-is_nH_0} W_{\tgamma_{n-1}} \ldots
    e^{-is_2H_0} W_{\tgamma_1} e^{-is_1H_0} \psi_0\; .
\label{eq:exp}
\ee
In the notation the star $(*)$ will always refer to truncated functions.
Each term $\psi_{t, \tgamma}$
 along the expansion procedure is 
 characterized by its order
$n$ and by a sequence $\tgamma\in\tGamma_n$. The main terms
are given by {\it non-repetitive} sequences
 that contain only potential labels, i.e. we define
\be
    \Gamma_k^{nr}:=\Big\{
    \gamma = (\gamma_1, \ldots , \gamma_k) \; : \;
   \gamma_j\in\{1, \ldots, M\}, \;
 \gamma_i\neq \gamma_j \; \mbox{if} \; i\neq j\Big\}  \subset \tGamma_n\; .
\label{tg}
\ee
The sum of the corresponding elementary wave functions is denoted by
\be
   \psi_{t,k}^{nr}:= \sum_{\gamma\in \Gamma_k^{nr}} \psi_{t,\gamma} \; .
\label{psinr}
\ee
The rate of collisions is $O(\lambda^2)$, thus the total
number of collisions is typically of order $k\sim \lambda^2t$.
We thus set
\be
     K := [\lambda^{-\delta}(\lambda^2 t)]\; 
\label{def:K}
\ee
($[ \; \cdot \; ]$ denotes integer part)
to be an upper threshold for the number of collisions
in the expanded terms. Here
 $\delta=\delta(\kappa)>0$ is 
a small positive number to be fixed later on.

\subsection{Structure of the proof.}

In Section 5 of \cite{ESYI} the Main Theorem was 
proved from three key theorems. For completeness, we repeat these
three statements here. 
Recall that the prime indicates restriction to
$\Lambda_L$ and hence  dependence on $L,M$

\begin{theorem} [$L^2$-estimate of the error terms]\label{7.1}
Let $t=O(\lambda^{-2-\kappa})$ and $K$ given by \eqref{def:K}.
If $\kappa < \kappa_0(d)$ and $\delta$ is sufficiently small (depending only
on $\kappa$),  then
$$
   \lim_{\lambda\to0}\lim_{L\to\infty}\bE' 
\Big\| \psi_t' -\sum_{k=0}^{K-1} \psi_{t,k}^{\prime \; nr} \Big\|^2_L =0 \; .
$$
In $d=3$ dimensions, one can choose $\kappa_0(3)=\frac{1}{500}$.
\end{theorem}

\begin{theorem} [Only the ladder diagram contributes] \label{thm:L2}
Let $\kappa<\frac{2}{34d+39}$,  $\e=\lambda^{2+\kappa/2}$,
$\eta=\lambda^{2+\kappa}$,
 $t=O(\lambda^{-2-\kappa})$,
 and $K$ given by \eqref{def:K}.
 For a sufficiently small 
positive $\delta$ and for  any $1\leq k\leq K$ we have
\be
   \lim_{L\to\infty} \bE' \|\psi_{t,k}^{\prime \; nr}\|^2_L=  V_\lambda(t, k)
    + O\Big( \lambda^{\frac{1}{3}-
(\frac{17}{3}d+\frac{13}{2})\kappa-O(\delta)}\Big)
\label{eq:L2bound}
\ee
\be  
\lim_{L\to\infty} \langle \wh \cO_L, \bE' \wh W^\e_{\psi_{t,k}^{\prime \; nr}}
\rangle_L=  W_\lambda(t, k, \cO)
    + O\Big( \lambda^{\frac{1}{3}-
(\frac{17}{3}d+\frac{13}{2})\kappa-O(\delta)}\Big)
\label{eq:Wbound}
\ee
as $\lambda\ll 1$. Here
\begin{align}
   V_\lambda(t, k): = & \frac{\lambda^{2k}e^{2t\eta}}{(2\pi)^2}
  \iint_{-\infty}^\infty \rd\a\rd\beta  \; e^{i(\a-\beta)t} 
 \int\Big(  \prod_{j=1}^{k+1}  \rd p_j \Big) \; 
|\wh\psi_0(p_1)|^2
\nonumber\\
&\times \prod_{j=1}^{k+1} \ov{ R_\eta(\a, p_j)} R_\eta(\beta, p_j)
  \prod_{j=1}^k |\wh B(p_{j+1}-p_j)|^2 \; 
\label{ladder} 
\end{align}
\begin{align}
W_\lambda(t, k, \cO): = & \frac{\lambda^{2k}e^{2t\eta}}{(2\pi)^2}
  \iint_{-\infty}^\infty\rd\a\rd\beta  \; e^{i(\a-\beta)t} \int \rd \xi
\int  \Big( \prod_{j=1}^{k+1} \rd v_j \;  \Big)
\wh\cO(\xi, v_{k+1})\overline{\wh W_{\psi_0}^\e}(\xi, v_1) 
  \nonumber \\
 & \times 
\prod_{j=1}^{k+1} 
  \ov{R_\eta\Big(\a, v_j +\frac{\e\xi}{2}\Big)}
   R_\eta\Big(\beta, v_j -\frac{\e\xi}{2}\Big)
  \prod_{j=1}^k |\wh B(v_j-v_{j+1})|^2\;.
\label{9.511}
\end{align}
\end{theorem}
The definition \eqref{9.511} does not apply literally
 to the free evolution term $k=0$; this term is defined
separately: 
\be
   W_\lambda(t, k=0,\cO)  : = \int \rd\xi\rd v\;
    e^{it\e v\cdot \xi}\; e^{2t\lambda^2 {\scriptsize \mbox{Im}}\,
 \theta (v)}\;
 \wh\cO(\xi, v)\ov{\wh W_0}(\e\xi, v) \; .
\label{xi0}
\ee
\begin{theorem} [The ladder diagram converges to the heat equation]
\label{thm:laddheat}
Under the conditions of Theorem~\ref{thm:L2} and setting
$t=\lambda^{-2-\kappa}T$, we have 
\be
  \lim_{\lambda\to0} \sum_{k=0}^{K-1} W_\lambda(t, k,\cO) = 
  \int \rd X  \int \rd v \; \cO(X, v)f(T, X, e(v))\; ,
\label{eq:laddheat}
\ee 
where $f$ is the solution to the heat equation \eqref{eq:heat}.
\end{theorem}

The main result of \cite{ESYI} was the proof of 
Theorem \ref{thm:L2}. In Sections~\ref{sec:stopping}--\ref{sec:cases}
 of this paper we prove
Theorem~\ref{7.1} and in Section~\ref{sec:wigner}
we prove Theorem~\ref{thm:laddheat}. 
We now explain the ideas behind these theorems. 
The actual estimates are somewhat weaker
than the heuristics predicts.

The proof of Theorem~\ref{7.1} consists in controlling
the wave functions of collision histories that contain $\vartheta$-labels
or repeated potential labels. The repeated potential labels
correspond to recollisions with the same obstacle. 
Immediate recollision with the same obstacle occurs
with an amplitude $O(\lambda^2)$. Over the total history
of the evolution, this would yield a large
contribution of order $\lambda^2 t\gg1$.
However, the wave functions of these
collision histories will be resummed with those
containing $\vartheta$-labels. Thanks to the choice of 
the renormalization counterterm, $\lambda^2\theta(p)$,
the contributions of the immediate recollisions and  the $\vartheta$-labels
cancel each other up to leading order. 
Each resummed term
thus has an amplitude $O(\lambda^4t)$.
The full propagator in the
 error term, $\Psi_N(t)$ \eqref{eq:PsiN}, however,
will be estimated by unitarity \eqref{unitar}. This estimate effectively
loses an extra $t$ factor. To compensate for it, we
have to continue the expansion up to two immediate recollisions
in the error term.

The amplitudes of the non-immediate recollisions are much
smaller  and they can be estimated individually. 
Heuristically, the probability of such recollisions can be understood  
in classical mechanics.
Since the mean free path is $\lambda^{-2}$, returning to
an already visited obstacle after visiting another obstacle
at distance $\lambda^{-2}$
has probability $O(\lambda^{2d})$. Another scenario
is when the particle collides with  obstacle $\gamma_1$,
then it bounces back from 
a nearby obstacle $\gamma_2$, $|\gamma_1-\gamma_2|=O(1)$,
and then it recollides with $\gamma_1$.
This situation is atypical and it is penalized
by $O(\lambda^4)$ because the time elapsed between these
collisions is $O(1)$ while the collision cross-section
is $O(\lambda^{2})$.  
In conclusion, the probability of a non-immediate recollision
among the total $k\sim \lambda^2 t$ collisions
is at most $O(k\lambda^4)$, thus the total effect
of these recollisions on our time scale is negligible, 
even when multiplied with the additional factor $t$ from 
the unitarity estimates for $\Psi_N(t)$.

This outline neglects the key analytic difficulty originated from the 
growth of the combinatorics 
of Feynman diagrams. The amplitude
of non-repetition wave functions can be written as a sum of $k!$
Feynman diagrams. Only one of them, the ladder diagram,
contributes to the heat equation. All other diagrams can be estimated
by  $O(\lambda^2)$ due to phase cancellations.
This estimate  is not sufficient to sum up
all  diagrams since their number, $k!\sim \exp(\lambda^{-const})$,  is 
exponentially large.
The size of a few combinatorially simple diagrams
is indeed $O(\lambda^2)$, but much stronger estimates
were  obtained in \cite{ESYI}
as the combinatorial complexity of the diagram increases.
This improvement  balances the increased
combinatorial factor for more complicated
diagrams and it  allows
  us to control the expansion for non-repetition wave functions for  time 
scale
$t\sim \lambda^{-2-\kappa}$.

In this paper, we
extend the classification scheme  to include {\it all} diagrams arising from 
the Duhamel expansion.
Thanks to the stopping rules,
this part involves only a few extra collisions. The main idea of this 
paper is to design
a  surgery of Feynman diagrams so that a general diagram can be decomposed 
into
a repetition and a
non-repetition  part: the repetition part involves only a few variables 
and the
integration can  be estimated accurately; the non-repetition  part is 
reduced to
Theorem \ref{thm:L2}. The errors from the 
surgery are controlled by the small factors from the repetition
part. This renders all repetition diagrams negligible.
Thus we prove that among all diagrams only the ladder diagrams without
repetition  contribute to the final heat equation.
In practice, the scheme used in this 
paper
is much more complicated than is stated here. But this description 
gives  a 
good first idea.

Finally, the proof of Theorem~\ref{thm:laddheat} is a fairly explicit
but delicate calculation involving singular integrals.
The proof shows how the long time evolution of
the Boltzmann equation emerges from the ladder diagrams.

\section{The stopping rules}
\label{sec:stopping}

We use the Duhamel expansion to identify the non-repetition
error terms to be estimated in Theorem~\ref{7.1}.
This method allows for the flexibility that
at every new term of the expansion we perform the
separation into elementary waves, $\psi_{*s,\wt\gamma}$,
and we can decide whether we want to stop (keeping the full propagator
as in \eqref{eq:PsiN}) or we 
continue to expand that term further.
This decision  will depend on the collision history, $\wt\gamma$,
and it will be given by a precise algorithm, the stopping rules.
The key idea is that
 once the collision
history $\tgamma$ is ``sufficiently'' atypical, i.e.
it contains either atypical recollision or too many collisions,
we stop the expansion for that elementary
wave function immediately to reduce
the number and the complexity of the expanded terms.

Not every recollision is atypical. An immediate second
collision with the same obstacle contributes to the
main term; this is  actually the reason why the 
dispersion relation had to be corrected with the self-energy 
$\lambda^2\theta(p)$.

\begin{figure}
\begin{center}
\epsfig{file=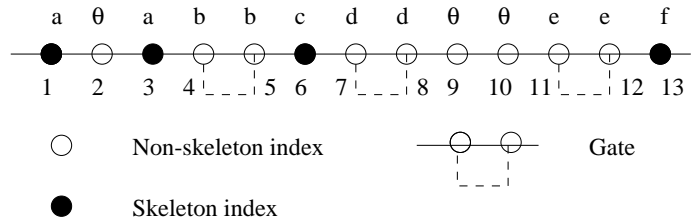, scale=.85}
\end{center}
\caption{Gates and skeletons}\label{fig:skel}
\end{figure}

In a sequence $\tgamma$
we thus identify the {\it immediate recollisions} 
 inductively starting from $\tgamma_1$
(due to their graphical picture, they are also called {\it gates}).
The gates must involve potential labels and not $\vartheta$.
For example,  the
sequence $\tgamma=(a, \vartheta, a, b,b, c, d,d,
\vartheta, \vartheta, e,e, f)$ has three
gates (see Fig.~\ref{fig:skel}).
 In the sequence $(a,b,b,c,c,c)$ there are  two gates. Any potential
label which does not belong to a gate will be called {\it skeleton
label}. The index $j$ of a skeleton label $\gamma_j$
in $\tgamma$ is called {\it skeleton index}. The set of skeleton indices is
$S(\tgamma)$.  Similar terminology is
 used for the gates.
 In the first example $1, 3,  6, 13$ are skeleton
indices and $a, a, c, f$ are skeleton labels, in the second example
$1, 6$ are skeleton
indices and $a, c$ are skeleton labels. The $\vartheta$ terms are never
part of the skeleton.

\noindent
This definition is recursive so we can
identify skeleton indices successively along the expansion procedure.
Notice that a skeleton index may become a gate index at
a later stage of the expansion, but never the other way around.

The formal definition is as follows. Let $I_n:=\{1, 2, \ldots , n\}$.

\begin{definition}[Skeleton labels and indices]\label{def:skeleton}
Let $\tgamma= (\tgamma_1, \tgamma_2, \ldots, \tgamma_n)\in\tGamma_n$
and let $\tgamma^*: = (\tgamma_1, \tgamma_2, \ldots ,
\tgamma_{n-1})$ be its truncation.
The set of skeleton  indices of $\tgamma$, $S(\tgamma)\subset I_n$,
is defined inductively (on the length of $\tgamma$)
as follows. If $\tgamma\in \tGamma_1$, then
$S(\tgamma) : =\{ 1\} $ if $\tgamma_1 \neq \vartheta$ and
$S(\tgamma) :=\emptyset$ otherwise. Furthermore, for
any $\tgamma\in \tGamma_n$, $n\ge 2$, let
$$
    S(\tgamma): =\left\{ \begin{array}{cll} S(\tgamma^*)& \qquad \mbox{if}
    \quad &\tgamma_n = \vartheta\\
                  S(\tgamma^*)\setminus \{ n-1\}& \qquad \mbox{if}
        \quad & n-1\in S(\tgamma^*)\quad \mbox{and}\quad\tgamma_n=\tgamma_{n-1} \\
                 S(\tgamma^*)\cup \{ n \} & \qquad \mbox{if} \quad &
    \tgamma_n \neq \vartheta \quad \mbox{and}\quad
             [\tgamma_n \neq \tgamma_{n-1} \quad \mbox{or}
    \quad n-1 \not\in S(\tgamma^*)]\; . \end{array} \right.
$$
Finally,  $\gamma_n$ is called skeleton label if $n\in S(\gamma)$.
\end{definition}

For any $\tgamma\in\tGamma_n$, let
$$
    k(\tgamma):= |S(\tgamma)|, \quad
$$
be the number of skeleton
indices in $\tgamma$,
let
$$
    I_n^\theta(\tgamma):=\{ j \; : \; \tgamma_j=\vartheta\}
$$
be the set of $\theta$-indices and $t(\tgamma): = |I_n^\theta(\tgamma)|$.
Denote the total $\lambda^2$-power collected from non-skeleton indices
by
\be\label{r}
    r(\tgamma):= \frac{1}{2}[n-k(\tgamma)] + t(\tgamma)
\ee
Notice that $r(\tgamma)$ is integer.

\medskip

The exact stopping rule requires somewhat tedious 
definitions of different types
of elementary wave functions. First we give these definitions
intuitively, 
state our final representation formula for $\psi_t$ 
 using these concepts,  then we
give the precise definitions and prove the formula.

\medskip

Sequences where the only repetitions in potential labels
occur within the gates are called {\it non-repetitive}
sequences. A special case is the set of non-repetitive
sequences, $\Gamma^{nr}_k$, without gates
and  $\theta$-labels.
 The repetitive sequences are divided into the following categories
(Fig.~\ref{fig:nest}).
If  two non-neighboring skeleton labels coincide,
then the collision history includes a 
{\it genuine (non-immediate) recollision}.
If a skeleton label coincides with a gate label, then we have a {\it triple
collision} of the same obstacle. If two neighboring skeleton labels coincide
and are not immediate recollisions because there are gates or $\vartheta$'s
in between, then we have a {\it nest}.

\bef
\bec
\epsfig{file=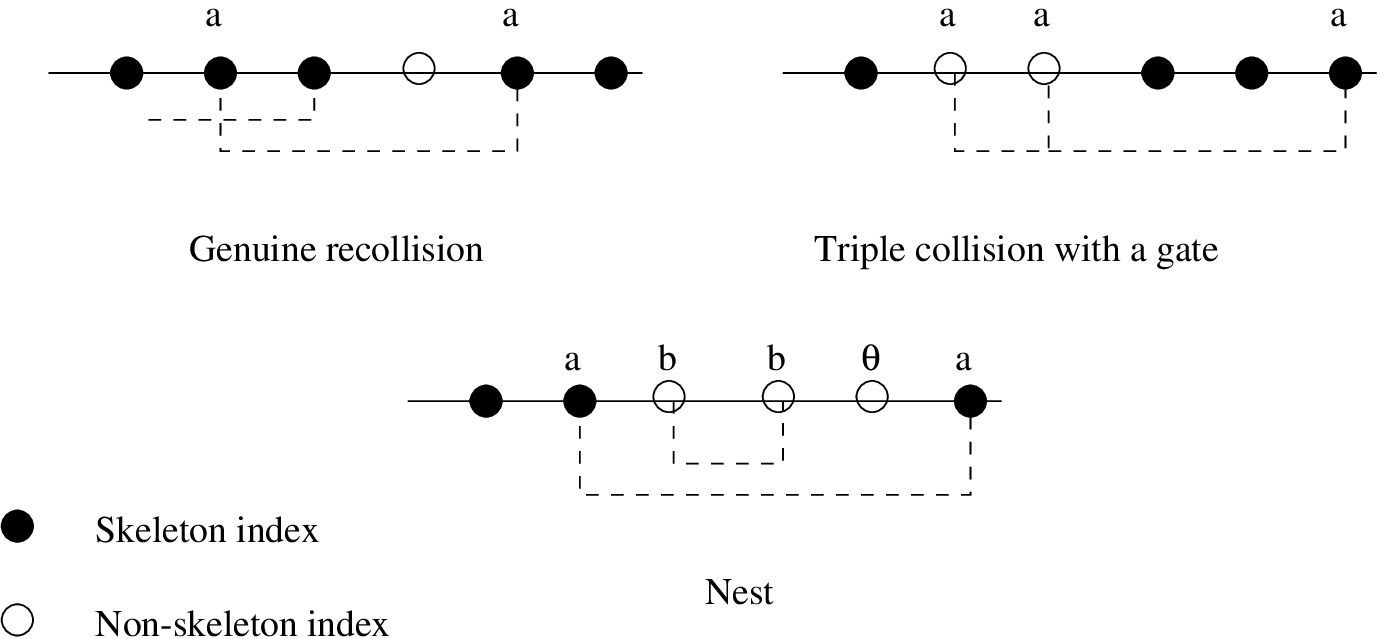,scale=.9}
\eec
\caption{Repetition patterns}\label{fig:nest}
\eef

We stop the expansion at an elementary truncated wave function (\ref{eq:trunc})
characterized by $\tgamma$,
if any of the following happens
(precise definitions will be given in Definition \ref{def:seqset}).

\medskip

$\bullet$  The number of skeleton indices  in $\tgamma$ reaches $K$
(see \eqref{def:K}).
We denote  the sum of the truncated elementary non-repetitive
wave functions up to time $s$ with at most one
$\lambda^2$ power from the non-skeleton labels or $\vartheta$'s
and with $K$ skeleton indices by
$\psi_{*s, K}^{(\leq 1),nr}$. The superscript $(\leq 1)$ refers
the number of collected $\lambda^2$ powers from
non-skeleton labels.

\medskip

$\bullet$  We have collected $\lambda^4$ from non-skeleton labels.
We denoted by
$\psi_{*s,k}^{(2),last}$ the sum of the truncated elementary wave functions
up to time $s$ with two
$\lambda^2$ power from the non-skeleton indices
(the word {\it last} indicates that the last $\lambda$ power was collected
at the last collision).

\medskip

$\bullet$  We observe a repeated skeleton label, i.e., a recollision
or a nest. The corresponding wave functions are denoted by
$\psi_{*s,k}^{(\leq 1),rec}, \psi_{*s,k}^{(\leq 1),nest}$.

\medskip

$\bullet$  We observe three identical
potential labels, i.e., a triple collision.
The corresponding wave functions are denoted by
$\psi_{*s,k}^{(\leq 1),tri}$.

\medskip

Finally, $\psi_{t,k}^{(\leq 1),nr}$ denotes the sum of non-repetitive
elementary wave functions without truncation (i.e. up to time $t$)
 with at most one
$\lambda^2$ power from the non-skeleton indices or $\vartheta$'s
and with $k$ skeleton indices.
In particular, the non-repetition
wave functions without gates and $\vartheta$'s 
(denoted by $\psi_{t,k}^{nr}$ above) contribute to this sum.
For notational consistence, we will rename $\psi_{t,k}^{(0),nr}:=
\psi_{t,k}^{nr}$ to explicitly indicate the number of $\lambda^2$-powers
collected from gates or $\vartheta$'s.

\medskip
This stopping rule gives rise to the following representation.

\begin{proposition}\label{prop:duh}[Duhamel formula] For any $K\ge 1$
we have
\be
\psi_t=e^{-itH}\psi_0 = \sum_{k=0}^{K-1} \psi_{t,k}^{(\leq 1),nr}\qquad
\qquad \qquad\qquad \qquad \qquad \qquad \qquad \qquad
\label{eq:duha}
\ee
$$
\qquad \qquad -i \int_0^t\rd s \; e^{-i(t-s)H}\Bigg\{
\psi_{*s, K}^{(\leq 1),nr}+\sum_{k=0}^{K}
\Big( \psi_{*s,k}^{(2),last} + \psi_{*s,k}^{(\leq 1),rec}+
\psi_{*s,k}^{(1),nest}+\psi_{*s,k}^{(1),tri}\Big)
\Bigg\} \; .
$$
\end{proposition}

{\it Proof of Proposition \ref{prop:duh}.}
We start with the precise definitions.
For $\tgamma\in \tGamma_n$ and $\ell<n$
we introduce the notation $\tgamma_{[1,\ell]}
:=(\tgamma_1, \ldots, \tgamma_\ell)$ to denote the
beginning segment, or truncation, of length $\ell$ of the sequence
$\tgamma$.

\begin{definition}[Sets of sequences]\label{def:seqset}
For $0\leq r\le 2$, $k\leq n$ we let
$$
        \tGamma_{n,k}^{(r)}: =  \{ \tgamma\in\tGamma_n\; : \;
        k(\tgamma)=k, \;  r(\tgamma)=r\}
$$
be the set of sequences with $k$ skeleton indices and
 $\lambda^{2r}$ collected from non-skeleton indices.
Let
$$
    \tGamma_{n,k}^{(r), nr}: = \{ \tgamma\in\tGamma_{n,k}^{(r)}
    \; :  \; [\tgamma_j =\tgamma_{j'}\neq \theta]
    \Longrightarrow [|j-j'|=1, j,j'\not\in S(\tgamma)]\}
$$
be the set of {\bf non-repetitive}
sequences. For $r=0$ we have $n=k$ and we set $\Gamma_k^{nr}: =  \tGamma_{k,k}^{(0), nr}$.
Let
$$
    \tGamma_{n,k}^{(r),last}: = \{ \tgamma\in\tGamma_{n,k}^{(r),nr}\; : \;
    \tgamma_n\not\in S(\tgamma)\}
$$
be the set of non-repetitive sequences whose last element is non-skeleton.
Let
$$
        \tGamma_n^{nr} : = \bigcup_{k\leq n} \bigcup_{r\leq 2} \tGamma_{n,k}^{(r), nr}
$$
be the set of all non-repetitive sequences of length $n$ and let
$$
        \tGamma^*_n: = \{\tgamma\in \tGamma_n\setminus \tGamma^{nr}_n\; : \; \tgamma_{[1, n-1]} \in
        \tGamma^{nr}_{n-1}\}
$$
be the set of sequences that are repetitive, but their proper truncations are  non-repetitive.
We let
$$
    \tGamma_{n,k}^{(r),tri}: =\tGamma^*_n\cap  \{ \tgamma\in\tGamma_{n,k}^{(r)}\; : \;
    \tgamma_n\in S(\tgamma), \;
    \exists j\leq n-2, \tgamma_j,\tgamma_{j+1}\not\in S(\tgamma),
    \tgamma_n= \tgamma_j = \tgamma_{j+1}\}
$$
be the set of {\bf triple-collision
sequences}, i.e. sequences whose last entry $\tgamma_n$
is a part of a triple collision with a gate.
Let
$$
    \tGamma_{n,k}^{(r), rec}: = \tGamma^*_n\cap
         \{ \tgamma\in\tGamma_{n,k}^{(r)}\setminus  \tGamma_{n,k}^{(r),tri} \; : \;
    \exists j\in S(\tgamma), j\le n-2, \tgamma_j = \tgamma_n,
    S(\tgamma)\cap \{ j+1, \ldots , n-1\}\neq \emptyset\}
$$
be the set of {\bf recollision sequences}.
Finally, let
$$
    \tGamma_{n,k}^{(r),nest}: = \tGamma^*_n\cap \{ \tgamma\in \tGamma_{n,k}^{(r)}
    \setminus ( \tGamma_{n,k}^{(r),rec} \cup \tGamma_{n,k}^{(r),tri}) \; : \;
    \exists j\in S(\tgamma), j\leq n-2, \tgamma_j=\tgamma_n\}
$$
be the set of {\bf nested sequences}.
In all cases we introduce the notation
$$
 \tGamma_{n,k}^{(\leq R),\#}: = \bigcup_{r=0}^R \tGamma_{n,k}^{(r),\#}
$$
where $\# =  tri, rec, nest, nr, last$ refers to the {\bf structure}
of the wave function.
\end{definition}

Notice that the last index $n$ of any $\tgamma \in  \tGamma^*_n$ is a skeleton index,
in particular  $\tGamma_{n,k}^{(r),last}\cap \tGamma^*_n =\emptyset$.
This index can create a repetition in  three different ways:
triple collision, recollision or nest.
It is  therefore clear from the definition,
that the sets
$\tGamma_{n,k}^{(r),tri}$,  $\tGamma_{n,k}^{(r),rec}$,
$\tGamma_{n,k}^{(r),nest}$ and $\tGamma_{n,k}^{(r),nr}$
for $0\leq r\leq 2$ are disjoint. Moreover, for triple collision and nested sequences
we have $r\ge 1$.
The next lemma shows that these sets include the appropriate
beginning segment of any infinite sequence.

\begin{lemma}\label{lemma:stop}
Let a positive integer $K$ be given.
Let $\tgamma =(\tgamma_1, \tgamma_2,\ldots )
\in\tGamma_\infty$ be an infinite sequence.
Then there exist a unique $k\leq K$
 and $n \in [k, k+4]$ such that the truncation of length $n$
of $\tgamma$, $\tgamma_{[1,n]}$ belongs to
the (disjoint) union
\be
     \tGamma(n,K):= \tGamma_{n,K}^{(\leq 1),nr} \cup
         \bigcup_{k=0}^{K}\Bigg(  \tGamma_{n,k}^{(1),tri}\cup \tGamma_{n,k}^{(\leq1),rec}
        \cup \tGamma_{n,k}^{(1),nest}\cup \tGamma_{n,k}^{(2),last}
        \Bigg)\; .
\label{eq:union}
\ee
\end{lemma}

{\it Proof.} We look at
the increasing family of truncated sequences $\tgamma_{[1,2]}$,
$\tgamma_{[1,3]}, \ldots$ inductively. If
$\tgamma_{[1,n]} \in \tGamma_{n,K}^{(r), nr}$ for some $n$ and $r\leq 1$,
then  it falls into the first set of (\ref{eq:union}).

Otherwise there is an $n\leq K+4$ such that $\tgamma_{[1,n-1]}$
is non-repetitive with $r\leq 1$, but $\tgamma_{[1,n]}$
is either repetitive or  $r(\tgamma_{[1,n]})=2$. If it is repetitive, then
$\tgamma_n$ is a skeleton index, so $r(\tgamma_{[1,n]})=r(\tgamma_{[1,n-1]})\leq 1$,
and the repetition can be a triple collision, recollision or nest with $k\leq K$.
If $\tgamma_{[1,n]}$ is non-repetitive, then the $r$ has increased from
$r(\tgamma_{[1,n-1]})=1$ to $r(\tgamma_{[1,n]})=2$ and $\tgamma_n$ is non-skeleton.
These four possibilities correspond to the remaining sets in (\ref{eq:union}).
The disjointness of these sets follows from their  definition.
$\Box$

\bigskip

For $0\leq r\leq 2$ and $\#= rec, nest,tri, last, nr$, let
$$
    \psi_{(*)t,k}^{(r), \#}: = \sum_{n=k+r}^{k+2r}\sum_{\tgamma\in
    \tGamma_{n,k}^{(r),\#}} \psi_{(*)t,\tgamma}
$$
be the wave function with $k$ skeleton labels, with $\lambda^{2r}$
total power collected from non-skeleton terms
and with recollision, nest, triple collision
or no repetition (with the last collision being skeleton or not)
specified  by $\#$.
The notation $(*)$ indicates that the same
definition is used  for the wave functions with or without truncation.
Finally we set
$$
    \psi_{(*)t, k}^{(\leq 1), \#}:=
    \psi_{(*)t, k}^{(0), \#}  + \psi_{(*)t, k}^{(1), \#} \; .
$$

\bigskip

We stop the expansion at
 the  elementary truncated wave function (\ref{eq:trunc})
characterized by $\tgamma\in \tGamma_n$, if $\tgamma$
falls into one of the sets in (\ref{eq:union}), but none
of its proper truncations $\tgamma_{[1,n']}$
fell into the appropriate sets (\ref{eq:union})
with $n$ replaced with $n'$.
Lemma \ref{lemma:stop} shows that the expansion is stopped for every term
for a unique reason.
This procedure proves Proposition \ref{prop:duh}. $\;\;\Box$

\section{Error terms}\label{sec:error}
\setcounter{equation}{0}

The content of Theorem~\ref{7.1} is that
the main contribution to the wave function $\psi_t$ in (\ref{eq:duha})
comes from the fully expanded non-recollision terms with $r=0$, i.e
$\psi_{t,k}^{(0), nr}$. Here we show that the contribution of all
other terms are negligible. Each error term in (\ref{eq:duha})
has a specific reason to be small.
The result can be summarized in the following Theorem which is
proven in  Sections~\ref{sec:error} and \ref{sec:cases}.
We recall that prime indicates restriction to $\Lambda_L$.

\begin{theorem}\label{thm:error}
We assume $t=\lambda^{-2-\kappa}T$, $T\in [0, T_0]$, 
 and $1 \leq k\leq K$.
If $\kappa < \frac{2}{34d+39}$, 
 then
\be
   \lim_{L\to\infty}
 \bE' \| \psi_{*t, k}^{\prime\; (r), \#} \|^2 
= o(\lambda^{4 + 2\kappa +2\delta }),
\qquad \lambda\to0,
\label{eq:noladerr}
\ee
for the following choices of parameters:
$\{ \#=rec, \; r=0,1\}$,   $\{\#=nest, tri, \; r=1\}$ or
  $\{ \#=last, \; r=2\}$.
Furthermore, for $k=K$ and $r=0,1$,
\be
    \lim_{L\to\infty} \bE' \| \psi_{*t, K}^{\prime\; (r), nr} \|^2 
= o(\lambda^{4 + 2\kappa +2\delta}) \; ,
\label{eq:laderr}
\ee
and 
for $k<K$,
\be
    \lim_{L\to\infty} \bE' \| \psi_{t, k}^{\prime\;(1), nr} \|^2 
= o(\lambda^{2\kappa+2\delta}) \; .
\label{eq:laderr1}
\ee
\end{theorem}

{\it Proof of Theorem~\ref{7.1} using Theorem
\ref{thm:error}. }  By (\ref{eq:duha}) and the unitarity of
the operator $e^{-i(t-s)H}$, we have
\be
   \bE' \Big\| \int_0^t \rd s \; e^{-i(t-s)H} \sum_{k\leq K}
 \psi_{*s, k}^{\prime\;(r), \#} \Big\|^2
   \leq t^2K \sum_{k=0}^{K}\; 
\sup_{0\leq s\leq t}\bE' \Big\| \psi_{*s, k}^{\prime\;(r), \#} \Big\|^2 \; ,
\label{unitar}
\ee
where the value of $r$ is determined by $\#$ according to
the terms on the right hand side of (\ref{eq:duha}).
The non-repetition terms with $r=1$ (first term on
the right hand side of  (\ref{eq:duha})) are fully expanded and
there is no need for unitarity. After a Schwarz inequality,
$$
    \bE' \Big\| \sum_{k=0}^{K-1} \psi_{t,k}^{\prime\; (1),nr}\Big\|^2
 \leq K \sum_{k=0}^{K-1} \bE'\| \psi_{t,k}^{\prime\;(1),nr}\|^2.
$$
Given Theorem~\ref{thm:error}, all these error terms are negligible
if we first take $L\to\infty$ and then
 $\lambda\to0$.
 $\;\;\Box$

\medskip

The natural way to prove  Theorem~\ref{thm:error} would be to 
rewrite
$\bE\|\psi\|^2$ into a sum of Feynman graphs, similarly to 
Proposition 7.2 in \cite{ESYI}, 
then
to identify a repetition subgraph of a few vertices (recollisions, nests etc.)
that renders the graphs small and remove them by graph surgeries
after extracting a small factor.
We then should sum up the remaining graphs on the core indices
for all possible permutations and lumps as in Section 9 of
\cite{ESYI}.
For the sake of technical simplifications, however,
at the price of a smaller $\kappa$ we follow a somewhat different path. 
We first identify the vertices in the graph (called {\it core indices})
that carry no complication whatsoever (no repetition, no gate, no $\theta$).
We then symmetrize the non-core indices in $\psi$ and $\bar \psi$, 
by a Schwarz inequality to reduce the number of repetition patterns.
For graphs with sufficient high combinatorial complexity
(with large joint degree, see Definition~\ref{def:deg} below)
 we simply neglect the possible gain from the repetition
patterns by removing them from the graph with a crude estimate.
The necessary small factor will come from Proposition 9.2
of \cite{ESYI} with $q$ being large.
For graphs with low complexity,  the gain comes from 
analyzing the repetition patterns case by case. Since 
there are not too many low-complexity graphs, we can neglect 
the possible gain from the complexity and still sum up
for all combinatorial patterns of the core indices.

\subsection{Feynman graphs and their values}\label{sec:feyndef}

In this section we collect the necessary definitions
from \cite{ESYI} to estimate the values of Feynman graphs.
More details can be found in  Sections~7.1 and 7.2 of \cite{ESYI}.

The Feynman graph is an oriented circle graph on $N\ge2$
vertices and with two distinguished vertices, denoted by $0$, $0^*$.
The number of vertices between $0$ and $0^*$ are $n$ and $n'$,
in particular $N=n+n'+2$.
The vertex set can  thus be identified with the set $\cV=\cV_{n,n'}:=
\{0, 1, 2, \ldots, n, 0^*, \tilde n', \wt{n'-1},
 \ldots ,\tilde 1 \}$ equipped with the circular ordering.
We set $I_n:=\{ 1, 2, \ldots n\}$
and $\wt I_{n'} :=\{ \wt 1, \wt 2, \ldots, \wt n'\}$.
The set of oriented edges, $\cL(\cV)$, can be 
partitioned into $\cL(\cV)=\cL\cup
\wt\cL$ so that $\cL$ contains the edges between $I_n\cup \{0,0^*\}$
and $\wt\cL$ contains the edges between $\wt I_{n'}\cup \{0,0^*\}$.

For $v\in \cV$ we use the notation $v-1$ and  $v+1$ for
the vertex right before and after $v$ in the circular ordering.
We also denote $e_{v-}=(v-1,v)$ and $e_{v+}=(v,v+1)$ 
the edge right before and after
the vertex $v$, respectively.  
For each $e\in\cL(\cV)$, we introduce a momentum $w_e$ and
a real number $\alpha_e$ associated to this edge. The
collection of all momenta is denoted by $\bw=\{ w_e \; :\; e\in \cL(\cV)\}$
and $\rd\bw = \otimes_e \rd w_e$ is the Lebesgue measure.
The notation $v\sim e$ will indicate
that an edge $e$ is adjacent to a  vertex $v$.

Let $\bP = \{ P_\mu \; : \; \mu \in I\}$ be a
 partition of the set $\cV \setminus\{0, 0^*\}=I_n\cup \wt I_{n'}$
into nonempty, pairwise disjoint sets,
where $I=I(\bP)$ is the index set to label the sets in
the partition. Let $m(\bP):= |I(\bP)|$. The sets $P_\mu$ are called
$\bP$-{\it lumps} or just {\it lumps}.  
We  assign an auxiliary variable, $u_\mu\in \bR^d$, $\mu \in I(\bP)$,
to each lump. 
The vector of auxiliary momenta is denoted by
 $\bu: = \{ u_\mu\; : \; \mu\in I(\bP)\}$.
We will always assume that they add up to zero
\be 
\sum_{\mu\in I(\bP)} u_\mu=0 \; 
\label{sumu}
\ee
and that they  satisfy $|u_\mu|\leq O(\lambda^{-2\kappa-4\delta})$.
 The set of all partitions of the vertex
set $\cV \setminus\{0, 0^*\}$ 
is denoted by $\cP_\cV$. When we wish to indicate
the $n,n'$ dependence and identify
 $\cV \setminus\{0, 0^*\}$ with $I_n \cup \wt I_{n'}$,
then the set of all partitions on $I_n \cup \wt I_{n'}$
will be denoted by $\cP_{n,n'}$ instead of $\cP_\cV$.

For any $P\subset\cV$, we
let
$$
L_+(P) : = \{ (v,v+1)\in\cL (\cV)\; : \; v+1\not\in P, \; v\in P\}
$$
denote the set of  edges that go out of $P$, with respect to the
orientation of the circle graph, and similarly
 $L_-(P)$ denote the set of  edges that go into $P$. 
We set $L(P):= L_+(P) \cup L_-(P)$.

For any $\xi\in\bR^d$ we define the following product of delta functions
 \be
   \Delta(\bP, \bw, \bu): = 
\delta \Big(  \xi+\sum_{e\in L_\pm( \{ 0^*\} )} \pm w_e\Big)
\prod_{\mu\in I(\bP)} \delta\Big( \sum_{e\in L_\pm(P_\mu)}
    \pm w_e - u_\mu\Big) \; .
\label{def:Delta}
\ee
The sign $\pm$ indicates that momenta $w_e$ is added or subtracted
depending whether the edge $e$ is outgoing or incoming, respectively.

\medskip

For each subset $\cG\subset \cV\setminus \{ 0, 0^*\}$,
 we define
\be
   \cN_{\cG}(\bw): = \prod_{e\sim 0}|\wh \psi_0(w_e )|
 \prod_{v\in \cV \setminus\{ 0, 0^*\}\setminus\cG} 
|\wh B(w_{e_{v-}} - w_{e_{v+}})|  \prod_{v\in \cG} 
\langle w_{e_{v-}} - w_{e_{v+}} \rangle^{-2d} \; .
\label{def:cN}
\ee
We also  define the restricted Lebesgue measure 
\be
  \rd\mu(w): = {\bf 1}(|w|\leq \zeta) \rd w\; ,
  \quad
  \zeta:= \lambda^{-\kappa-3\delta}, 
 \qquad \rd \mu(\bw) : = \otimes_e \rd \mu (w_e) \; .
\label{def:mu}
\ee
On the support of $\Delta$
this restriction will not substantially influence our
 integrals (see (7.9)--(7.10) of \cite{ESYI}).

With these notations, we define,
 for any $\bP\in \cP_\cV$ and $g=0,1,2,\ldots$, the 
{\bf $E$-value of the partition}
\be
    E_g(\bP, \bu,\balpha): =\lambda^{N-2} \sup_{\cG \; : \; |\cG|\leq g}
    \int \rd\mu(\bw)\prod_{e\in \cL(\cV)}   \frac{1}{|\alpha_e- \om(w_e)+i\eta|}
   \; \Delta(\bP, \bw, \bu) \cN_{\cG}(\bw)  \; .
\label{def:E}
\ee
The $E$-value depends also on the parameters $\lambda, \eta$, 
but we will not specify them in the notation.
We will also need a {\bf truncated version} 
 of this definition:
\be
    E_{*g}(\bP, \bu,\balpha): =\lambda^{N-2}\sup_{\cG \; : \; |\cG|\leq g}
    \int \rd\mu(\bw)\prod_{e\in \cL(\cV)\atop
      e\not\in L( \{ 0^*\} )}   \frac{1}{|\alpha_e- \om(w_e)+i\eta|}
    \; \Delta(\bP, \bw, \bu) \cN_\cG(\bw)\; .
\label{def:E*}
\ee

Let $\bP\in\cP_{n,n'}$ be a partition
on the set $I_n \cup \wt I_{n'}$. The lumps of a partition
containing only one vertex will be called {\it single lumps.}
The vertices $0$ and $0^*$ 
 will not be considered single lumps.
Let $G=G(\bP)$ be the set of edges that go into  a single lump 
 and let $g(\bP):= |G(\bP)|$ be its cardinality.
In case of $n=n'$, we will use the shorter 
notation $\cV_n=\cV_{n,n}$, $\cP_n=\cP_{n,n}$
etc. We will always have
\be
|n-n'|\leq g(\bP) \leq 4, \qquad n,n'\leq K .
\label{nn'}
\ee

We also introduce a function $Q$ that will represent 
the momentum dependence of the observable.
We can assume, for convenience, that $\|Q\|_\infty\leq 1$.
We define
\begin{align}
   \cM(\bw) :=  & \prod_{e \in \cL\cap G}[- \ov{\theta(w_e)}]
 \prod_{e \in \wt\cL\cap G}[- \theta(w_e)] \; 
\prod_{e \in \cL\setminus G\atop e\not\sim 0^*}  \ov{\wh B(w_e- w_{e+1})} \;
\prod_{e \in \wt\cL\setminus G\atop e\not\sim 0}  \wh B(w_e- w_{e+1}) \;  
\label{def:cM}  
\\
& \times\ov{\wh\psi(w_{e_{0+}})}\wh\psi(w_{e_{0-}}) Q\Big[ \frac{1}{2}
  (w_{e_{0^*-}} +w_{e_{0^*+}}) \, \Big] \nonumber
\end{align}
where 
  $e+1$ denotes the edge succeeding $e\in \cL(\cV)$ in
the circular ordering.

Let $\a, \beta\in\bR$, $\bP\in \cP_{n,n'}$, and define
\be
   V(\bP,\a,\beta): =\lambda^{n+n'+g(\bP)}
\int \rd\mu(\bw)
    \prod_{e\in\cL} \frac{1}{\alpha- \ov\om(w_e) - i\eta} \prod_{e\in\wt\cL}
        \frac{1}{\beta- {\om}(w_e) + i\eta}
\label{def:Vlong}
\ee
$$
    \times  \Delta(\bP, \bw, \bu\equiv 0) \cM (\bw)\; .
$$
The truncated version, $V_{*}(\bP,\a,\beta)$,
is defined analogously
but the $\alpha$ and $\beta$ denominators that
 correspond to $e \in L( \{ 0^*\} )$ are removed.

We set $Y:=\lambda^{-100}$ and define   
\be
       V_{(*)}(\bP): = \frac{e^{2t\eta}}{(2\pi)^2}\iint_{-Y}^Y
        \rd\alpha\rd\beta \; e^{it(\alpha-\beta)}  V_{(*)}(\bP,\a,\beta)
\label{def:Vshort}
\ee
and
\be
        E_{(*)g}(\bP,\bu):= \frac{e^{2t\eta}}{(2\pi)^2}
        \iint_{-Y}^Y \rd\alpha\rd\beta \;  E_{(*)g}(\bP,\bu,\balpha) \; ,
\label{def:Eshort}
\ee
where
 $\balpha$ in $ E_{(*)g}(\bP,\bu,\balpha)$
is defined as $\alpha_e = \alpha$ for $e\in\cL$ and
 $\alpha_e:=\beta$ for $e\in\wt\cL$. The notation $(*)$ 
indicates the same formulas with and without truncation.
We will call these numbers 
the {\it $V$-value} and
{\it $E$-value of the partition $\bP$}, or sometimes, 
of the corresponding Feynman graph.
Strictly speaking, the $V$- and $E$-values
 depend on  $\xi$ through $\Delta$ and the $V$-value depends
on the choice of $Q$ as well.
When necessary, we will make these dependencies explicit in the notation,
e.g. $E_\xi$ or
$V_\xi(\bP; Q)$.
The $E$-value is a convenient estimate for the $V$-value of
the graph (see (7.14) of \cite{ESYI})
\be
        \big| V_{(*)}(\bP)\big|\leq (C \lambda)^{g}\;
        E_{(*)g}(\bP,\bu\equiv 0) \; 
\label{eq:VleqE}
\ee
with $g=g(\bP)$.
We will use the notation $E_{(*)g}(\bP):=E_{(*)g}(\bP, \bu \equiv 0)$.

For the graphical representation
of the Duhamel expansion we will really need 
\be
 V_{(*)}^\circ(\bP):= \frac{e^{2t\eta}}{(2\pi)^2}\iint_\bR \rd\alpha\rd\beta \;
     V_{(*)}(\bP,\alpha,\beta)\; ,
\label{def:circ}
\ee
i.e. a version of
 $V_{(*)}(\bP)$ with unrestricted $\rd\alpha\, \rd\beta$ integrations.
 (The circle superscript in $V^\circ$  will refer
to the unrestricted version of $V$). 
 The difference
between the restricted and unrestricted $V$-values are
negligible even when we sum them up for all partitions
(Lemma 7.1 of \cite{ESYI}).

Sometimes we will use the numerical labelling of the edges.
We will label the edge between $(j-1, j)$
 by $e_j$, the edge between $(\wt j, \wt{j-1})$
by $e_{\tilde j}$ and we set
$e_{n+1}:=(n, 0^*)$,
$e_{\wt{n'+1}}:=(0^*, \tilde n')$, $e_1=(0,1)$
 and $e_{\wt 1}:=(\wt 1, 0)$.
Therefore the
edge set $\cL=\cL(\cV_{n,n'})$ is identified with the index
set $I_{n+1} \cup \wt I_{n'+1}$ and
we set 
\be
p_j: = w_{e_j}, \qquad\tp_j: = w_{e_{\wt j}}.
\label{iden}
\ee

\subsection{Resummation for core indices}\label{sec:coreres}

We need to identify 
the non-repetitive potential labels in a  sequence.

\begin{definition}[Core of a sequence]\label{def:core}
Let $\tgamma \in \tGamma_n$,
 then
 the set of {\bf core indices} of $\tgamma$ is defined as
$$
  I_n^{core}(\tgamma):= \Big\{ j\in S(\tgamma) \; : \; 
\tgamma_j \neq \tgamma_i,
  \; \forall\, i\neq j\Big\}
$$
and we set $c(\tgamma)= |I_n^{core}(\tgamma)|$.
The corresponding $\tgamma_j$ labels are called {\bf core labels}.
The subsequence of core labels form an element in $\Gamma_c^{nr}$, i.e. 
a sequence
of different potential labels.
The elements of
$$
I_n^{nc}(\tgamma):= I_n\setminus [I_n^{core}(\tgamma) \cup I_n^\theta(\tgamma)]
$$
are called {\bf non-core potential indices}.
\end{definition}

{\it In other words,   the core indices are those skeleton indices (Definition
\ref{def:skeleton}) that
do not participate in any recollision, gate, triple collision or
nest.} Given the  stopping rules (Section \ref{sec:stopping}), 
the number of non-core potential
indices and $\theta$-indices together is at most 4.
The number of core indices $c=c(\tgamma)$ is related to the number of
 skeleton indices $k=k(\tgamma)$ as follows
\be
c: = \left\{  \begin{array} {c@{\quad \mbox{if}\quad}l}
   k & \#=nr, last \\
   k-1 & \#=triple\\
   k-2 & \#=nest,rec.
\end{array}\right.
\label{def:c}
\ee

For any $\tgamma\in \tGamma_n$, the index set $I_n$ is partitioned as
$I_n= I_n^{core}\cup I_n^{nc}\cup I_n^\theta$ into core indices,
 non-core potential
indices and $\theta$-indices. 
Let $\t=\t(\tgamma):=(\t_1, \t_2, \ldots ,\t_c) \in \Gamma_c^{nr}$
 denote the core labels of the sequence $\tgamma$.
We also introduce the
notation $\t_{[a,b]}:= (\t_a, \t_{a+1}, \ldots ,\t_b)$
if $a\leq b$, and $\t_{[a,b]}=\emptyset$ otherwise.

Now let $\tgamma\in \tGamma(n,K)\cup \tGamma^{(1),nr}_{n,k}$ 
(see (\ref{eq:union}) and Definition \ref{def:seqset}
for the notation). 
We recall that
 the total number of gates and $\theta$'s is given by $r$.
The  possible values of $r$ are determined by
 $\# = rec, nest,  triple, last$
according to (\ref{eq:union}) or $r=1$ if $\#=nr$ and  $k<K$.
 We will refer to a gate or $\theta$ index
as a gate/$\theta$-index in short.

We rewrite each error term in (\ref{eq:duha}) by first summing over
core labels $\t$. For fixed number of core indices $c$ we
sum over all possible locations of non-core indices.
If a non-core index is inserted between the $(w-1)$-th and
$w$-th core indices, we characterize its location by $w$.

The locations of non-core indices
within the sequence
are given by a {\bf location code} $w$. For example, if $\#=last$,
then $w\in I_{c+1}$ encodes that the first gate/$\theta$-index
is located between the $(w-1)$-th and $w$-th core indices.
The location of the second gate/$\theta$-index need not be encoded because
it is fixed to be after the last core index.
If $\#=rec$ and $r=0$, then $w\in I_c$ encodes that
the first recollision label is between the $(w-1)$-th and $w$-th core indices.
The most complicated case is $\#=rec$, $r=1$, when the
 code $w$ consists of two numbers, $w=(w_1, w_2) \in I_c\times I_{c+1}$,
where $w_1$ and $w_2$ encode the location of the recollision
and gate/$\theta$, respectively. If $w_1=w_2$, an extra binary
code determines whether the gate/$\theta$ is immediately
before or after the recollision.
The set of possible location codes therefore depends on $\#$, $c$ and $r$,
and it will be denoted by $W=W_c^{(r), \#}$.

The detailed description of the set $W$ in the other cases is
obvious but lengthy and we omit the formal details.
 The precise structure of $W$ is not
 important, but we remark that its cardinality satisfies $|W|\leq (c+1)^2$
in all cases.
We thus have
the following resummation formula:
\be
    \psi_{(*)t, k}^{\prime \; (r), \#} = \sum_{\t \in \Gamma_c^{nr}}
    \sum_{w\in W}\psi_{(*)t, \t, w}^{\prime \; (r), \#} \; ,
\label{eq:core}
\ee
where  $\psi_{(*)t, \t, w}^{\prime \; (r), \#}$ is the wave function
with core labels $\t$, with structure $\#$, with $r$
gates/$\theta$-indices and
with location of non-core indices given by $w$.
We note
that the wave function $\psi_{(*)t, \t, w}^{\prime \; (r), \#}$
includes a summation over the non-core labels with
the restriction that they are distinct from the core
labels $\t$.

Having specified the locations of the $r$ gates/$\theta$-indices
within the sequence of core indices,  we
introduce another code $h\in \{ g, \theta\}^r$, called {\bf gate-code},
to specify whether there is a gate or a $\theta$ at
the given location. This gives the decomposition
\be
    \psi_{(*)t, \t, w}^{\prime \; (r), \#} =
    \sum_{h\in \{ \theta, g\}^r} \psi_{(*)t, \t, w}^{\prime \; h, \#} \;
\label{eq:hcore}
\ee
with the obvious definition of $\psi_{(*)t, \t, w}^{\prime \; h, \#}$.

\subsection{Symmetrization of the non-core indices}\label{sec:symm}

Starting from (\ref{eq:core}), we can use the Schwarz inequality
\be
\begin{split}
    \bE' \|\psi_{(*)t, k}^{\prime \; (r), \#} \|^2  & =
    \sum_{w, w'\in W}\bE' \Big\langle  \sum_{\t \in \Gamma_c^{nr}}
    \psi_{(*)t, \t, w}^{\prime \; (r), \#} \;  ,
    \sum_{\t'\in \Gamma_c^{nr}} 
\psi_{(*)t, \t', w'}^{\prime \; (r), \#} \Big\rangle
\label{eq:core1} \\
& \leq |W| \sum_{w\in W} \sum_{\t, \t'\in  \Gamma_c^{nr}}
    \bE' \Big\langle
    \psi_{(*)t, \t, w}^{ \prime \; (r), \#} \;  ,
     \psi_{(*)t, \t', w}^{\prime \;(r), \#} \Big\rangle \; ,
\nonumber
\end{split}
\ee
where $c$ is given  by $k$ and $\#$ according to
(\ref{def:c}), and recall that $W$ depends on $\#, c, r$.

Notice that any non-core 
potential label appears in pair (in a gate, nest or recollision) and none
of them appear more than six times by the stopping rules. 
Since the first, third and fifth moments of 
the random variables $v_\gamma$
are zero, the expectation in (\ref{eq:core1}) is nonzero only if $\t'$
and $\t$ are paired, i.e. if there is a permutation $\s \in \fS_c$ such
that $\t'=\s(\t)$, meaning $\t_i'= \t_{\s(i)}$. Therefore
\be
\begin{split}
    \bE' \|\psi_{(*)t, k}^{\prime \;(r), \#} \|^2 &\leq 
|W| \sum_{w\in W} \sum_{\s\in \fS_c}
    \sum_{\t \in  \Gamma_c^{nr}}
      \bE' \Big\langle
    \psi_{(*)t, \t, w}^{\prime \;(r), \#} \;  ,
     \psi_{(*)t, \s(\t), w}^{\prime \;(r), \#} \Big\rangle \; 
\label{eq:core2} \\
&\leq |W| \sum_{w\in W} \sum_{\s\in \fS_c}
    \sum_{h, h'\in \{ g, \theta\}^r} \sum_{\t \in  \Gamma_c^{nr}}
      \bE'\Big\langle
    \psi_{(*)t, \t, w}^{\prime \;h, \#} \;  ,
     \psi_{(*)t, \s(\t), w}^{\prime \;h', \#} \Big\rangle \; .
\end{split}
\ee

\medskip

Note that each wave function $\psi_{(*)t, \t, w}^{\prime \;(r), \#}$
 has been further
decomposed into a sum over $h$-codes according to (\ref{eq:hcore}). However,
we did not estimate the $h\neq h'$ cross 
terms by an additional Schwarz inequality.
 The term with a gate must cancel the term with
a $\theta$ exactly at the same location, i.e. $\psi^g$ and $\psi^\theta$
would not individually be  negligible, but their sum is of smaller order.

Notice also that each wave function $\psi_{(*)t, \t, w}^{\prime \; h, \#}$
may
involve  summations over one or two further non-core potential labels.
We use the convention that the recollision or nest label is denoted
by $\nu$, the label of the gate or triple is denoted by
$\mu$. In case of a second gate, $\#=last$, its label will be
  $\mathring{\mu}$.
 According to the non-repetition rules,
 ($\nu, \mu, \mathring{\mu}$) may not coincide
with each other or with any element of $\t$. In the products
$\ov{\psi}_{(*)t, \t, w}^{\prime \; h, \#} 
\psi_{(*)t, \s(\t), w}^{\prime \;h', \#}$
there is no repetition among $\t$ and $\s(\t)$ indices other
than the ones prescribed by $\s$. However, if the additional non-core labels
within  $\psi_{(*)t, \s(\t), w}^{\prime \; h', \#}$ are denoted by
 $\nu', \mu'$ or $\mathring{\mu}'$, then there may be a few coincidences
between primed and non-primed non-core labels.
Those coincidences are allowed 
 that do not violate the non-repetition rules requiring that
$\nu, \mu, \mathring{\mu}$ are  distinct and their
 primed counterparts are also distinct among themselves.

\bigskip

Once the number of core indices
$c$, a location-code $w$ and a gate-code $h$ are fixed, this defines
a unique insertion of the non-core indices into the
sequence of core indices $I_c$. The core and non-core
indices in the given order can be identified with $I_n$,
where $n$, the total number of collisions,
is given by $n= k + r + |\{ j\; : \; h_j=g\}|$
and $k$ is given by (\ref{def:c}).
This naturally defines an embedding map
 $s=s_w^h: I_c\mapsto I_n$.
Similarly, $\wt s_w^{h'}: \wt I_c\mapsto \wt I_n$ can be defined.
The precise
definition depends on $\#$, $r$ and $h$ in a natural way. For illustration,
 we describe the
most complicated case; $\#=rec$, $r=1$ and $h=g$.
In this case $n=c+4$. For definitiveness,
we consider the case when the recollision precedes the gate
 $w_1\leq w_2$. In this case the complete collision sequence is
 $(\t_{[1,w_1-1]}, \nu, \t_{[w_1, w_2-1]}, \mu, \mu,
 \t_{[w_2, c]}, \nu)$. Let
 $$
    s_w^{h=g}(j): =\left\{  \begin{array} {c@{\quad \mbox{if}\quad}l}
   j& j< w_1 \\
   j+1 & w_1\leq j <w_2\\
   j+3 & w_2\leq j \leq c \; .
\end{array}\right.
$$
With this notation, the pairing of core indices, originally determined by
$\s\in \fS_c$,
 is given by the pairs $\{ s_w^h(j), \wt s_w^{h'}(\s(j))\}$
as subsets of the full index set $I_n\cup \wt I_{n'}$.

Given $(\#, c, \s, w, h, h')$,
we define the partition $\bD_0=\bD_0(\#, c, \s, w, h, h')$
of $I_n\cup\wt I_{n'}$
by lumping {\it only} those indices that {\it are required} to
carry the same potential label by the prescribed structure $\#$
and the gates. The vertices with $\theta$ always remain a single lump,
the remaining vertices
are paired.

For example, if $\#=rec$, $r=1$, $w=(w_1, w_2)$, $h=h'=g$, we obtain
$$
  \bD_0:= \Big\{ \{ s(j), \wt s(\s(j))\}_{j\in I_c},
  \{ w_1, n\} , \{ w_2+1, w_2+2\}, \{ \wt w_1, \wt n\}, \{\wt{w_2+1},
  \wt{w_2+2}\} \Big\}
$$
(see Fig.~\ref{fig:grec1})
and all other cases are similar.
\bef\bec
\epsfig{file=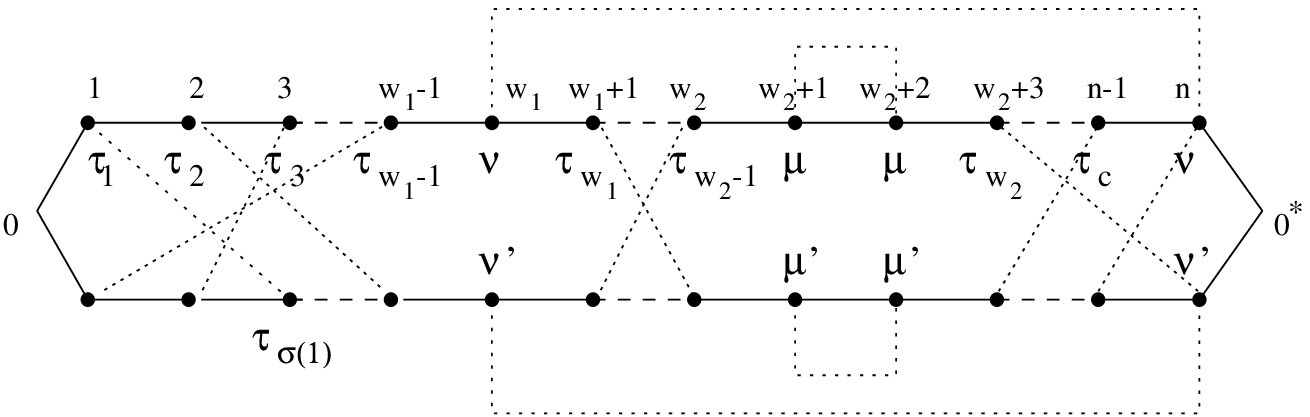,scale=1}
\eec
\caption{Symmetrized recollision with a gate. 
${\bf D}_0$ consists of the paired vertices}\label{fig:grec1}
\bec
\epsfig{file=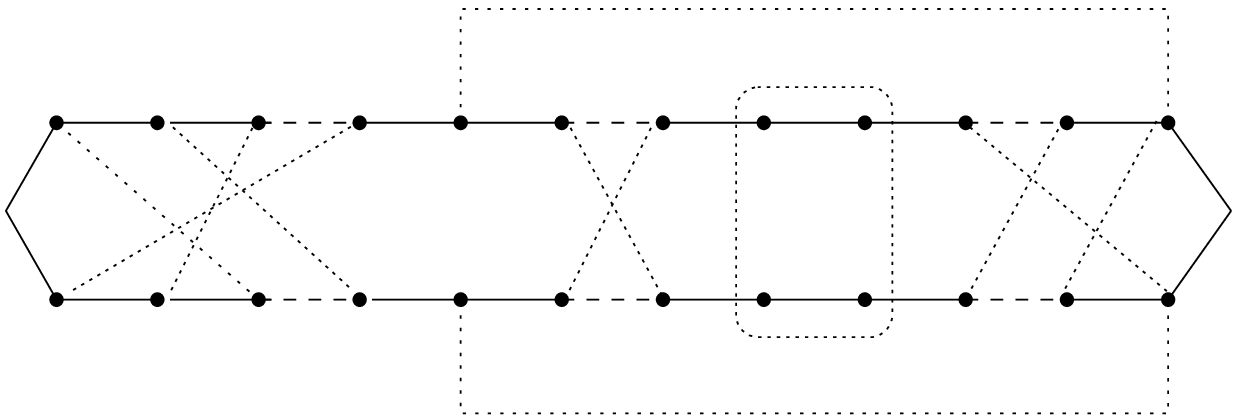,scale=1}
\eec
\caption{Partition ${\bf D}$ lumps some non-core elements of ${\bf D}_0$}
\label{fig:grec2}
\eef
The elements of $\bD_0$ consisting of pairs of core indices,
$\{ s(j), \wt s(\s(j))\}_{j\in I_c}$, are called {\it core
elements} of $\bD_0$ and those
 elements of $\bD_0$ that contain non-core indices will be
called {\it non-core elements}. In the example above,
the last four elements are the non-core elements of $\bD_0$
describing the two gates and two recollisions.
The non-core potential labels $\nu,\mu, \nu',\mu'$
will correspond to these non-core elements, respectively.

Some of the non-core elements may be lumped together according
to the possible coincidence between the $\{ \nu, \mu\}$
and $\{ \nu', \mu'\}$ since the non-repetition rule do not
prevent it. This procedure defines new partitions $\bD$
that we will call {\it derived partitions}, denoted by $\bD \succ \bD_0$.
In this particular case, there are seven
possible lumpings of the four non-core elements
without violating the
non-repetition rule within the sets $(\t, \nu, \mu)$ and $(\t, \nu', \mu')$.
Figure 10 shows the partition $\bD$ derived from
the above partition $\bD_0$ when $\mu=\mu'$ but
$\nu\neq \nu'$.

In general $\bD$ is defined
by lumping together a few non-core elements of $\bD_0$
under the constraint of the non-repetition rule
(Fig.~\ref{fig:grec2}). The single elements of $\bD_0$
that correspond to $\theta$ are never lumped.
Note that  each of these pairings gives rise to the appearance of exactly
four or six identical potential labels; higher moments do not appear.
The number of such quartets and sextets is denoted by $\varrho_4(\bD)$
and $\varrho_6(\bD)$.
Clearly $\varrho_4(\bD)\leq 2$ and $\varrho_6(\bD)\leq 1$. 

Let $\bD^*\subset \bD$ denote the collection of non-single elements of $\bD$.
Note that for each element of $\bD^*$ one selects a distinct potential label.
The quantity (\ref{eq:core2}) contains a summation over all
such potential labels. We will use the connected graph formula
(Lemma 6.1 from \cite{ESYI}) for the index set $\bD^*$.

Let $\bA \in\cA(\bD^*)$ be a partition of the set $\bD^*$. We
define $\bP(\bA, \bD)\in\cP_{n,n'}$
 to be the partition of $I_n\cup \wt I_{n'}$
whose lumps are given by the equivalence relation that
two elements of $I_{n}\cup \wt I_{n'}$ are
$\bP(\bA, \bD)$-equivalent
if their $\bD$-lump(s) are $\bA$-equivalent. The single lumps of $\bD$ remain
single in $\bP$ (these are the $\theta$ indices).

\medskip

We recall  the definition of $V_{(*)}(\bP)$ and  $V_{(*)}^\circ(\bP)$
 from \eqref{def:Vshort} and \eqref{def:circ}.
Since we will compute the $L^2$-norm, 
the momentum shift at $0^*$ is chosen to be 
 $\xi=0$ (see \eqref{def:Delta}) and
the $Q$ function in \eqref{def:cM}, representing the observable,
will be $Q\equiv 1$. 
Furthermore, when defining Feynman graphs, we
 will always assume the following range of parameters
(see (7.25) of \cite{ESYI}) unless stated otherwise:
\be
\eta = \lambda^{2+\kappa},
 \quad t=\lambda^{-2-\kappa}T, \quad T\in [0, T_0],
\quad K= [\lambda^{-\delta}(\lambda^2t)], \quad k\leq K, \quad
\zeta=\lambda^{-\kappa-3\delta}, \quad g\leq 8\;
\label{param}
\ee
with a sufficiently small $\delta>0$ that is independent of $\lambda$
but depends on $\kappa$. We recall that $\eta$ is the regularization
of the propagator, $K$ is the upper treshold for the number of skeleton
indices, $k$, in the expansion, $\zeta$ is the momentum cutoff
(see \eqref{def:mu})
 and $g$ is the number of exceptional
vertices where the standard $|\wh B(w_{in}-w_{out}) |$ potential decay
is not present (this happens for the single lumps).
All estimates will be uniform in $\xi$
and in $T\in [0,T_0]$.

For  each fixed $(\#,c,\s,w,h,h')$,
by  using  \eqref{momm} and the connected graph formula, 
similarly to Proposition 7.2 of \cite{ESYI} we  obtain
\be
     \lim_{L\to\infty}\sum_{\t \in  \Gamma_c^{nr}} \bE' \Big\langle
    \psi_{(*)t, \t, w}^{\prime \;h, \#} \;  ,
     \psi_{(*)t, \s(\t), w}^{\prime \; h', \#} \Big\rangle \;
     =   \sum_{\bD\succ \bD_0}\underline{m}^{\varrho(\bD)}
     \sum_{\bA\in\cA(\bD^*)} c(\bA) V_{(*)}^\circ( \bP(\bA, \bD)) \, 
\label{eq:core4}
\ee
with $\underline{m}^{\varrho(\bD)}:= m_4^{\varrho_4(\bD)}
m_6^{\varrho_6(\bD)}$ (we recall $m_k= \bE \, v_\gamma^k$ and \eqref{momm}).
The summary of the results in this section is

\begin{proposition} Let $k\leq K$,
let $\#= rec, nest, triple, last $ and $r$ be one of the possible values
allowed by (\ref{eq:union}) or $r=1$ if $\#=nr$, $k<K$.
 Let $c$ be given by (\ref{def:c}), and let
$W=W^{(r), \#}_c$ be the set of location codes.
Then
\be
    \lim_{L\to\infty}
\bE' \|\psi_{(*)t, k}^{\prime \;(r), \#} \|^2 \leq
    |W| \sum_{w\in W} \sum_{\s\in \fS_c}
   \sum_{h,h'\in \{g,\theta\}^r} \sum_{\bD\succ\bD_0}
\underline{m}^{\varrho(\bD)}
\sum_{\bA\in \cA(\bD^*)}c(\bA) V_{(*)}^\circ( \bP(\bA, \bD)) \; .
\label{eq:decS}
\ee
$\;\;\Box$
\end{proposition}

\subsection{Splitting into high and low complexity regimes}

Given $(\#, c, \s, w, h, h')$,
we consider the partition $\bD_0$ of $I_n\cup \wt I_{n'}$
as defined above and let
$\bD \succ \bD_0$. Note that the collection
$\bD^*$ contains all  core elements of
$\bD_0$, i.e. all
pairs of core indices $\{ s(j), \wt s(\s(j))\}_{j\in I_c}$.
The restriction of a partition $\bA\in \cA(\bD^*)$
 onto  these core elements
can therefore be naturally identified with a partition
of $I_c$ using the map
$\{ s(j), \wt s(\s(j))\}_{j\in I_c}\mapsto j\in I_c$.
We denote this restricted partition by $\wh\bA$.
In the sequel we shall therefore view $\wh \bA\in \cA_c$,
i.e. as a partition on $I_c$.

The restricted partition $\wh\bA$ together with $\s$
also generates a partition
$\bP(\wh\bA,\s)$ on the set $I_c\cup \wt I_c$.
The lumps $P_\mu$ of  $\bP(\wh\bA,\s)$ are given by $\wh A_\mu
\cup \sigma(\wh A_\mu)$, where $\wh A_\mu$ are the lumps of $\wh \bA$.
Notice that the $(s,\wt s)$-image of the restriction
of $\bP(\bA, \bD)\in \cP_{n, n'}$ onto  the set of core indices
$I_n^{core}\cup \wt I_{n'}^{core}$ is exactly $\bP(\wh\bA,\s)$.
Since the cardinality of non-core elements of $\bD^*$ is at most 4,
 for any given $\bD$ and
$\wh\bA$ there can exist at most $(c+4)^4$ partitions, $\bA \in \cA(\bD^*)$,
whose restriction onto the core elements is $\wh\bA$.

We recall the definition of {\it joint degree} from \cite{ESYI}:
\begin{definition}\label{def:deg}
(i) Let $\bA\in \cA_k$ be a partition of $I_k=\{ 1, 2 \ldots, k\}$.
  Set $a_\nu:=|A_\nu|$,
$\nu\in I(\bA)$, to be the size of the $\nu$-th lump.  Let
$$
        S(\bA): = \bigcup_{\nu\in I(\bA)\atop a_\nu \ge 2} A_\nu
$$
be the union of nontrivial lumps. The cardinality of this set,
$s(\bA): = |S(\bA)|$, is called the {\bf degree of the partition} $\bA$.

(ii) Let  $\bA\in \cA_k$ and $\s\in \fS_k$. The number
\be
   q(\bA, \sigma):=\max\Big\{ {\rm deg}(\sigma), \frac{1}{2} s(\bA)\Big\}
\label{def:q}
\ee
is called the {\bf joint degree} of the pair $(\sigma, \bA)$
of the permutation $\s$  and partition $\bA$.
\end{definition}

The sum (\ref{eq:decS}) will be split into two parts and estimated differently.
In the regime of high combinatorial complexity, i.e., when the joint
degree $q(\wh \bA, \s)$ of $\sigma$ and  $\wh\bA$
is bigger than a threshold $q\ge 1$ (to be determined later),  then we
can use the method of Section 9 (especially 
Proposition 9.2) from \cite{ESYI} robustly. 
This will be explained in Section
\ref{sec:compl}.
For low combinatorial complexity we use the special
structure given by the recollisions, nests, triple collisions
or gates (Section \ref{sec:smallcomb}).
The threshold $q$ will be chosen  differently for the estimates
\eqref{eq:noladerr}--\eqref{eq:laderr} and for \eqref{eq:laderr1}.

The precise estimate is the following
\be
   \lim_{L\to\infty}
   \bE' \|\psi_{(*)t, k}^{\prime\; (r), \#} \|^2 
\leq (I) + (II) + O(\lambda^5)
 \label{I+II}
\ee
with
\be
   (I) := |W|
   (c+4)^4 \sum_{w\in W}\sum_{h,h' } \sum_{\s\in \fS_c} \sum_{\bD\succ\bD_0}
   \underline{m}^{\varrho(\bD)}
   \!\!\!\!\!\sum_{\bA'\in \cA_c\atop
     q( \bA', \s)\ge q }
   \sup_\bA \Big\{ | V_{(*)}( \bP(\bA, \bD)) c(\bA) | \; : \;
\wh\bA =\bA' \Big\}
\label{eq:I}
\ee
where the supremum is over all possible $\bA\in \cA(\bD^*)$
whose restriction $\wh\bA$
is the given partition $\bA'$;
and
\be
 (II) :=   |W| \sum_{w\in W} \sum_{\s\in \fS_c} \Bigg|
   \sum_{h,h'\in \{g,\theta\}^r} \sum_{\bD\succ\bD_0}
\underline{m}^{\varrho(\bD)}
\sum_{\bA\in \cA(\bD^*)\atop q( \wh\bA, \s)< q} V_{(*)}
( \bP(\bA, \bD)) c(\bA) \Bigg| \; .
\label{eq:II}
\ee
We recall that $\bD_0$ is determined by $(\#, c, \s, w, h, h')$.
The error term $O(\lambda^5)$ comes from replacing 
$V_{(*)}^\circ(\cdots)$ with $V_{(*)}(\cdots)$;
see Lemma 7.1 of \cite{ESYI}.

\subsection{Case of high combinatorial complexity}
\label{sec:compl}

Here we estimate the term (I) in  (\ref{eq:I}).
Clearly $\underline{m}^{\varrho(\bD)}\leq \langle m_4\rangle^2 \langle
m_6\rangle \leq C$.
We estimate $V_{(*)}(\bP(\bA, \bD))$ by
using \eqref{eq:VleqE}.
 Then, by applying 
Operation I from Appendix \ref{sec:genestcir}, we
 break up all the lumps $P_\mu$ in the partition  $\bP(\bA, \bD)$
that involve elements from non-core indices, $I_n^{nc}\cup \wt I_{n'}^{nc}$,
in such a way that all non-core indices must form single lumps.
Let $\bP^*(\bA, \bD)$ denote this new partition.
Note that the projection of $\bP^*(\bA, \bD)$ onto the core indices
is unchanged. The number of application of Operation I is at most 6.
Using Lemma \ref{lemma:breakup}, we can estimate
$E_{(*)g}(\bP(\bA, \bD))$ in terms of $\sup_\bu E_{(*)g}(\bP^*(\bA, \bD),\bu)$
with an additional factor of at most $\Lambda^{6}$, where
$$
     \Lambda: = [CK\zeta]^d = O(\lambda^{-2d\kappa- O(\delta)})\; .
$$

Then we apply Operation II (Appendix \ref{sec:genestcir})
to remove all single lumps with non-core
indices and use (\ref{eq:nontr}) from Lemma \ref{lemma:remove}.
After removing the non-core indices, the remaining vertex set can
naturally be identified with $I_c\cup \wt I_c$ by using the $(s,\wt s)$ maps,
and the partition $\bP^*(\bA, \bD)$ restricted to core indices $I_n^{core}\cup
\wt I_{n'}^{core}$
is identified with $\bP(\wh \bA,\s)$.
Lemma \ref{lemma:remove} is applied at most $8$ times, therefore
we obtain that for any $\sigma$, $\bD$ and $\bA'$ in the sum (\ref{eq:I}):
\be
|V_{(*)}( \bP(\bA, \bD))|
   \leq  C\Lambda^{6}(\lambda\eta^{-1})^8
 \sup_{\bu, g\leq 8}  E_{(*)g}(\bA', \sigma, \bu)\; .
\label{eq:8}
\ee
The application of Operation II
 is schematically shown on Fig.~\ref{fig:est}.

\bef\bec
\epsfig{file=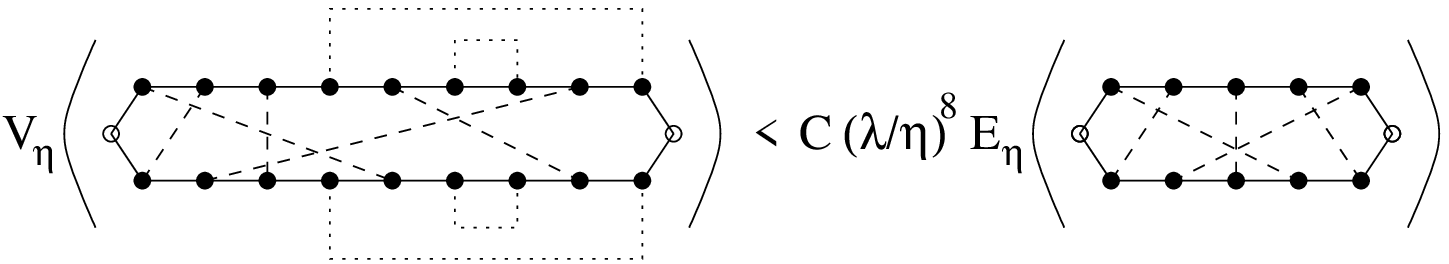, scale=.9}
\eec
\caption{Estimate after removing all non-core indices}\label{fig:est}
\eef

The summations over $h, h'\in \{ g,\theta\}^r$
and $\bD\succ \bD_0$ in (\ref{eq:I})
 contribute with at most a constant factor  since $r\leq 2$ and
the number of different $\bD$'s is at  most 7.
The cardinality of $W$ can be bounded by $(c+1)^2$ and $c\leq K  \leq C
\lambda^{-\kappa-\delta} $.
Therefore
we obtain
$$
   (I)\leq  C \lambda^{-8 - \kappa(16+2d)-O(\delta)}
   \sum_{\s\in \fS_c} \sum_{\bA'\in \cA_c\atop
     q(\bA', \s)\ge q }   \sup_{\bu,g\leq 8}
  E_{(*)g}(\bA ', \sigma, \bu)  |c(\bA')| \; .
$$
Using Proposition 9.2 from \cite{ESYI}, 
 we  have
$$
   (I)\leq C
\lambda^{q[ \frac{1}{3}-(\frac{17}{3}d+\frac{13}{2})\kappa
 -O(\delta)]-8-(16+2d)\kappa }\; .
$$
We immediately see, that the contribution of (I) to
the error term in \eqref{eq:laderr1}
 in  Theorem \ref{thm:error}
satisfies the announced bound with a sufficiently small $\delta$ if
\be
   \kappa < \frac{2q-48}{ (34d+39)q +112+12d} \; .
\label{eq:qlow1}
\ee
The bounds (\ref{eq:noladerr})--\eqref{eq:laderr} are satisfied if
\be
   \kappa < \frac{2q-72}{(34d+39)q +112+12d} \; .
\label{eq:qlow}
\ee

\subsection{Case of small combinatorial complexity}\label{sec:smallcomb}

Here we control the term (II) in \eqref{eq:II}.
First we estimate the combinatorics.

\begin{lemma}\label{lemma:smallcomb}
 For any  $q\in \bN$,  $c\leq K$
 and structure type $\#$,
 we have
$$
     \sup_{w, h, h'}\sum_{\sigma \in \fS_c}
     \sup_{\bD\succ\bD_0}\sum_{\bA \in \cA(\bD^*)\atop q( \wh\bA,\s)<q}
     |c(\bA)| \leq (CqK)^{3q+3} \; ,
$$
where we recall that $\bD_0$ depends on $(\#, c, \s, w, h, h')$.
\end{lemma}

{\it Proof.}  The bound 
\be
    \# \{ \sigma\in \fS_k\; : \; \ell(\sigma)=\ell\} \leq (Ck)^{k-\ell+1} \; 
\label{eq:permnumb}
\ee
(see (8.14) from \cite{ESYI})
 shows that the number of permutations
$\sigma\in \fS_c$ with ${\rm deg}(\sigma) < q$
 is bounded by $(CK)^{q}$ using $c\leq K$.
The number of  $\bA$'s whose restriction yields the same $\wh\bA$ is
at most $(c+3)^4\leq (CK)^4$.  The number of $\wh\bA\in \cA_c$ with $s(\wh\bA) < 2q$
is bounded by $c^{2q-1}\leq (CK)^{2q-1}$. Finally, $|c(\bA)|\leq \prod_j^* a_j^{a_j-2}
\leq (2q)^{2q-2}$
 $\;\;\Box$.

\bigskip

The individual terms in (\ref{eq:II})
are estimated in the following Proposition
whose proof will be given in Section \ref{sec:cases}.

\begin{proposition}\label{prop:i-iv} 
We assume \eqref{param} and $\kappa<\frac{2}{34d+39}$.
 Let
 $\sigma\in \fS_c$, $w\in W_c^{(r), \#}$,
 $h, h'\in \{ g, \theta\}^r$, where $\#$ and $r$ vary
 in the different cases and let
$\bD\succ\bD_0(\#, c, \sigma, w, h, h')$.

\medskip
\noindent
1) Let $\bA\in \cA(\bD^*)$ such that
$q(\wh\bA, \sigma )< q$, where $q$ is a fixed number. Then the following
individual estimates hold.

(1a) [Many collisions] Let $\#=nr$, $r=0,1$ and $c=K$,  then
\be
     |V_{*} (\bP(\bA, \bD))|\leq  C^q\lambda^{\frac{\delta}{2}K}  \; .
\label{eq:manyev}
\ee

(1b) [Recollision]. Let $\#=rec$, $r=0,1$, then
\be
    |V_{*} (\bP(\bA, \bD))|\leq 
C^q\lambda^{6-4d\kappa(q+3)} \; .
\label{eq:recv}
\ee

(1c) [Triple collision] Let $\#=triple$, $r=1$, then
\be
    |V_{*} (\bP(\bA, \bD))|\leq  C^q\lambda^{6-4d\kappa(q+3)}\; . 
\label{eq:triplev}
\ee

\medskip\noindent
2) Now let
 $\bA'\in \cA_c$ be given. Then the following estimates hold:

(2a) [Non-repetition with a gate] Let $\#=nr$, $r=1$, then
\be
     \sup_{\sigma, w} \Big|\sum_{h,h'\in \{ g, \theta\}^r}
     \sum_{\bD\succ\bD_0} \sum_{\bA\in \cA(\bD^*) \atop \wh\bA=\bA'}
 V (\bP(\bA, \bD)) c(\bA)\Big|\leq C\lambda^{\frac{1}{3}- (\frac{17}{3}d
+ 8)\kappa-O(\delta)} \; .
\label{eq:nrev}
\ee

(2b) [Last] Let $\#=last$, $r=2$, then
\be
     \sup_{\sigma, w} \Big|\sum_{h,h'\in \{ g, \theta\}^r}
     \sum_{\bD\succ\bD_0} \sum_{\bA\in \cA(\bD^*) \atop \wh\bA=\bA'}
 V_{*} (\bP(\bA, \bD)) c(\bA)\Big|\leq C
\lambda^{6-(14d+6)\kappa-O(\delta)} \; .
\label{eq:lastev}
\ee

(2c) [Nest] Let $\#=nest$, $r=1$, then
\be
     \sup_{\sigma, w} \Big|\sum_{h,h'\in \{ g, \theta\}}
     \sum_{\bD\succ\bD_0} \sum_{\bA\in \cA(\bD^*) \atop \wh\bA=\bA'}
 V_{*} (\bP(\bA, \bD)) c(\bA)\Big|\leq
  C\lambda^{6-(10d+8)\kappa-O(\delta)} \; .
\label{eq:nestev}
\ee
\end{proposition}

Combining  Lemma
\ref{lemma:smallcomb} with these estimates, and  using $|W|\leq K^2$,
 we see that
$$
  (II) \leq  (Cq K)^{3q+7} 
 \lambda^{\frac{1}{3}-(\frac{17}{3}d+8)\kappa-O(\delta)}\;
$$
for the case $\#=nr$, $r=1$ (case (2a) above), and 
$$
  (II) \leq  (CqK)^{3q+7} 
 \lambda^{6-4d\kappa(q+3)-O(\delta)}\;
$$
for all other cases (with $q\ge 2$).
So the contributions of the error terms from (II) to
$ \bE' \|\psi_{(*)t, k}^{\prime \; (r), \#} \|^2 $ (see (\ref{I+II}))
satisfy the bound \eqref{eq:laderr1} if
\be
\kappa < \frac{1}{9q+17d+51}\; ,
\label{eq:qup1}
\ee
and they satisfy
(\ref{eq:noladerr})-\eqref{eq:laderr} if
\be
\kappa < \frac{2}{(4d+3)q+(12d+9)}
\label{eq:qup}
\ee
and $\delta$ is sufficiently small.
Combining this with (\ref{eq:qlow1})--(\ref{eq:qlow}) and optimizing,
we obtain that there exists  $\kappa_0(d)>0$ such that for
any  $\kappa<\kappa_0$, the systems of 
inequalities \eqref{eq:qlow1}--\eqref{eq:qup1}
and   \eqref{eq:qlow}--\eqref{eq:qup} have solutions for $q$.
For $d=3$, the optimal $\kappa_0(d)$ 
is a bit above $\frac{1}{500}$.
 This finishes the proof of Theorem \ref{thm:error}. $\;\;\;\Box$.

\section{Proof of Proposition \ref{prop:i-iv}}\label{sec:cases}
\setcounter{equation}{0}

In each case except (1a),
 the corresponding  Feynman graph has a specific subgraph
of a few vertices (recollision, nest etc.) that renders the value small.
We shall prove that this subgraph gains
 at least a factor
$\lambda^{4+2\kappa + 2\delta}$ 
required  in  Theorem \ref{thm:error}. Then we remove all repetition
patterns from the graph, we use the robust bounds 
\begin{align}
   \sup_{\s\in \fS_k} \sup_\bu E ( \s, \bu) & \leq C|\log \lambda|^2 \; 
\label{eq:ladlogweak}
\\
   \sup_{\s\in \fS_k}
  \sup_\bu E_{*} (\s, \bu) & \leq C\lambda^2 |\log \lambda|^2\;
\label{eq:Etrunc}
\end{align}
from Lemma 10.2. of \cite{ESYI}
to conclude the estimate.

\subsection{Many collisions}

The estimate in case (1a)  will come from the fact that
any graph can be  robustly  estimated by the ladder graph and
the value of the ladder of length $L$ always carries a factor $1/L!$.
 This effect is the best seen
in the time integral form. We first change $V_{*}(\bP)$ back to
$V_{*}^\circ(\bP)$  with an error 
smaller than $O(\lambda^{10K- O(1)})$ by Lemma~7.1 from \cite{ESYI}.
We then  apply the $K$-identity (formula (6.2) in \cite{ESYI})
to the definition of $ V_{*}^\circ(\bP)$  given in \eqref{def:circ} 
to obtain
$$
   V_{*}^\circ(\bP)  = \lambda^{n+n'+g}  \iint
   \rd\bp\rd\tbp \; \ov{K(t, \bp,n)}
   K(t, \tbp,n') \Delta(\bP, \bw, \bu\equiv 0)\cM(\bw)
$$
with $\bP=\bP(\bA, \bD)$ and 
$$
     K(t, \bp,n) := (-i)^{n-1} \int_0^t [\rd s_j]_1^{n} \prod_{j=1}^{n}
    e^{-is_j\om(p_j)} \; .
$$
We recall that $g= g(\bP(\bA, \bD))$
denotes the number of single lumps, or, equivalently, 
the number of $\theta$ labels in $h$ and $h'$.
Note that we use the labelling $\bw$ and $\bp,\tbp$ in parallel,
keeping in mind the relabelling
convention from \eqref{iden}
(see also  Section~7.2 in \cite{ESYI} for more details).

We use a Schwarz inequality:
$$
   |V_{*}^\circ(\bP) |\leq \lambda^{n+n'+g}  \iint \rd\bp\rd\tbp \; \Big[
   |K(t, \bp,n)|^2+ |K(t, \tbp,n')|^2\Big] \Delta(\bP, \bw, \bu\equiv 0)
|\cM(\bw)|\; .
$$
By using  Operation I,
  we can break up  $\bA$ into single lumps.
Since $s(\wh\bA)\leq 2q$, we have $s(\bA)\leq 2q+4$, thus Operation I
will be applied at most $2q+3$ times.

If $r=0$,
i.e. the original graph was a non-repetition graph, and thus $n=n'=k$,
then  the trivial partition $\bA_0$
corresponds to a partition $\bP$ with a  complete pairing. Thus  both
$\bp$ and $\tbp$ momenta can be used as independent variables and
$$
   |V_{*}^\circ(\bP) |\leq \Lambda^{2q+3} 
(C\lambda)^{2k+g} \int\, \rd\bp\,
 |K(t, \bp, k)|^2 |\wh\psi_0(p_1)|^2\prod_{j=1}^k
 |\wh B(p_j-p_{j+1})|^2 \; .
$$
The estimate (\ref{eq:manyev}) is then completed
 by the bound \eqref{eq:faktbound} from the following
Lemma with any $1/2 <a<1$. The proof will be given below.

\begin{lemma}\label{lemma:Klemma}
For any $0\leq a <1$, $t=T\lambda^{-2-\kappa}$,
 there exists a constant $C_a$ such that
$$
 I(k):=  \int \rd \bp  |K(t, \bp, k)|^2 |\wh\psi_0(p_1)|^2\prod_{j=1}^k
 |\wh B(p_j-p_{j+1})|^2 \leq 
\frac{(C_aT\lambda^{-2-\kappa a})^{k-1}}{[(k-1)!]^a}
  |\log\lambda|^2 \; .
$$
In particular,
\be I(k)
\leq (C_a\lambda^{-2+\delta a})^{k-1}
\label{eq:faktbound}
\ee
if  $k\ge T\lambda^{-\kappa-\delta}$
and $\lambda \ll 1$.
\end{lemma}

{\it Proof.} This lemma is essentially the same as
   Lemma 3.1 in \cite{EY}. The only differences are
that here we estimate the truncated value, so $K$ has
one less time integration and the individual integrals
are performed by using Lemma \ref{le:opt}. The details
are left to the reader. $\;\;\Box$

\medskip

Finally, if $r=1$ and $w\in W_c^{(1), nr}$ is the
 location of the gate/$\theta$-index
among the core indices,  then  $p_w=p_{w+1}$ or
 $p_w=p_{w+2}$ (depending whether we have a  $\theta$ or a gate) is forced
 by $\Delta$ and similarly for $\tp_w$. In this case, the estimates
in Lemma \ref{lemma:Klemma} are
 worse by a factor of $t$.
This factor can be absorbed into the main
term $\lambda^{\delta a K}$.  This
completes the proof of (\ref{eq:manyev}). $\;\;\;\Box$

\bigskip

\subsection{Recollision and triple collision}

For the proof of (\ref{eq:recv}), we break up the partition $\bA$ into the 
trivial partition $\bA_0$ using Operation I. Since $s(\wh\bA)\leq 2q$, 
we have $s (\bA)\leq 2q+4$, thus  Operation I
will be applied at most $2q+3$ times.
Clearly
\be
     |V_{*}(\bP(\bA, \bD))| \leq \Lambda^{2q+3} \lambda^g
     \sup_\bu  E_{*}(\bP(\bA_0, \bD_0), \bu)
\label{eq:extr}
\ee
by using (\ref{eq:VleqE}) and Lemma~9.5 from \cite{ESYI}.
The single lumps are  removed 
by Operation II from $E_{*}(\bP(\bA_0, \bD_0), \bu)$, at the price $\lambda/\eta$;
the total contribution of one $\theta$-removal
 is $(\lambda^2/\eta)\sim \lambda^{-\kappa}$.
   The possible gates are eliminated
by Operation IV at the expense of $\lambda^2\eta^{-1}|\log\eta| \sim
\lambda^{-\kappa}|\log\lambda|$ each. Since $r\leq 1$, we lose
at most a factor $\lambda^{-2\kappa} |\log\eta|^2$ in this way
(see Fig.~\ref{fig:estreccon}; the double line denotes truncated propagators).

The vertex set of the remaining graph is naturally identified
with $I_{c+2}\cup \wt I_{c+2}\cup\{0, 0^*\}$, the permutation $\sigma$
provides a pairing between the elements $I_{c+2}\setminus\{ w_1, c+2\}$
and $\wt I_{c+2}\setminus\{ \wt{w_1}, \wt{c+2}\}$, furthermore $\{  w_1, c+2\}$
and $\{ \wt{w_1}, \wt{c+2}\}$ each form a lump (see picture). We
denote this partition by $\bP^*$ and by using Proposition \ref{prop:rec}
below, we will obtain
 (\ref{eq:recv}). $\;\;\Box$

\bef\bec
\epsfig{file=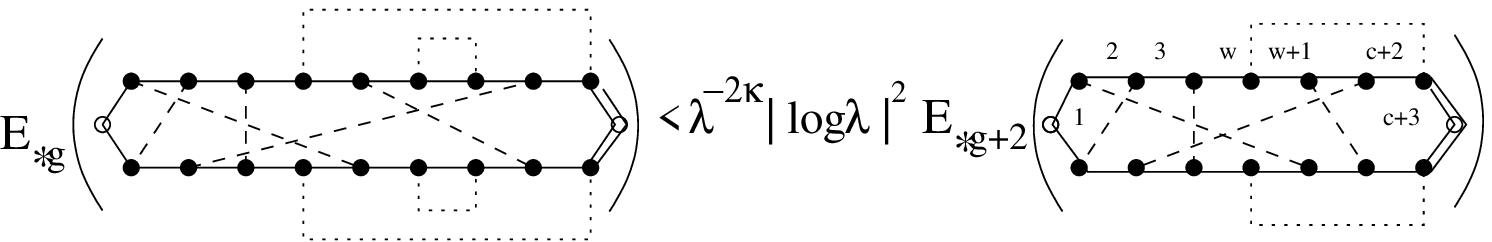, scale=1}
\eec
\caption{Removal of gates from a recollision}\label{fig:estreccon}
\eef

Later we need to estimate asymmtric recollision graphs as well,
so we formulate the following proposition in a more general setup:

\begin{proposition}\label{prop:rec}
Consider the Feynman graph on the vertex set $\cV_{k}$, $k\ge 3$,
choose numbers $a, b, a', b' \in I_k$ such that
$b-a\ge 2$, $b'-a'\ge 2$. Let $\sigma$ be a bijection between
$I_k\setminus \{ a, b\}$ and $\wt I_k\setminus \{ a', b'\}$.
Let $\bP^*$ be the partition on the set $I_k\cup \wt I_{k}$
consisting of the lumps $\{ j, \sigma(j)\}$, $j\in I_k\setminus \{ a, b\}$
and $\{ a, b\}$, $\{a', b'\}$ (Fig.~\ref{fig:reccon}).
Then
\be
      \sup_{\bu, g\leq 8} E_{*g}(\bP^*, \bu)
      \leq C\lambda^{6-3\kappa} \zeta^{4d} \; .
\label{eq:recesttr}
\ee
\end{proposition}

\bef\bec
\epsfig{file=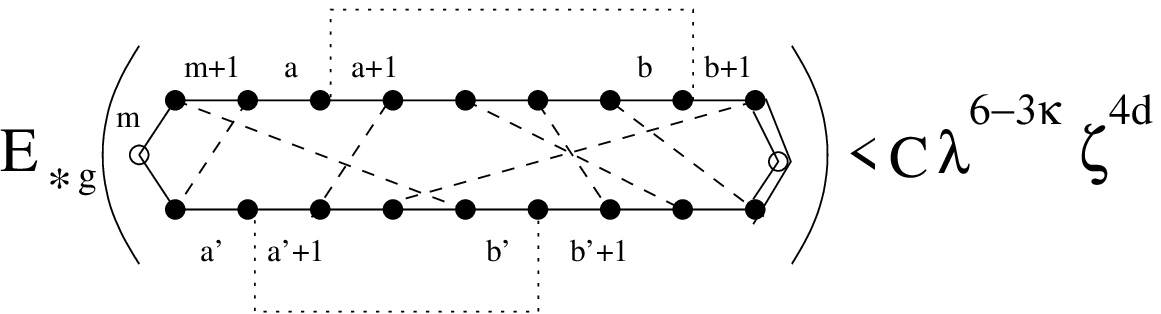,scale=1}
\eec
\caption{Estimate of a two-sided recollision graph}
\label{fig:reccon}
\bec
\epsfig{file=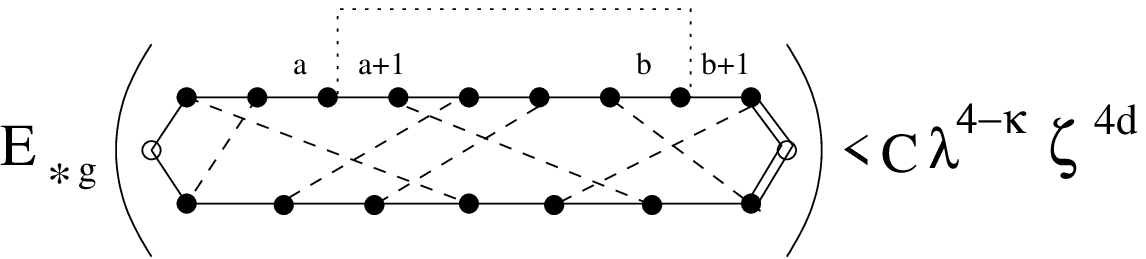,scale=1}
\eec
\caption{Estimate of a one-sided recollision graph}
\label{fig:rec1con}
\eef
We also need a ``one-sided'' version of this estimate (Fig.~\ref{fig:rec1con}).

\begin{proposition}\label{prop:rec1}
Consider the Feynman graph on the vertex set $\cV_{k, k-2}$, $k\ge 3$,
choose numbers $a, b \in I_k$ such that
$b-a\ge 2$. Let $\sigma$ be a bijection between
$I_k\setminus \{ a, b\}$ and $\wt I_{k-2}$.
Let $\bP^*$ be the partition on the set $I_k\cup \wt I_{k-2}$
consisting of the lumps $\{ j, \sigma(j)\}$, $j\in I_k\setminus \{ a, b\}$
and $\{ a, b\}$.
Then
\be
     \sup_{\bu, g\leq 8} 
E_{g}(\bP^*, \bu) \leq C\lambda^{2-\kappa}\zeta^{4d}
  \; ,
\label{eq:halfrecest}
\ee
and
 for the truncated version
\be
      \sup_{\bu, g\leq 8} E_{*g}(\bP^*, \bu) \leq C\lambda^{4-\kappa}
     \zeta^{4d}  \; .
\label{eq:halfrecesttr}
\ee
\end{proposition}

{\it Proof of Propostion \ref{prop:rec}.}
 We use $\bp=(p_1, \ldots , p_{k+1})$ and their tilde-counterparts to
denote the momenta to express
$$
   E_{*g}(\bP^*, \bu) = 
  \sup_{\cG\; : \;|\cG|\leq g}
\int_{-Y}^{Y} \rd \alpha\rd\beta\; \Xi_\cG(\alpha, \beta)
$$
with
$$
   \Xi_\cG(\alpha, \beta):=\lambda^{2k}
 \int \rd\mu(\bp)\rd\mu(\tbp) \prod_{j=1}^{k} \frac{1}{|\a-\ov\om(p_j)-i\eta|}
   \frac{1}{|\beta-\om(\tp_j)+i\eta|}
$$
$$
  \times \delta(p_{k+1}-\tp_{k+1}) \delta\Big(p_{a+1}-p_a
   + (p_{b+1}-p_{b}) -u_a\Big)  \delta\Big(-\tp_{a'+1}+\tp_{a'}
   - (\tp_{b'+1}-\tp_{b'}) -\tu_{a'} \Big)
$$
$$
   \times \prod_{j=1\atop j\neq a, b}^{k}
    \delta\Big(p_{j+1}-p_j
   - (\tp_{\s(j)+1}-\tp_{\s(j)}) - u_j\Big)\; \cN_\cG(\bw)\, ,
$$
where the $\bu$-momenta are labelled as
 $\bu = ( u_1, \ldots , u_{b-1}, u_{b+1}, u_{k},
\tu_{a'})$ and we used the identification from  \eqref{iden}
 between the
$\bw$ and $\bp,\tbp$ notations.

Without recollision, the momenta $\bp$ formed a spanning set
of all momenta, and similarly for $\tbp$.
Since now there is a delta function among the $\bp$ momenta,
we need to exchange one tilde-momentum (out of $\tp_{a'},
\tp_{a'+1}, \tp_{b'}, \tp_{b'+1}$)
with a non-tilde momentum (out of $p_a, p_{a+1}, p_b, p_{b+1}$).
We will call them {\it exchange momenta}.

For the moment, we choose $\tp_{b'}$ and $p_b$ to be the exchange momenta and
we partition the set of all $\bp, \tbp$ momenta into two subsets of
size $k+1$ each:
$$
    A: = \{ p_1, p_2, \ldots , p_{b-1}, p_{b+1}, \ldots p_{k+1}, \tp_{b'}\}
$$
$$
    B:= \{ \tp_1, \tp_2, \ldots , \tp_{b'-1}, \tp_{b'+1}, \ldots 
\tp_{k+1}, p_{b}\}\, .
$$

It is straightforward to check that
 all $A$-momenta can be uniquely
expressed in terms of linear combinations of the $B$-momenta
(plus the $\bu$-momenta) and conversely.  In particular
$$
    p_{b-1} = p_{b} - (\tp_{\s(b-1)+1}-\tp_{\s(b-1)}) -u_{b}
$$
$$
   \tp_{b'-1} = \tp_{b'} - (p_{m+1} -p_m) +
   u_m\quad \mbox{with} \qquad m:= \s^{-1}(b'-1) \; .
$$
The letters on the pictures
 indicate the indices of the correspoding $p$ or $\tp$
momenta.

We perform a Schwarz estimate to separate $A$ and $B$-momenta at the expense
of squaring the propagators, but we
 keep the denominators with $p_1, \tp_1,p_{b-1}, \tp_{b'-1} p_{b}, \tp_{b'}$
common and only on the first power:
\be
     \prod_{j=1}^{k} \frac{1}{|\a-\ov\om(p_j)-i\eta|}
   \frac{1}{|\beta-\om(\tp_j)+i\eta|}  \leq \frac{1}{2}\big[ (a) + (b)\big]
\label{abest}
\ee
$$
   (a):=\prod_{j=1, b-1, b}
   \frac{1}{|\a-\ov\om(p_j)-i\eta|} \frac{1}{|\beta-\om(\tp_{j'})+i\eta|}
   \prod_{j=2\atop j\neq b-1, b}^{k} \frac{1}{|\a-\ov\om(p_j)-i\eta|^2}
$$
$$
    (b):=\prod_{j=1, b-1, b}
   \frac{1}{|\a-\ov\om(p_j)-i\eta|} \frac{1}{|\beta-\om(\tp_{j'})+i\eta|}
\prod_{j=2\atop j\neq b'-1, b'}^{k}
\frac{1}{|\beta-\om(\tp_j)+i\eta|^2}  \; .
$$
(with a little abuse of notations we used $j'$ for $1$,  $b'-1$ and $b'$
when $j=1, b-1$ and $b$, respectively).

Since  common factors can be explicitly expressed both in terms of
$A$ and $B$-momenta, we can compute the integral of (a) by first
integrating all $B$-momenta that removes all delta functions, then estimating
the $A$-momentum integrals. Similar procedure works for (b). The result is
(with $m:= \s^{-1}(b'-1)$)
\begin{align}
   \bE_{*g}&(\bP^*,\bu)\leq  \lambda^6 \sup_{|\cG|\leq g}
 \iint_{-Y}^Y
\rd \alpha
\rd\beta
   \int \rd \mu(\tp_{b'})\Big(\prod_{j=1\atop j\neq b}^{k+1}
   \rd \mu(p_j)\Big) \cN_\cG(\bw)\; \frac{1 }{|\a-\ov\om(p_1)-i\eta|}
\label{eq:scww} \\
& \times  
   \frac{1 }{|\beta-\om(p_1)+i\eta|} 
     \frac{1}{|\a-\ov\om(p_{b-1})-i\eta|}
   \frac{1}{|\beta-\om(\tp_{b'} - (p_{m+1} -p_m) +
   u_m)+i\eta|}
\nonumber
 \\
  & \times  \frac{1}{|\a-\ov\om(p_{b+1}-p_a+p_{a+1}- u_a)-i\eta|}
   \frac{1}{|\beta-\om(\tp_{b'})+i\eta|}\prod_{j=2\atop j\neq b-1, b}^{k}
    \frac{\lambda^2}{|\a-\ov\om(p_j)-i\eta|^2} \; . \nonumber
\end{align}

We recall the key technical bound Lemma~10.8 from \cite{ESYI}
to estimate integrals with shifted denominators.
We also recall the notation $\tri q\tri := \eta + \min \{ |q| , 1\}$. 
We start with estimating
$$
    \frac{\lambda^2}{|\a-\ov\om(p_{b+1})-i\eta|^2} \leq 
\frac{\lambda^2\eta^{-1}}{|\a-\ov\om(p_{b+1})-i\eta|}
$$
and integrating out $p_{b+1}$ by using (10.25) from \cite{ESYI}.
We collect a point singularity $\tri p_a-p_{a+1}+u_a\tri^{-1}$
and a factor $C\lambda^2\eta^{-1}\zeta^{d-3}|\log \eta|^2$.
This argument works if $b<k$; the $b=k$ case is even easier
since the denominator with $p_{k+1}$ is not present.
Then we perform the $\tp_{b'}$   integration again by (10.25) from \cite{ESYI},
and we collect
a new point singularity
$\tri p_{m+1}-p_{m}+u_m\tri^{-1}$ and a factor $C\zeta^{d-3}|\log\lambda|^2$.
Note that these two  point singularities
are not identical, because $m$, as an inverse
image of $\sigma$, is not equal to $a$.

Next we integrate out $p_{b-1}$ yielding $C|\log\lambda|$ from 
\eqref{eq:logest}.
If $p_{b-1}$ appears in one of the point singularities,
then we use the bound
\be
   \sup_{\a,r}\int \frac{\rd \mu(p)}{|\a - \om(p)
 +i\eta|}\frac{1}{\tri p-r\tri }
  \leq  C\zeta^{d-2}|\log\eta| \;
\label{1dee}
\ee
(see  (A.3) and  (A.7) from \cite{ESYI})
 to collect $C\zeta^{d-2}|\log \lambda|$ 
and  the point singularity disappears.
If $p_{b-1}$ appears in both point singularities, then we 
separate them by the telescopic estimate
(A.1) of \cite{ESYI}  before applying \eqref{1dee}.

Now we integrate out all $p_j$'s with $j\neq 1,  b-1, b, b+1$ in decreasing
order
with the successive integration scheme (10.7)--(10.9) from Section~10.1.2 of
\cite{ESYI}.
The factor $\cN_\cG(\bw)$ provides the
necessary  $|\wh B(p_j-p_{j-1})|^2$ terms with at most eight exceptions,
namely when  $j-1\in \cG$ or $\sigma(j-1)\in\cG$ (recall $|\cG|\leq 8$).
At each exceptional index  
$|\wh B(p_j-p_{j-1})|^2$ is replaced with $\langle p_j-p_{j-1}\rangle^{-2d}$
and we use \eqref{eq:2aint} instead of (10.8) of \cite{ESYI}.
Thus the successive production of the factors $(1+ C\lambda^{1-12\kappa})$
 breaks at these indices and we obtain a uniform constant $C$ instead.
The successive scheme also breaks at 
the indices $j= b-1, b, b+1$ that have already been integrated
out. Furthermore, it also may break at $j=m+1, a+1$,
i.e., at the indices where the point singularities are first affected
(unless $b-1\in \{ m+1, a+1\}$ and the point singularity
has already been integrated out).
At each of these indices we use \eqref{1dee}
 and collect 
$C\lambda^2\eta^{-1}\zeta^{d-2}|\log\eta|^2$
instead of the constant factor from (10.7)--(10.9) of \cite{ESYI}.

Since there are at most 13  exceptional indices, so we collect at most
$$
C^{13}[\lambda^2\eta^{-1}\zeta^{d-2}|\log\eta|^2]^2
(1+ C\lambda^{1-12\kappa})^{K}.
$$ 
The other factors of $\cN_\cG$, that are not explicitly
used in the successive integration, are estimated by supremum norm,
except $|\wh\psi_0(p_1)|^2$.
Finally the $\rd\a$, $\rd\beta$ integrals contribute with an
additional $C|\log\lambda|^2$. The last $p_1$-integral
is finite by the factor $|\wh\psi_0(p_1)|^2$. Collecting
these estimates and recalling that \eqref{1dee} has been
used twice,
 the result is \eqref{eq:recesttr}. $\;\;\Box$

\bigskip

{\it Proof of Proposition \ref{prop:rec1}.}
This proof is  very similar to the previous
one but the estimate is weaker
since (10.25) from \cite{ESYI} can be used only once.
We just indicate that the set of $A$ and $B$ momenta are as follows:
$$
    A: = \{ p_1, p_2, \ldots , p_{b-1}, p_{b+1}, \ldots , p_{k+1}\}\; ,\qquad
    B:= \{ \tp_1, \tp_2, \ldots , \tp_{k-1}, p_{b}\} \; ,
$$
and we leave the details to the reader. $\;\; \Box$

\bigskip

The case of the triple collision, (\ref{eq:triplev}),
can be easily reduced to the case of a recollision
(Fig.~\ref{fig:tripcon}). We first use the
analogue of the estimate (\ref{eq:extr}). Clearly $g=0$
in case of a triple collision. Then we remove half of each of the two gates
 (Lemma \ref{lemma:removehalfgate})
and we collect a factor $\lambda^2|\log\lambda|^2$.
 The  resulting Feynman graph
has either a recollision
 or a gate at the end. In the first case we
apply  Proposition \ref{prop:rec}.
In the second case, we remove half of each gates by
a second application of Lemma \ref{lemma:removehalfgate}, then
the estimate (\ref{eq:2tru}) from
 Lemma \ref{lemma:remove}  together 
with (\ref{eq:ladlogweak}) can be applied.
$\;\;\Box$

\bef\bec
\epsfig{file=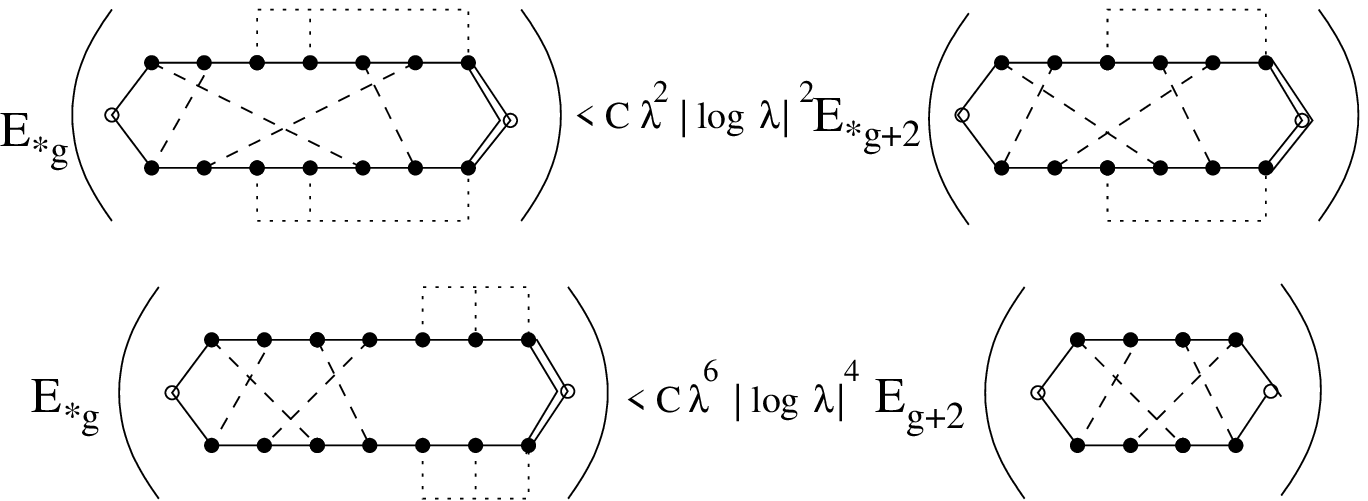, scale=1}
\eec
\caption{Estimate of triple collisions for $w<c+1$ and $w=c+1$}
\label{fig:tripcon}
\eef

\subsection{Cancellation with a gate}

The key mechanism behind the estimates \eqref{eq:nrev}--\eqref{eq:nestev}
is the cancellation between a gate and a $\theta$-label, 
We first present estimates on general graphs.

\subsubsection{Cancellation between a gate and $\theta$}

Fix $n, n'$ integers and consider a partition $\bP\in \cP_{n,n'}$
with no single lump
on the set $I_n\cup \wt I_{n'}$ within the vertex set $\cV=\cV_{n,n'}$.
Let $1\leq m \leq n+1$ an integer. We define two new cyclically ordered
sets:
\be
  \cV' : = \{0, 1, 2, \ldots, m-1, \clubsuit, m, \ldots , n, 0^*, \tilde n', \wt{n'-1},
 \ldots ,\tilde 1 \}
\label{eq:V}
\ee
$$
   \cV'':=\{0, 1, 2, \ldots, m-1, \diamondsuit, \heartsuit, m, \ldots,
   n, 0^*, \tilde n', \wt{n'-1},
 \ldots ,\tilde 1 \}
$$
with additional elements $\cs, \ds, \hs$. For the result of this
section it  would make no difference if the extra elements
were inserted into the sequence of tilde-variables.

These sets can be naturally identified with $\cV_{n+1,n'}$ and $\cV_{n+2,n'}$
and we will use this identification with the obvious choice of
the relabelling map.
We define two partitions on these sets, $\bP'\in \cP_{n+1, n'}$ and
$\bP''\in \cP_{n+2, n'}$, simply by adding the single lump $\{ \cs \}$ to $\bP$
in the first case and the double lump $\{ \ds, \hs \}$
in the second case. This will correspond to adding a $\vartheta$ 
label or a gate  whose potential labels have been paired
to the original partition $\bP$, respectively. The following
lemma shows that the $V$-value of these two partitions cancel each other
up to the lowest order (Fig.~\ref{fig:gatecan1con}).

\begin{lemma}\label{lemma:gatecancel}
With the notations above and assuming $\lambda^3\ll \eta\ll \lambda^2$, we have
\be
     \Big|  V_{(*)} (\bP') + V_{(*)} (\bP'')\Big|
     \leq C\lambda^2\eta^{-1/2} E_{(*)g=0} (\bP) \; .
\label{eq:gatecancel}
\ee
\end{lemma}

\bef\bec
\epsfig{file=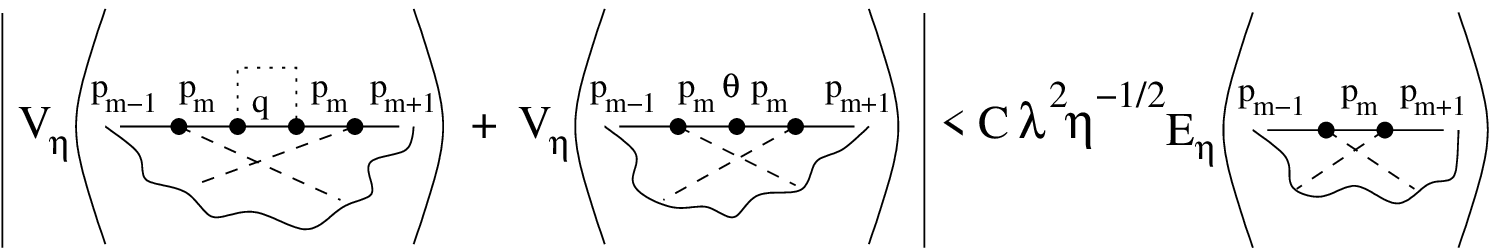, scale=.9}
\eec
\caption{Cancellation of a gate and $\theta$ (partition is the same elsewhere)}
\label{fig:gatecan1con}
\eef

{\it Proof of Lemma \ref{lemma:gatecancel}.} Introduce the notations
 $\bp:= (p_1, \ldots , p_{n+1})$,
$\tbp := (\tp_1, \ldots , \tp_{n'+1})$,
$$
    \int\rd\bp: = \iint \rd p_1\rd p_2 \ldots \rd p_{n+1}
$$
and similarly for $\int\rd\tbp$. Then we have
$$
    V_{(*)}(\bP') + V_{(*)}(\bP'') = 
    \lambda^{n+n'+g(\bP)} \frac{e^{2t\eta} }{(2\pi)^2}
\iint_{-Y}^{Y} \rd\a\rd\beta \; e^{i(\alpha-\beta)t}
    \int \rd\bp\; \rd \tbp \; \Delta(\bP, \bw, \bu\equiv 0)
  \cM(\bw)
$$
\be
    \times\Omega(\alpha, p_m)
    \prod_{j=1}^{n+1 ,\; (n)} \frac{1}{\alpha -\ov\om(p_j)-i\eta}
    \prod_{j=1}^{n'+1, \; (n')} \frac{1}{\beta -\om(\tp_j)+i\eta}\; ,
\label{eq:twosum}
\ee
with
\be
    \Omega(\alpha, p_m):=\Bigg[  \int
    \frac{|\wh B(q-p_m)|^2 \rd q}{\alpha -\ov\om(q)-i\eta}
-\ov\theta(p_m) \Bigg]
    \frac{\lambda^2}{\alpha -\ov\om(p_m)-i\eta} \; .
\label{eq:fact}
\ee
The expression $n+1, (n)$ on the product sign indicates that
for the truncated values (*) the last 
fraction is not present, i.e. $j$ runs up to $n$.
Notice that the sum of the contributions of a gate and a $\vartheta$ inserted
between $m$ and $m-1$ yields an additional factor $\Omega(\alpha, p_m)$
in the $V$-value of the $\bP$ partition.
Using (\ref{A4}) and 
(\ref{zhold}) with $\e=\eta$, $\e'\to0+0$,
 we have
$$
    \Big|  \int
    \frac{|\wh B(q-p_m)|^2\rd q}{\alpha -\ov\om(q)-i\eta}-\ov\theta(p_m) \Big|
    \leq C\Big(\eta^{1/2} +\eta^{-1/2}| \a - \lambda^2\Theta(\a)-e(p_m)|\Big) \; .
$$
Therefore, using (\ref{eq:holder}) and $\lambda^3\ll \eta\ll\lambda^2$, we have
\be
   | \Omega(\alpha, p_m)|
  \leq C\lambda^2\eta^{-1/2}\Big(1 + \frac{\lambda^2 |\Theta(\alpha)- \theta(p_m)|}
  { | \a - \om(p_m) +i\eta|}\Big) \leq C\lambda^2\eta^{-1/2}
 \label{eq:canest}
\ee
uniformly in $\a$ and $p_m$. $\;\;\Box$

\bigskip

The proof shows that the cancellation between the gate and the 
$\vartheta$ is completely
local in the graph.
 In particular, if we fix $L$ locations, maybe with multiplicity,
 between the elements of  $V_{n,n'}$, and
we consider all possible $2^L$ combinations of insertions of
gates and $\vartheta$'s at these locations, then we gain a factor
$\lambda^2\eta^{-1/2}$ from each location.

More precisely, let $\underline{\upsilon}\in \bN^{\cV_{n,n'}}$
be a given sequence of integers, $\upsilon_0, \upsilon_1, \ldots,
\upsilon_{0^*}$, 
labelled by the elements of $\cV_{n,n'}$. The number $\upsilon_j$ indicates
how many gates or $\vartheta$'s 
 are inserted between the $j$-th and $(j-1)$-th vertex.
Let $|\underline{\upsilon}|: 
= \sum_j \upsilon_j$ be the total number of insertions. A sequence
$S\in \{ g, \vartheta\}^{|\underline{\upsilon}|}$ 
encodes whether the insertion is gate or $\vartheta$.

Fix a sequence $\underline{\upsilon}$ and
 for any $S\in \{ g, \vartheta\}^{|\underline{\upsilon}|}$ we define
 the extended set $\cV_S$  consisting of $\cV_{n,n'}$ and we
 insert extra single or double symbols for $\vartheta$ 
and gate indices (determined by $S$)
at the locations given by $\underline{\upsilon}$. In the example above, we have $\underline{\upsilon}= (0,0,\ldots, 0, 1, 0 \ldots
0)$ (i.e. $q_m=1$, the rest is zero), $|\underline{\upsilon}|=1$,
 and $\cV'$ corresponds to $S=\{\vartheta\}$,
 while $\cV''$ corresponds to $S=\{ g \}$.
Given  a partition $\bP\in \cP_{n,n'}$,
we also define the extended partition $\bP_S$ on $\cV_S$
 by simply adding the single symbols
as single lumps and the double symbols as paired lumps to $\bP$.

\begin{lemma}\label{lemma:manygate}
With the notations above, for any fixed 
$\bP\in \cP_{n,n'}$ and $\underline{\upsilon}\in \bN^{\cV_{n,n'}}$
\be
\Big| \sum_{S\in  \{ g, \vartheta\}^{|\underline{\upsilon}|}} V_{(*)} (\bP_S)\Big|
\leq \Big(C\lambda^2\eta^{-1/2} \Big)^{|\underline{\upsilon}|}E_{(*)} (\bP)  \; .
\label{eq:manygate}
\ee
\end{lemma}

{\it Proof.} Notice that the sum of the contributions of a gate
and a $\vartheta$ inserted at the same place between $m$ and $m-1$
yields a factor of $\Omega(\alpha, p_m)$, while the same insertion
between $\wt m$ and $\wt {m-1}$ yields a factor $\ov{\Omega(\beta,
\tp_m)}$. These insertions are independent of each other, thanks
to the summation over all possible $S$-combinations. So $\sum_S
V(\bP_S)$ is represented by an expression similar to
(\ref{eq:twosum}), where a total factor
$$
   \prod_{m=1, \ldots , n, 0^*} [\Omega(\alpha, p_m)]^{q_m}
   \prod_{m=\wt n', \ldots , \wt 1, 0} [\ov{\Omega(\beta, \tp_m)}]^{q_m}
$$
is inserted. The uniform estimate (\ref{eq:canest}) for each $\Omega$
factor gives (\ref{eq:manygate}). $\;\;\;\Box$.

\medskip

Next we will apply
Lemmas \ref{lemma:gatecancel}--\ref{lemma:manygate}
to prove \eqref{eq:nrev}--\eqref{eq:nestev}.
  The 
difficulty is that these estimates hold only if 
the gate remains  isolated even after the lumping procedure,
otherwise the gain comes from the artificial recollision
introduced by the lump. Recall that the lumping procedure
has two steps. The original partition $\bD_0$ may lump nontrivially,
yielding the derived partition $\bD$, due to a few possible coincidences
among the gate or recollision labels of $\psi$ and $\ov{\psi}$.
Then the non-single elements of $\bD$ (denoted by $\bD^*$)
lump into a coarser partition imposed by $A\in \cA(\bD^*)$
due to the connected graph formula. 

We will estimate the value of individual graphs only. The 
number of terms in the summations
in \eqref{eq:nrev}--\eqref{eq:nestev} is bounded by $O((c+4)^4)
\leq CK^4$.
This extra factor $CK^4$ will be added to
the factors gained in the cases discussed below
to obtain
 \eqref{eq:nrev}--\eqref{eq:nestev}.

\subsubsection{Non-repetition graphs with a gate}

To prove  \eqref{eq:nrev}, we first note that the non-repetition
rules in  $\psi^{(1),nr}_{t,k}$  force $\bD$
to be identical with $\bD_0$ unless $h, h'= g$. In this latter case
there is a gate both in the expansion of $\psi$ and $\ov{\psi}$, 
and either $\bD=\bD_0$
or $\bD$ lumps the two gate-lumps in $\bD_0$ together.

We first consider the case $\bD \neq \bD_0$. Then
$\bD$ has a lump 
consisting of all four gate indices, $\{ w, w+1, \wt w, \wt{w+1}\}$.
The corresponding Feynman graphs can be identified with certain 
nontrivial lumpings of  non-repetition graphs
 on $I_{c+2}\cup \wt I_{c+2}$, where
the indices $w$ and $w+1$ are lumped together.
More precisely, for
a given $\sigma\in \fS_c$, $w \in I_c$, $h=h'=g$, $\bA'\in \cA_c$, 
\be
   \sum_{\bD \neq \bD_0} \sum_{\bA \in \cA(\bD^*) \atop \bA = \bA'}
   V(\bP(\bA, \bD))c(\bA) = 
   \sum_{\bA \in \cA_{c+2}, \wh\bA=\bA'\atop w\equiv w+1 (mod \;\bA)}
   V(\bA, \wt\sigma) c(\bA)\; ,
\label{rew}
\ee
where $\wt\sigma \in \fS_{c+2}$ is the natural extension of $\sigma$
where two new elements, $w, w+1$,  are added to the base set
$I_c$, the indices are shifted by
the embedding map $s_w^h$ (Section \ref{sec:symm})
   and  $\sigma(w)=w$,  $\wt{\sigma}(w+1)=\wt{w+1}$.
The summation has at most $c+1$ terms and it expresses
the choice of joining  $w, w+1$ to one of the existing
lumps in $\bA'$ or keeping the lump $\{ w, w+1\}$ separate in $\bA$.
Because of lumping $w$ and $w+1$, the partition $\bA$ is
non-trivial, and $q(\bA, \wt\sigma)\ge 1$ (see \eqref{def:q}).
After estimating $|V(\cdots)|$ by $E(\cdots)$,
 each term on the right hand is estimated by the bound
\be
     \sup_{ \bu} E_{(*)g}
   (\bA, \sigma,\bu) \leq C|\log \lambda|^2
   \Big(\lambda^{\frac{1}{3}-(\frac{17}{3}d+\frac{3}{2})\kappa-O(\delta)} 
     \Big)^{q(\bA, \sigma)}  \; 
\label{eq:jointdeg}
\ee
that holds
whenever $\sigma\in \fS_k$, $\bA\in \cA_k$ and $\kappa
 \leq \frac{2}{34d+9}$ (see (9.4) from \cite{ESYI}).

The estimate \eqref{eq:jointdeg}, combined with the combinatorial
bound on $c(\bA)$ and on the summation
 (Lemma~\ref{lemma:smallcomb}),
gives \eqref{eq:nrev} in the case when $\bD\neq \bD_0$.

Now we focus on 
 the case $\bD =\bD_0$. If at least  
one of the gate-lumps, $\{ w, w+1\}$ or $\{ \wt w, \wt{w+1}\}$,
do not remain isolated in $\bA$, then  we can repeat
the argument above since $\bP(\bA, \bD)$ has a non-trivial lump
of size at least 4.

Finally, we can assume that the gate lumps remain
isolated in $\bA$. We fix $\bA'$ and consider the sum
of four terms corresponding to $h, h'\in \{g, \theta\}$.
The partition $\bA$ is defined by adding the gate lump(s) to $\bA'$.
Note that $c(\bA)$ is the same for all these four cases
since replacing a $\theta$ index with an isolated gate lump
adds only a single lump to $\bA$.
For these partitions, we 
apply Lemma \ref{lemma:manygate} (with the choice 
$\upsilon_w=\upsilon_{\wt w}$, all other $\upsilon$'s are zero)
to obtain  a factor $(C\lambda^{1-\frac{\kappa}{2}})^2$.
The remaining non-repetition graphs
 bounded by $O(|\log\lambda|^2)$ by 
using Proposition~9.2 from \cite{ESYI} with $q=0$.
This completes the proof of \eqref{eq:nrev}. $\;\;\Box$

\subsubsection{Last gate}

First we consider the case when the lumps  $\{ w, w+1\}\subset I_n$
and $\{ \wt w, \wt{w+1}\} \subset \wt I_{n'}$
of the two first gates remain isolated in $\bP=\bP(\bA, \bD)$.
As in the previous section, by applying 
 Lemma \ref{lemma:manygate}, we can sum up the two times two
 possibilities for the first components of the code $h$ and $h'$,
i.e. sum up four Feynman diagrams that differ only by the choice
of gate or $\theta$ at the $w$ or $\wt w$ position. 
We collect $(\lambda^{1-\kappa/2})^2$.
The remaining two gates will be isolated from the rest in $\bP$
by Operation I, then half of each gate is removed by Operation III,
collecting $\Lambda^2\lambda^2|\log\lambda|^2$. Finally, (\ref{eq:2tru})
can be used to remove the remaining two halves of the gates,
collecting $\lambda^2$. By (\ref{eq:ladlogweak}),
 the untruncated $E$-values of
the remaining graph are bounded by $O(|\log\lambda|^2)$.
 The total estimate is
 $C\lambda^{6-(4d+1)\kappa-O(\delta)}$.

 If only one of the lumps, $\{ w, w+1\}$ or
 $\{ \wt w, \wt{w+1}\}$,  remains isolated in $\bP$, we can still apply
Lemma \ref{lemma:gatecancel} to obtain a cancellation of order
$\lambda^{1-\kappa/2}$ from adding up the $V$-values of those
pairs of graphs that differ only by changing this gate to $\theta$.
 Note that the value $c(\bA)$
is again the same for these two graph.

Next we consider those lumps among  $\{ w, w+1\}$ and
 $\{ \wt w, \wt{w+1}\}$  that do not remain isolated in $\bP$.
By breaking up lumps via Operation I,
 we can ensure that every such gate is either lumped exactly
with one other gate or with a core index pair.
For definiteness, let  $\{w, w+1\}$ from $I_n$ be such a gate.
We also isolate all other gates from the rest. To do that,
Operation I is used at most four times at the total expense
of $\Lambda^4$. We distinguish three cases:

\bigskip

{\it Case 1.} The gate $\{ w, w+1\}$
is lumped with a core index-pair $(j, \wh\sigma(j))$,
where $\wh\sigma$ is the natural extension of $\s$ 
from $I_c$ to $I_n$. If $j$ is next
to $w$ or $w+1$, say $j=w+2$ (Fig.~\ref{fig:gate21con}),  
then both vertices of the gate can be removed
by Operation III since the momenta between $(w-1, w)$ and $(w, w+1)$ do not
appear in any delta function, and we gain $\lambda^2|\log\eta|^2$ from
this gate.
We can now remove the two gates at the end 
(using Operation III and (\ref{eq:2tru}) as above), collect
an additional $\lambda^4|\log\lambda|^2$. The remaining 
gate at $\{\wt w, \wt{w+1}\}$
 can be removed by Operation IV
at the expense of $\lambda^{-\kappa}|\log\lambda|$. 
By Proposition~9.2 from \cite{ESYI},
 the remaining graphs are bounded by $O(|\log\lambda|^2)$.
 We thus collect
 $C\lambda^{6-(8d+1)\kappa-O(\delta)}$.

\bef\bec
\epsfig{file=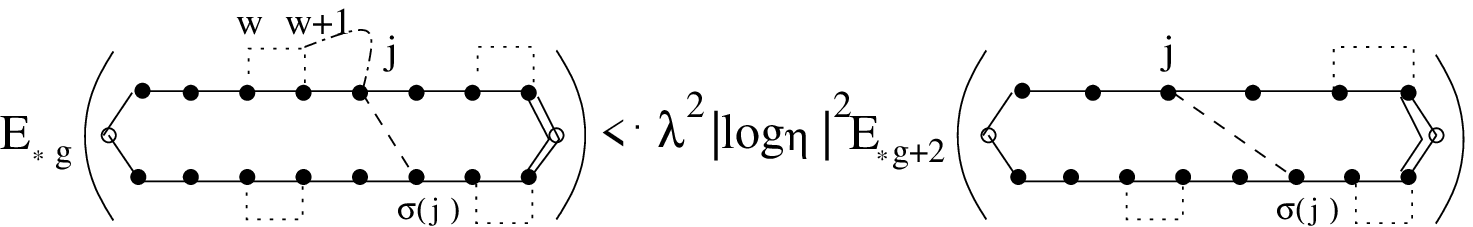, scale=1}
\eec
\caption{Case 1. Removal of a gate lumped to an adjacent core index}
\label{fig:gate21con}
\eef

If $j$ is not next to $w$ or $w+1$ (Fig.~\ref{fig:gate22con}),
 then we remove $w+1$
by Operation III,  break up the lump into $\{j, w\}$ and $\{ \wh \s(j)\}$
by Operation I and
remove the single lump $\{ \wh \s(j)\}$ by Operation II.
The total price for these steps is $\Lambda\lambda^2\eta^{-1}|\log\eta|^2$.
We can now again remove the two gates at the end,
collect $\lambda^4|\log\lambda|^2$
and we end up with a graph with a one sided recollision, so
(\ref{eq:halfrecest}) from Proposition \ref{prop:rec1}
applies. We collect 
$C\lambda^{6-(14d+2)\kappa-O(\delta)}$ in this case.

\bef\bec
\epsfig{file=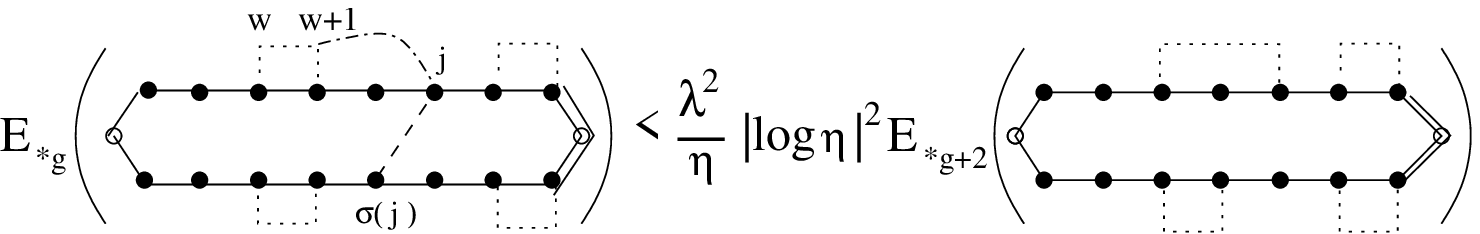, scale=1}
\eec
\caption{Case 1. Removal of a gate lumped to a non-adjacent core index}
\label{fig:gate22con}
\eef

\medskip

{\it Case 2.}  The gate $\{ w, w+1\}$ is lumped with the other gate
in $I_n$, i.e. with   $\{ n-1, n\}$.
If $w+1$ and $n-1$ are neighbors
(Fig.~\ref{fig:gate31con}), then we can remove three vertices
$n-2=w+1, n-1, n$ by Operation III.
We collect $\lambda^{3}|\log\eta|^3$.
On the $\wt I_{n'}$ side, we remove the gate that is not adjacent
to $0^*$ by Operation IV  at the
expense of $\lambda^{-\kappa}|\log\lambda|$ as above.
The other gate is adjacent to $0^*$; first we remove
its leg not adjacent to $0^*$ by Operation III, collecting
$\lambda|\log\lambda|$. Finally, we remove the two remaining
vertices, originally with indices $n-3$ and $n'$ that now form
single lumps and they are both adjacent to $0^*$. By using
\eqref{eq:2tru}, we gain a factor $\lambda^2$.
Altogether we thus collect $C\lambda^{6-(8d+1)\kappa-O(\delta)}$.

\bef\bec
\epsfig{file=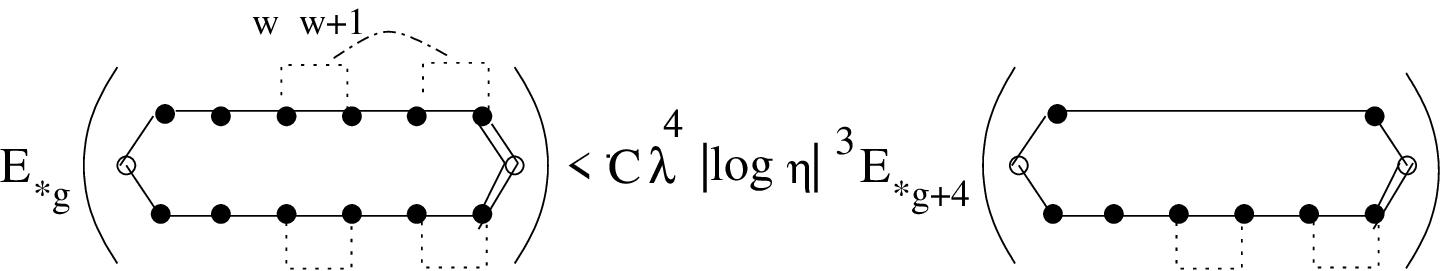, scale=1}
\eec
\caption{Case 2. Removal of two adjacent
lumped gates}
\label{fig:gate31con}
\eef

If $w+1$ and $n-1$ are not neighbors (Fig.~\ref{fig:gate32con}),
then we remove $w+1$ and $n$, gaining $\lambda^2|\log\eta|^2$
and  the remaining partition again has a one-sided recollision.
The other gate adjacent to $0^*$ can be removed by collecting
$\lambda^2|\log\lambda|^2$, the remaining gate collects 
$\lambda^{-\kappa}|\log\lambda|$,
and finally we use (\ref{eq:halfrecest}). The result is again
 $\lambda^{6-(12d+1)\kappa-O(\delta)}$.

\bef\bec
\epsfig{file=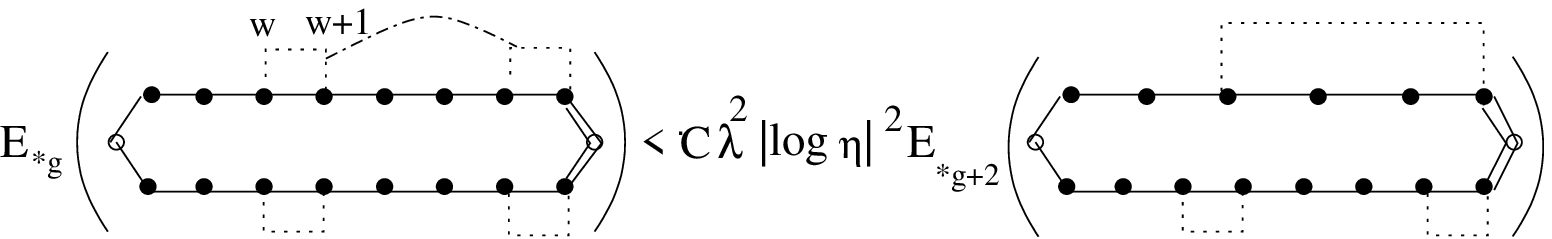, scale=1}
\eec
\caption{Case 2. Removal of half of each gates lumped on the same
side}
\label{fig:gate32con}
\eef

\begin{figure}
\begin{center}
\epsfig{file=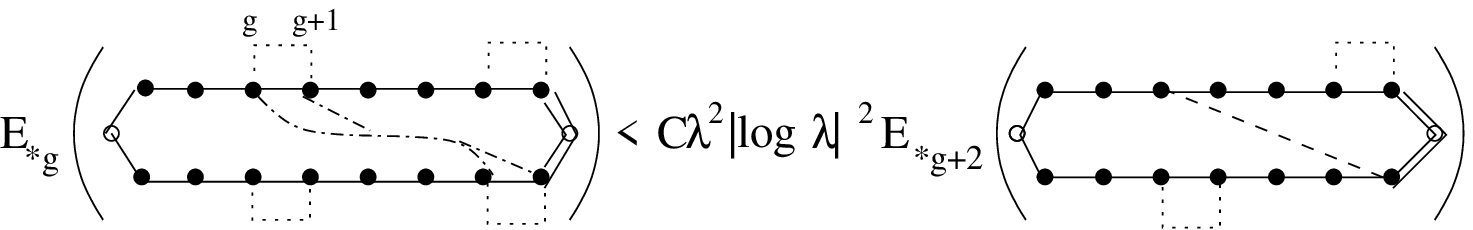,scale=1}
\end{center}
\caption{Case 3. Removal of two opposite lumped gates}\label{fig:gate1con}
\end{figure}

{\it Case 3.} Finally, if neither $\{ w, w+1\}$ nor $\{ \wt w, \wt{ w+1} \}$
falls into Case 1 or 2, then each of them either remains isolated and collects
$\lambda^{1-\kappa/2}$ from Lemma \ref{lemma:gatecancel}
 or  is lumped with another gate on the
 ``opposite side'' (Fig.~\ref{fig:gate1con}).
For definiteness, assume $\{ w, w+1\}$ is lumped
 with $\{ \wt\ell, \wt{\ell+1} \}$  in $\wt I_{n'}$,
where $\wt\ell$ is either $\wt w$ or $\wt{n'-1}$.
Then we simply remove half of each gates, say $w+1$ and
$\wt {\ell+1}$ using Operation III,
 gain $\lambda^2|\log\eta|^2$ and we extend the set of
core indices to include $w$ and extend the permutation $\sigma$ by adding
$\sigma(w)=\ell$. Therefore we effectively 
gained $\lambda|\log\eta|$ from each such
gate. Finally, after having gained either $\lambda^{1-\kappa/2}$
or $\lambda|\log\lambda|$ from each gate,
 we gain $\lambda^2|\log\lambda|^2$ from the remaining
truncated graph \eqref{eq:ladlogweak}--\eqref{eq:Etrunc} and we thus collect
at least $C\lambda^{6-(8d+1)\kappa-O(\delta)}$. This completes
the proof of (\ref{eq:lastev}). $\;\;\;\Box$.

\subsubsection{Nest}

 The procedure is very similar to the analysis of the last gate, so
we just outline the steps. We point out that the main
reason why the nested graphs are small is the cancellation
between the gate and $\theta$ inside the nest. This is a
different mechanism than the one used in \cite{EY}.

First we consider the cases when both nests are independent, i.e.
no part of the nest in $I_n$ is lumped with any part of the
nest in $\wt I_{n'}$.  We will show that one can gain
at least $ \lambda^{3-(2d+2)\kappa-O(\delta)}$
from each such nest. The total gain from both nests is then
$\lambda^{6-(4d+4)\kappa -O(\delta)}$.

 Consider the gate $\{ n-2, n-1\}$ inside the nest.
If this gate remains an isolated lump in $\bP$, then it cancels
the same graph with $\theta$ up to order $\lambda^{1-\kappa/2}$
by Lemma \ref{lemma:gatecancel}. After this cancellation,
 the outer shell of the nest,
$\{ n-3, n\}$,
becomes a gate in the reduced graph that is adjacent with $0^*$.
After separating it from the rest by Operation I,  if necessary,
its removal yields $\lambda^2$ because of the truncation.
Therefore we gain at least 
$\Lambda\lambda^{3-\kappa/2}\leq \lambda^{3-(2d+\frac{1}{2})\kappa-O(\delta)}$
 from this nest.

If the gate $\{ n-2, n-1\}$ is not isolated in $\bP$, then we
again distinguish a two cases.

 If $\{ n-2, n-1\}$ is lumped with a
 core index $j<n-3$ but not
with its outer shell, $\{ n-3, n\}$,
 then we can always create a one-sided recollision
in such a way that $n-2$ will be
paired with $\sigma(j)$ in the extended permutation, while $n-1$ is removed
 (Operation III, gain $\lambda|\log\lambda|$)
and $j$ is removed (Operation II, lose $\lambda^{-1-\kappa}$).
The net result is $\lambda^{-\kappa}|\log\lambda|$ and the outer shell
of the nest, $\{ n-3, n\}$, becomes a genuine recollision. We may have
to separate this from the rest of the graph by an Operation I
before applying  (\ref{eq:halfrecesttr}).
This will
effectively give $\Lambda\lambda^{3-2\kappa}\zeta^{4d}|\log\lambda|^{O(1)}
\leq \lambda^{3-(6d+2)\kappa-O(\delta)}$ 
for this nest. In this calculation we used only $\lambda^{3-\kappa}$
from (\ref{eq:halfrecesttr}), 
the additional $\lambda$
is due to the truncation on the ``other side'' and will be counted
there.

If the gate $\{ n-2, n-1\}$ is lumped with its outer shell, then the
momenta between $(n-3, n-2)$, $(n-2, n-1)$ and $(n-1, n)$
can be freely integrated,
we can remove the nest completely and collect $\Lambda\lambda^4\leq C
\lambda^{4-2d\kappa}$
from this nest. We may have to use Operation I once to separate
the nest from the rest of the graph.

So far we have treated the cases when the two nests  are independent.
In the remaining cases some parts of the nests are lumped with each other.

If the gate $\{ n-2, n-1\}$ is lumped with the
 other gate $\{ \wt{n-2}, \wt{n-1}\}$,
then half of each gate is removed, say $n-1,\wt{n-1}$
by Operation III  (gain $\lambda^2|\log\lambda|^2$),
the other halves are separated from the rest of the
graph (Operation I) and then connected by an extending the permutation
$\sigma (n-2)=n-2$ (i.e. we include $n-2$ among the core indices).
The resulting graph has two recollisions, so (\ref{eq:recesttr}) applies
and the total size is $\Lambda\lambda^{8-3\kappa}\zeta^{4d}
|\log\lambda|^{O(1)}$.

Finally, if $\{ n-2, n-1\}$ is lumped with the 
outer shell $\{ \wt{n-3}, \wt n\}$
of the other nest, then we isolate and
remove the gate $\{\wt{n-2}, \wt{n-1}\}$
 inside the other nest (price: $\Lambda\lambda^{-\kappa}
|\log\lambda|$), remove $\wt n$ and $n-1$ each as one half of a gate, gaining
$\lambda^2$ and the remaining graph is a one-sided 
truncated recollision graph after a possibly isolating the recollision
lump $\{n-3, n\}$ from the rest.
Thus (\ref{eq:halfrecesttr}) gives 
$\Lambda\lambda^{4-\kappa}\zeta^{4d}|\log\lambda|^{O(1)}$,
after a possible application of Operation I.
The total size is  $O(\lambda^{6-(8d+2)\kappa-O(\delta)})$.
Collecting the various cases, we obtain (\ref{eq:nestev}). $\;\;\;\Box$

\section{Convergence of the ladder diagrams  to the heat equation}
\label{sec:wigner}
\setcounter{equation}{0}

We start the proof of Theorem~\ref{thm:laddheat}
by noticing that
$$
    W_\lambda(t, k, \cO) = \int V^\circ_{\e\xi}
  \big( \bA_0, \wh\cO(\xi, \cdot)\big)\rd \xi, \qquad k\ge1
$$
with $\bA_0$ being the trivial partition on $I_k$,
where we chose the function $Q(v)$ in the definition
of $V^\circ$ to be $\xi$-dependent, namely $Q(v)= Q_\xi(v):=\wh\cO(\xi, v)$
(see \eqref{def:circ}
for the definition
of $V^\circ$).
First we note that the $\rd\xi$ integration
can be restricted to the regime $\{|\xi|\le\lambda^{-\delta}\}$
with a negligible error (even after summation over $k$):
\be
    \sum_{1\leq k<K}W_\lambda(t, k, \cO) = \sum_{1\leq k<K}
 \Xi^\circ_k +o(1)\; , \qquad
\Xi^\circ_k:=\int^* V^\circ_{\e\xi}
  \big( \bA_0, \wh\cO(\xi, \cdot)\big)\rd \xi \; ,
\label{xires}
\ee
where   use the notation 
$$
   \int^* \big(\cdots\big)\rd\xi : = \int \big(\cdots\big){\bf 1}(|\xi|\leq 
\lambda^{-\delta} )\rd \xi\; .
$$
 To see \eqref{xires}, we first recall that
 replacing
$V^\circ (\cdots)$ with $V(\cdots)$ yields a negligible
error even after summing up for all $k$
(Lemma~7.1 of \cite{ESYI}). The linearity of the estimate
in $\| Q_\xi\|_\infty = \sup_v |\wh\cO(\xi, v)|$ guarantees
the integrability in $\xi$ since $\cO$ is a Schwarz function.
We then use 
the estimate
$$
   \Big| V_{\e\xi}
  \big( \bA_0, \wh\cO(\xi, \cdot)\big) \Big| \leq \|Q_\xi\|_\infty
  \sup_{\xi, \bu} E(\sigma=id, \bu)
$$
and the uniform bound  \eqref{eq:ladlogweak} and finally we conclude 
\eqref{xires}
by the arbitrarily fast decay of $  \|Q_\xi\|_\infty$
in $\xi$.

\medskip
From the definition of $\Xi^\circ_k$, we have
\begin{align}
\Xi_k^\circ= & 
\lambda^{2k}
\iint_\bR \frac{\rd \alpha \rd \beta}{(2\pi)^2}
 \; e^{it(\alpha-\beta)+ 2 t \eta} \int^*  \rd \xi 
\int  \prod_{j=1}^{k+1} \rd \mu(v_j) \;  
\wh\cO(\xi, v_{k+1})\overline{\wh W_0}(\e\xi, v_1) 
  \nonumber \\
 & \times 
\prod_{j=1}^{k+1} \Big[ 
  \ov{R_\eta\Big(\a, v_j +\frac{\e\xi}{2}\Big)}
   R_\eta\Big(\beta, v_j -\frac{\e\xi}{2}\Big)|\wh B(v_j-v_{j+1})|^2\Big] \;.
\label{9.51}
\end{align}
To simplify the notation, in \eqref{9.51} we followed the convention, that
 $|\wh B(v_j-v_{j+1})|=1$ for $j=k+1$ because of the non-existence of
 $v_{k+2}$.
Similar convention will be followed later, also for
$|\wh B(v_{j-1} - v_{j})|^2 =1$ if $j=1$. Note also  
that the measure $\rd v_j$ has been changed to $\rd\mu(v_j)$
by using the support properties of $\wh B$ and $\psi_0$
\eqref{suppM}.

The estimates of the error terms were performed with the 
choice $\eta= \lambda^{2+\kappa}$.
However, $\Xi_k^\circ$,  given by \eqref{9.51}, is clearly
{\it independent} of $\eta$;
this follows from the $K$-identity (formula (6.2) in
\cite{ESYI}). Therefore we can change the  value of 
$\eta$ to $\eta:=\lambda^{2+4\kappa}$
for the rest of this calculation and we define
$$
        R(\a, v):=  R_{\eta}(\a, v) \; , \qquad \mbox{with} 
\quad \eta:=\lambda^{2+4\kappa} \; .
$$

We also  recall, that the restriction of the $\rd\alpha\rd\beta$
integration in \eqref{9.51} to any set that contains 
$\{ \a, \beta \, : \,|\a|, |\beta|\leq Y = \lambda^{-100}\, \}$
 results in
negligible errors, even after the summation over $k$
(Lemma~7.1 of \cite{ESYI}).
We will consider the set  $D:= \{ (\a, \beta) \, : \,
 |\a + \beta|\leq 2Y, |\a-\beta|\leq 2Y\,\}$.
We denote by $\Xi_k$ the version of $\Xi_k^\circ$
given by  formula \eqref{9.51} with the $\rd\a\rd\beta$ 
integrals  restricted to $D$,
$$
    \Xi_k : = \lambda^{2k} \iint_{D} \frac{\rd \alpha \rd \beta}{(2\pi)^2}
 \; \Big[ \mbox{Integrand from \eqref{9.51}} \Big] \; ,
$$
then
$$
   \sum_{1\leq k\leq  K} |\Xi_k^\circ- \Xi_k| = o(1) \; .
$$
We also remind the reader  that this
 argument literally does not apply
to the trivial $k=0$ case, when the $\rd\a\; \rd\beta$ integral in
\eqref{9.51} gives free evolutions and this term should be computed
directly:
\be
     \Xi_0: = \int^* \rd\xi\rd v\;
    e^{it\e v\cdot \xi}\; e^{2t\lambda^2 \mbox{Im}\, \theta (v)}\;
 \wh\cO(\xi, v)\ov{\wh W_0}(\e\xi, v) + o(1)\; ,
\label{xi01}
\ee
where the error term comes from the error term in $\ov{\theta}(v+\e\xi/2)
- \theta(v-\e\xi/2) = 2 i \cI(v) + O(\e\xi)$. 
By using $t\lambda^2 \to \infty$, the bound  
$$
    \mbox{Im}\;\theta(v) \leq -c_1\min \{ |p|^{d-2}, |p|^{-1} \}\; ,
$$
(from Lemma~3.2 of \cite{ESYI})
 and the decay of the observable,
 one easily obtains that 
 $|\Xi_0| = o(1)$ anyway.

\medskip

To evaluate the integral \eqref{9.51}, we need the following
 crucial technical lemma which is proven 
 in the Appendix.

\begin{lemma}\label{lemma:opt1}
Let $\kappa < 1/8$, define $\gamma: = (\a +\beta)/2$ and
let  $\eta$ satisfy
 $\lambda^{2+ 4\kappa}\leq \eta\leq \lambda^{2+\kappa}$.
Then for $|r|\le \lambda^{2+\kappa/4}$ we have,
\begin{align}
 \Omega: &=\int \frac {\lambda^2 f(p) }{\Big( \a - \bar \om(p-r)
  - i\eta \Big)
     \Big(\beta - \om(p+r)  
+i\eta \Big)} \; \rd p \nonumber \\
&     =   -2\pi i\lambda^2\int   \frac{ f(p)\; \delta(e(p)-\gamma)}{ (\a-\beta)
 +  2p \cdot r -
       2 i [\lambda^2  {\cal I} (\gamma)+\eta ]} \, \rd p + 
O(\lambda^{1/2-4\kappa}) \|  f\|_{4d, 1} \; .
\label{eq:opt1}
\end{align}
\end{lemma}

\medskip

Now we compute $\Xi_k$ by applying Lemma \ref{lemma:opt1}.
 Denote  $a:=(\a+\beta)/2$ and $b:=\lambda^{-2}(\a-\beta)$.
By using $\e= \lambda^{2+\kappa/2}$, $\eta=\lambda^{2+4\kappa}$,
 we have
\begin{align}\label{9.3}
\lambda^2 \int  \rd v  \;  & \Upsilon(\xi, v)\; 
\ov{R\Big(\a, v +\frac{\e\xi}{2}\Big)}
   R\Big(\beta, v -\frac{\e\xi}{2}\Big)  \\
& = \int
    \frac{-2\pi i \Upsilon(\xi, v)\;
     \delta(e(v)-a)}{  b +  \lambda^{\kappa/2} v\cdot \xi -
       2 i [ {\cal I} (a)+\lambda^{4\kappa}]} \; \rd v
       + O(\lambda^{1/2-4\kappa})\|\Upsilon\|_{4d,1} \; .
\nonumber
\end{align}
In the applications, $\Upsilon(\xi, v)$ will always be
supported on $|v|\leq \zeta$, therefore the measure $\rd v$
can be freely changed to $\rd\mu(v)$.

We now replace the  product of $k+1$ factors in 
the restricted version of \eqref{9.51} one by one. 
We need a $\lambda^2$ factor for each application of  (\ref{9.3}),
thus we need $\lambda^{2k+2}$  in \eqref{9.51}. But \eqref{9.51} contains
only $\lambda^{2k}$,  the 
missing $\lambda^2$ comes from the change of variables
 $\rd\alpha \rd \beta = \lambda^2 \rd a \rd b$.
The domain of integration, $(\a, \beta)\in D$,
is replaced by the domain 
$D^*:=\{ (a, b)\; : \; |a|\leq Y, |b|\leq 2\lambda^{-2}Y\}$.

For any $\ell=1, 2, \ldots k+1$, we introduce the notation 
\begin{align}
  \cF_{k,\ell}: = &     \int^*
    \rd \xi
  \int_{D^*} \frac{\rd a  \rd b}{(2\pi)^2} \Big( 
 \prod_{j=1\atop j\neq \ell}^{k+1} \int
 \rd \mu(v_j) \Big) 
 \prod_{j=1}^{\ell-2}  \Bigg(
   \Bigg|  \frac{2\pi i \; F_j(\xi, v_j) |\wh B (v_j-v_{j+1})|^2
\;\delta(e(v_j)-a)}{ b + \lambda^{\kappa/2} v_j\cdot \xi -
       2 i [\cI (a)+\lambda^{4\kappa}]}\Bigg|\Bigg)
\nonumber\\
&
 \times \Bigg|\frac{ 2\pi i F_\ell (\xi, v_{\ell-1})\; 
  \delta(e(v_{\ell-1})-a)}{ b + \lambda^{\kappa/2} 
 v_{\ell-1}\cdot \xi -      2 i [\cI (a)+\lambda^{4\kappa}]}  \Bigg| 
\nonumber\\
&\times \Bigg| 
\int \rd \mu(v_\ell) \Upsilon_\ell(v_{\ell-1}, v_\ell, v_{\ell+1})
 \Bigg[ \lambda^2 \ov{R\Big(\a, v_\ell +\frac{\e\xi}{2}\Big)}
   R\Big(\beta, v_\ell -\frac{\e\xi}{2}\Big) 
\nonumber\\
&\quad  - 
   \frac{ - 2\pi i \; \delta(e(v_\ell)-a)}{ b + \lambda^{\kappa/2} 
  v_\ell\cdot \xi -
       2 i [\cI (a)+\lambda^{4\kappa}]}\Bigg] \Bigg|
\nonumber\\ 
&
  \times 
\prod_{j=\ell+1}^{k} \Bigg( \lambda^2
  \Bigg |R\Big(\a, v_j +\frac{\e\xi}{2}\Big)
  R\Big(\beta, v_j -\frac{\e\xi}{2}\Big) F_j(\xi, v_j)\;
  |\wh B (v_j-v_{j+1})|^2
 \Bigg|  \Bigg) 
\nonumber\\
& 
\times 
 \lambda^2
  \Bigg |R\Big(\a, v_{k+1} +\frac{\e\xi}{2}\Big)
  R\Big(\beta, v_{k+1} -\frac{\e\xi}{2}\Big)
   F_{k+1}(\xi, v_{k+1}) 
 \Bigg| 
\label{fkj}
\end{align}
with
$$
 \Upsilon_\ell(v_{\ell-1}, v_\ell, v_{\ell+1}):= \left\{
\begin{array}{lll}
\wh W_0(\e\xi, v_1) \, |\wh B(v_1-v_2)|^2 & \mbox{for} & \ell=1\\
|\wh B(v_{\ell-1}-v_\ell)|^2|\wh B(v_\ell-v_{\ell+1})|^2 &\mbox{for} & 2\leq \ell\leq k\\
|\wh \cO(\xi, v_{k+1})| \, |\wh B(v_1-v_2)|^2\; &\mbox{for} & \ell=k+1\; 
\end{array}
\right.
$$
and
$$
    F_j(\xi, v): = \left\{ \begin{array}{lll} 
   \ov{\wh W}_0(\e\xi,v)  & \mbox{for} & j=1\\
  1 &\mbox{for} & 2\leq j\leq k\\
 \wh\cO(\xi, v)  &\mbox{for} & j=k+1\; .
\end{array}
\right.
$$
The formula \eqref{fkj} is literally valid for $2\leq \ell \leq k$.
For $\ell=1$ the first product and the factor in the second line
are absent, for 
 $\ell = k+1$ the factors in the last two lines are absent.
We also recall the convention made after \eqref{9.51} about
the intepretation of $|\wh B(v_j-v_{j+1})|$ for $j=k+1$.

With these notations and by introducing $\tau:= \lambda^2 t
=\lambda^{-\kappa}T$, and $W(k):=W_\lambda( t,k, \cO)$,  
we obtain the following telescopic estimate from
\eqref{9.51}
\begin{align}\label{9.4}
 \Bigg| \sum_{k< K} W(k)
 & -   \sum_{k< K}
 \int^*  \rd \xi  
   \int_{D^*} \frac{\rd a  \rd b}{(2\pi)^2} 
 \; e^{i\tau b+ 2 t \eta} \nonumber \\
&\times\Bigg(\prod_{j=1}^{k+1} \int
   \frac{-2\pi i F^{(j)}(\xi, v_j) \; 
\delta(e(v_j)-a) }{ b + \lambda^{\kappa/2} v_j\cdot \xi -
       2 i [ \cI (a)+\lambda^{4\kappa}]} |\wh B(v_j-v_{j+1})|^2\rd \mu(v_j)
  \Bigg)  \Bigg| \nonumber \\
&   \leq  \sum_{k< K}\sum_{\ell=1}^{k+1} \cF_{k,\ell} +o(1)\; .
\end{align}
Now we explain how to estimate $\cF_{k,\ell}$ for the general 
case ($2\leq \ell\leq k$), the modifications for the
two extrema are straighforward.

First we estimate $\ov{\wh W}_0$ by supremum norm and estimate
all denominators in the first two lines by their imaginary
part:
\be
   \Big| \frac{1}{  b + \lambda^{\kappa/2} v_j\cdot \xi -
       2 i [\cI (a)+\lambda^{4\kappa}]} \Big| \leq \frac{1}{2\cI(a)} \; .
\label{btriv}
\ee
 Then the $v_1, v_2, \ldots, v_{\ell-2}$ variables are integrated out in
this order, by using  \eqref{eq:opt},
yielding a total factor 1 from the product
in the first line of \eqref{fkj}.  By recalling \eqref{1edef}
and the estimate
$$
    \cI(a) = -\mbox{Im}\; \Theta(a) \ge c_1 \min \{ |e|^{\frac{d}{2}-1},
  e^{-1/2} \}
$$
from Lemma~3.2 of \cite{ESYI},
the integral of $v_{\ell-1}$
is estimated trivially by
\be
   \frac{1}{2\cI(a)} \int \delta(e(v_{\ell-1})-a) \; \rd\mu(v_{\ell-1})
  \leq \frac{Ca^{1/2}}{\cI(a)} \leq \langle a \rangle \; .
\label{l-1}
\ee
This estimate is used if $a\leq \zeta^2/2$, otherwise the
integral is zero by the support of $\rd\mu$, so we obtain
a factor $O(\lambda^{-2\kappa - O(\delta)})$.
In the regime $|b|\ge \lambda^{-\kappa}$, we have
$|b- \lambda^{\kappa/2}v_{\ell-1}
\cdot \xi |\ge |b|/2$ using $|v_{\ell-1}|\leq \zeta$ and $|\xi|\leq
\lambda^{-\delta}$. The estimate \eqref{btriv}
 can thus be changed to $2|b|^{-1}$,  
improving estimate 
\eqref{l-1} to $\leq Ca^{1/2}/|b|\leq C\zeta /|b|$.

The integral $\rd\mu(v_\ell)$ in \eqref{fkj} is estimated
by
\be
  O(\lambda^{1/2-4\kappa})  \sup_{v_{\ell-1}, v_{\ell+1}} \big\| 
    \Upsilon_\ell (v_{\ell-1}, \, \cdot \, , v_{\ell+1}) \big\|_{4d, 1}
  \leq  O(\lambda^{1/2-4\kappa}\zeta^{4d})  = O(\lambda^{1/2-(4d+4)\kappa
-O(\delta)})
\label{Upest}
\ee
by using  \eqref{9.3}  and the fact
that all $v_j$ variables satisfy $|v_j|\leq \zeta$.
For $\ell=1$ and $\ell= k+1$ we also used that the initial data and
the observable are Schwarz functions.

For the $\rd\mu(v_j)$, $j=\ell+1, \ell+2, \ldots , k$,
 integrals we separate the 
resolvents by Schwarz inequality,
$$
    \prod_{j=\ell+1}^k \big| R(\alpha, v_j + \ldots) R(\beta, v_j -
 \ldots)\big| \leq  
\prod_{j=\ell+1}^k \big| R(\alpha, v_j + \ldots)|^2 +\prod_{j=\ell+1}^k
 |R(\beta, v_j -
 \ldots)\big|^2 \; ,
$$
 and
 we use
the successive integration scheme (see 
 Section~10.1.2 of
\cite{ESYI})
to collect a constant factor. 

Before we integrate out the last momentum  variable, $v_{k+1}$, 
we perform the $\rd a\, \rd b$ integration.
We can change back the $a, b$ variables to $\alpha, \beta$,
we perform $\rd\alpha\, \rd\beta$ integrals  to collect
a $C|\log \lambda|^2$ factor since $D\subset \{ |\alpha|, |\beta|\leq
2Y\}$.
This argument applies unless $\ell =k+1$ and 
the last line in \eqref{fkj} is absent. In this case, however,
$\ell\ge 2$ (the $k=0$ case is treated separately, see \eqref{xi0}),
and then the denominator with $v_{\ell-1}$ in the second line
of \eqref{fkj} is present. We use the $a, b$ variables. Recall
that $\delta( e(v_{\ell-1})-a)$ restricts the domain
of the $\rd a$ integration to $|a|\leq \zeta^2/2$, giving
a contribution $O(\zeta^2)$.
 The domain of the $b$
integral is larger, $|b|\leq 2\lambda^{-2} Y$, but 
in the regime $|b|\ge \lambda^{-\kappa}$ we have
collected an additional $|b|^{-1}$ factor in \eqref{l-1},
thus the $\rd b$ integration contributes at most with a factor
$O(\lambda^{-\kappa})$.

Finally we integrate $v_{k+1}$  by using the integrability of $F_{k+1} = \cO$
and the $\xi$ integral gives a factor $O(\lambda^{-3\delta})$.
By collecting these estimates, we arrive at
$$
  \sum_{1\leq k< K}\sum_{\ell=1}^{k+1} \cF_{k,\ell} 
  \leq C\lambda^{1/2-(4d+9)\kappa - O(\delta)} 
$$
and that is negligible, since $\kappa<1/(8d+18)$.

\bigskip

Now we focus on the main term on the left hand side
 of \eqref{9.4}. First we extend the
$\rd b$  integration  from $|b|\leq 2\lambda^{-2}Y$ to
$\bR$. It is easy to see that
the error is negligible; all denominators can be 
bounded by $|b|/2$ and the result from the $\rd b$ integral
in the region $|b|\ge 2\lambda^{-2}Y$, 
$$ 
     \int_{|b|\ge 2\lambda^{-2}Y} \frac{\rd b}{|b|^{k+1}} 
    \leq \lambda^{2k} \; ,
$$
is negligible even after 
multiplying the $C^k$ from the $\rd v_j$ integrals.
We can also extend the $\rd a$ integration from $[-Y, Y]$
to $\bR$, since, due to the factor $\delta(e(v_j)-a)$
and the cutoff in $v_j$, the integrand is zero for $|a|\ge Y$.

Now we write
$$
   \frac{-i}{b + \lambda^{\kappa/2} v_j\cdot \xi 
- 2i[\cI(a)+\lambda^{4\kappa}] }
 = \int_0^\infty e^{-i\tau_j(b+ \lambda^{\kappa/2}\ldots)} \rd \tau_j
$$
and perform the $\rd b$ integration. We obtain
\begin{align}
\sum_{k<K} W(k)
= & \sum_{k< K} \int^* \rd \xi 
   \int_\bR \frac{\rd a}{2\pi}\;  e^{-2\tau \cI(a)- \tau \lambda^{4\kappa}}
 \Big( \prod_{j=1}^{k+1}\int_0^\infty \rd t_j\Big)
\delta\Big( \tau-\sum_{j=1}^{k+1} \tau_j \Big) \; e^{-i\lambda^{\kappa/2}
 (\sum \tau_jv_j)\cdot\xi}
 \nonumber \\
&
   \times \Big(\prod_{j=1}^{k+1} \int \rd \mu(v_j) \; 2\pi
  |\wh B(v_j-v_{j+1})|^2 \delta(e(v_j)-a)\Big) \wh\cO(\xi, v_{k+1})
  \overline{\wh W}_0(\e\xi, v_1) +o(1) \; .\nonumber 
\end{align}
We now
replace $\rd \mu(v_j)$ with $\rd v_j$ and  
remove the cutoff in $\xi$ to perform the
Fourier transform. We  also replace $W_0$ with $F_0$
and  remove the $k< K$ cutoff from the summation:
\begin{align}
  \sum_{k<K} W(k)
= &\sum_{k=0}^\infty \int \rd X
   \int_\bR \rd a \;  
e^{-2\tau\cI(a)} \Big( \prod_{j=1}^{k+1}\int_0^\infty \rd \tau_j\Big)
\delta\Big( \tau-\sum_{j=1}^{k+1} \tau_j \Big) 
 \nonumber\\
&
   \times\int \rd v_{1} \, \delta(e(v_{1})-a)
  \Bigg(\prod_{j=2}^{k+1} \int \rd v_j \; 2\pi
  |\wh B(v_j-v_{j-1})|^2 \delta(e(v_j)-a)\Bigg) \nonumber\\
&
\times  \cO(X, v_{k+1})
  F_0\Big( X -(2\pi)^{-1}
\lambda^{\kappa/2} (\sum_j \tau_jv_j), \, v_1 \, \Big)  +o(1) 
\label{markov}
\end{align}
with initial data
$$
    F_0(X, v):= \delta(X) |\wh\psi_0(v)|^2\; .
$$
The replacement of $\rd\mu(v_j)$ with $\rd v_j$ is justified since
$\wh\psi_0(v_1)$ is compactly supported and thus all other
$v_j$'s are restricted to a compact energy range
 by  the delta functions
$\prod_j \delta(e(v_j)-a)$.
The removal of the $\xi$-cutoff is
 allowed since  the integrand of the $\rd\xi$-integral
 can be majorized by
\be
  \sup_v |\wh \cO(\xi, v)| \int_\bR
   \frac{\rd a}{2\pi} \;  e^{-2t\cI(a)} 
     \sum_{k\leq K} \frac{\big[2\cI(a)t\big]^k}{k!}\;  \int \rd v_1 \; 
  \delta(e(v_1)-a) \overline{\wh W}_0(\e\xi, v_1)\leq 
    \sup_v |\wh \cO(\xi, v)| 
\label{unifint}
\ee
whose integral vanishes in the regime $|\xi|\ge \lambda^{-\delta}$
due to the assumptions on $\wh\cO$.
Here we used \eqref{eq:opt} to perform the $v_j$ integrations
successively and the time integration yielded $\tau^k/k!$.
 The replacement of $\overline{\wh W}_0(\e\xi, v)$ 
with $|\wh \psi_0(v)|^2$ comes from the uniformly
integrable bound \eqref{unifint} and from the  uniformity of the limit
$\| \wh\psi_0(\cdot \pm \e\xi) - \wh \psi_0(\cdot)\|\to 0$ 
as $\xi$ runs over any compact set. Finally, the removal of the
$k\leq K$ cutoff in the summation follows from the same
majorization as \eqref{unifint} together with
$$
        \sum_{k> K} \frac{\big[2\cI(a)\tau\big]^k}{k!} 
\leq \frac{(C\tau)^K}{K!} \leq (C\lambda^{\delta})^K\to 0 \; .
$$

\bigskip

For a fixed energy $e>0$
we consider the continuous time Markov process $\{ v(t) \}_{t\ge 0}$
on the
energy surface $\Sigma_e$ with generator \eqref{Lgen}. 
This process is exponentially mixing with the uniform
measure on $\Sigma_e$ being the unique invariant measure
(see Lemma~\ref{lemma:erg} in the Appendix).
Let $\cE_e^\psi$ denote the expectation value with
respect to this process starting from the initial state $\psi=\psi_0$
given by the normalized measure
$$
   \frac{|\wh\psi(v)|^2\,\delta(e(v_1)-e)
 \,\rd v}{\big[ |\wh\psi|^2\big](e)} \; 
$$
on $\Sigma_e$
(for the notations, see Section~\ref{sec:intro}).
Let $\cE_e$ denote the expectation with respect to the equilibrium.
Let $\rd\mu_\psi(e)=\big[ |\wh\psi|^2\big](e)\rd e$ 
be the energy distribution of $\psi$. The coarea formula,
$$
   \int_0^\infty \big[|\wh\psi|^2\big](e) \, \rd e =
\int |\wh\psi(v)|^2 \rd v  
$$
and  $\psi\in L^2$ guarantee that $\rd\mu_\psi$ is
absolutely continuous.

{F}rom \eqref{markov} and $\tau = \lambda^{-\kappa}T$ we have
$$
    \sum_{k<K} W(k)=
\int_0^\infty  \cE_e^\psi \cO \big( \lambda^{\kappa/2}
x(\tau),
 v(\tau) \big)\, \rd \mu_\psi(e) + o(1)\; \quad
\mbox{with}\quad x(\tau): =\int_0^\tau
  \frac{1}{2\pi}  v(s)\rd s \; .
 $$
Due to the exponential mixing and the continuity of
$\cO$, the replacement of $\cE_\psi$ with the equilibrium measure $\cE_e$
gives a negligible error since $\tau\to\infty$. By the central limit theorem
for additive functionals of exponentially mixing Markov chains,
$\lambda^{\kappa/2}x(\tau)$ 
converges to a centered Gaussian random variable with covariance
matrix
$$
        \cE_e\big[  \lambda^{\kappa/2}x(\tau)\otimes 
\lambda^{\kappa/2}x(\tau)\big] = \frac{\lambda^\kappa}{(2\pi)^2}
        \iint_0^\tau \cE_e \big[ v(s)\otimes v(s')\big]
    \rd s\rd s' \to 2 T D(e)\; .
$$
Since the equilibrium measure is uniform, the covariance matrix
is diagonal, $D_{ij}(e) = D_e \, \delta_{ij}$.
The diffusion coefficient, $D_e$, 
is  finite and positive.
This proves Theorem~\ref{thm:laddheat}.
 $\;\;\Box$

\appendix

\section{Mixing properties of the Boltzmann generator}

The Boltzmann velocity process with
 generator $L_e$ introduced in \eqref{Lgen} enjoys very good
statistical properties. The proof uses standard arguments
which we only indicate below.

\begin{lemma}\label{lemma:erg}
 For each $e>0$
the Markov process  $\{ v(t) \}_{t\ge 0}$ with generator $L_e$
is uniformly exponentially mixing. The unique invariant
measure 
is the uniform distribution,  $[\, \cdot \, ](e)/[1](e)$,
on the energy surface $\Sigma_e$.
\end{lemma}

{\it Sketch of the proof.}
 Let $\cP^t(u, A)$ be the transition kernel for any 
 $u\in \Sigma_e$, $A\subset \Sigma_e$.
Since the transition rate $\sigma(u,v)$ is continuous on $\Sigma_e$ and
$0\in \mbox{supp} \, (\wh B)$
holds, there exists an open set $S\subset \bR^d$,
$\mbox{diam}(S)\leq \sqrt{2e}$ and there exists
and $\delta = \delta(e)>0$
such that $\sigma(u,v)\ge \delta$ whenever $u-v\in S$, $u, v\in \Sigma_e$.
Since the state space $\Sigma_e$ is compact it follows
that the transition kernel $\cP^t$ satisfies a uniform Doeblin-type condition,
$$
  \inf_{u\in \Sigma_e} \int_0^1 \cP^t(u, A)\; \rd t \ge C(e)|A|   
  \qquad A\subset \Sigma_e \; ,
$$
with some $e$-dependent positive constant, 
where $|A|$ is the restriction of the
 Lebesgue measure (on $\Sigma_e$) of the set $A$. 
It is clear that 
 the Markov process $\{ v(t)\}_{t\ge 0}$ 
is irreducible and aperiodic, therefore it is
uniformly exponentially mixing. Moreover,
the rate of the mixing
is uniform as $e$ runs through a compact energy interval
since in this case $C(e)$ is uniformly separated away from zero.
It is easy to see that
the uniform measure on $\Sigma_e$ is  invariant 
and by exponential mixing it is the only invariant measure. $\;\;\Box$

\section{Estimates on Propagators}
\setcounter{equation}{0}

\subsection{Proof of Lemma \ref{le:opt}.}\label{sec:propag}

The following lemma proves
\eqref{eq:logest} and \eqref{eq:2aint}.
The proof of \eqref{eq:3aint} is analogous but easier and will be omitted.

\begin{lemma}\label{le:A1}  Let  $\kappa<1/6$
and  $\eta$ satisfying $\lambda^{2+4\kappa}\leq\eta\leq \lambda^{2+\kappa}$.
For any $0\leq a \leq 1$  we have the following approximation result
\be\label{A4}
\int \left| \frac 1 {\alpha- \om(p)+ i \eta}
- \frac 1 {\alpha- e(p) -\lambda^2\Theta(\alpha) + i \eta}
 \right|^{2-a} \; |h (p-q)| \rd p
\ee
$$
\le  C \lambda^{1-6\kappa} 
 \int \frac {|h(p-q)|} {|\alpha- e(p) -\lambda^2\Theta(\alpha) +
 i \eta|^{2-a}}
  \rd p \; .
$$
Moreover, for any $0\leq a <1$ we have
\be\label{eq:A}
\int  \frac {|h(p-q)|\; \rd p} {|\alpha- e(p)-\lambda^2\Theta(\alpha)
 + i \eta|^{2-a}}
\leq 
\frac{ C_a  \| h\|_{2d, 0} \lambda^{-2(1-a)} }
{ \langle \alpha \rangle^{a/2} \langle |q| -\sqrt{2|\alpha|}\rangle }
\ee
and
\be\label{eq:Alog}
\int  \frac {|h(p-q)|\; \rd p} {|\alpha- e(p) -\lambda^2\Theta(\alpha)
+ i \eta|}
\leq 
\frac{ C_a  \| h\|_{2d, 0} \, \big| \log \lambda \big|\;
 \log \langle\alpha\rangle }
{ \langle \alpha \rangle^{1/2} \langle |q| -\sqrt{2|\alpha|}\rangle } \; .
\ee
\end{lemma}

\noindent
{\it Proof.} 
 To prove (\ref{A4}), 
we  rewrite it as
\be\label{A5}
\begin{split}
&  \int \left| \frac 1 {\alpha- \om(p)+ i \eta}
- \frac 1 {\alpha- e(p) -\lambda^2\Theta(\alpha)
 + i \eta} \right|^{2-a}  \; |h (p-q)|\; \rd p \\
= &  \int \Big| \frac {\lambda^2 (\Theta(e(p))- \Theta(\alpha))}
 {(\alpha- \om(p)+ i \eta)
(\alpha- e(p) - \lambda^2 \Theta(\alpha)+ i \eta) } \Big|^{2-a}
\; |h (p-q)|\;\rd p  \; .
\end{split}
\ee
{F}rom the H\"{o}lder continuity,
\be   
|\Theta(\alpha) - \Theta(\alpha')| \leq C|\alpha-\alpha'|^{1/2} \; 
\label{eq:holder}
\ee 
(Lemma 3.1 from \cite{ESYI}),
 we can bound \eqref{A5} by
$$
\int \Big | \frac {|e(p)-\alpha|^{1/2}}
 {(\alpha- \om(p) + i \eta)}
\frac {\lambda^2 }{(\alpha- e(p)
- \lambda^2 \Theta(\alpha)+ i \eta) }\Big|^{2-a} \;|h (p-q)|  \rd p\; .
$$
Since
$$
|e(p)-\alpha|^{1/2} \le |\om(p)-\alpha |^{1/2}
+O(\lambda )\; ,
$$
this integral is bounded by
$$
(\lambda^2 \eta^{-1/2}+\lambda^3 \eta^{-1})^{2-a} \int
 \frac {|h(p-q)| \; \rd p} {|\alpha- e(p)
- \lambda^2 \Theta(\alpha) + i \eta|^{2-a}}\; .
$$
Notice that in these estimates we used that the imaginary
part of $\om(p)$ is negative.

To prove the estimates \eqref{eq:A} and \eqref{eq:Alog},
 we rewrite the integrals by
the co-area formula $(0\leq a \leq 1)$:
\begin{equation}\label{A9}
\int
 \frac {|h(p-q)| \rd p} {|\alpha- e(p)
- \lambda^2 \Theta(\alpha) + i \eta|^{2-a}}
=    \int_{0}^{\infty}
 \frac{\rd s}{|\a - s -\lambda^2 \Theta(\alpha) +i\eta|^{2-a}} \Xi(s)
\end{equation}
with
$$
   \Xi(s) : = \int_{|p|=\sqrt{2s}} \frac{|h(p-q)|\rd p}{|\nabla e(p)|}
   = \frac{1}{\sqrt{2s}} \int_{|p|=\sqrt{2s}} |h(p-q)|\rd p\; . 
$$
Using the decay properties of $h$, we have
$$ 
 \Xi(s) \leq \frac{\| h\|_{2d,0}}{ \langle |q|-\sqrt{2s}\rangle } 
\cdot \frac{\sqrt{s}}{\langle s \rangle}\; ,
$$
so
$$
\int
 \frac {|h(p-q)| \rd p} {|\alpha- e(p)
- \lambda^2 \Theta(\alpha) + i \eta|^{2-a}}
\leq   C\| h\|_{2d,0} \int_{0}^{\infty}
 \frac{\sqrt{s}\,\rd s}{\langle s\rangle\;\langle |q|-\sqrt{2s}\rangle
|\a - s -\lambda^2 \Theta(\alpha) +i\eta|^{2-a}}\; .
$$
For $a=1$ the last integral can be directly estimated by
$C\, \big|\log\eta\big|\, \langle \alpha \rangle^{-1/2}
\langle |q|-\sqrt{2|\alpha|}\rangle \log\langle \alpha\rangle$, yielding
\eqref{eq:Alog}.

To prove \eqref{eq:A}, i.e.
for $a<1$, we recall that  that $\Theta=\cR - i\cI$
with non-negative real $\cI$ and $\cR$.
The last integral is estimated by 
$$
I:=\int_{0}^{\infty}
 \frac{\sqrt{s}\,\rd s}{\langle s\rangle\;\langle |q|-\sqrt{2s}\rangle
\;\big[(\wt\a - s)^2 + (\lambda^2 \cI (\wt\alpha) +\eta)^2\big]^{1-a/2}}
$$
with $\wt\alpha :=\alpha -\lambda^2\cR (\alpha)$.
We used the H\"older continuity of $\cI$, 
$\lambda^2\cI(\alpha) = \lambda^2\cI(\wt\alpha) + O(\lambda^3)$ and
the fact that the error can be absorbed into $\eta$.

First we assume that $|\wt\alpha| \leq 1$, then 
the estimate on $\Theta(\alpha)$ from Lemma 3.2. of \cite{ESYI}
 yields $\cI(\wt\alpha)\leq c_1|\wt\alpha|^{1/2}$.
The $I$ integral can be estimated
$$
  I\leq \frac{1}{\langle |q|  \rangle}
 \int_0^2 \frac{\sqrt{s}\,\rd s}{
\;\big[(\wt\a - s)^2 + (\lambda^2|\wt\alpha|^{1/2} +\eta)^2\big]^{1-a/2}}
+ \int_2^\infty  
\frac{\sqrt{s}\, \rd s}{\langle s\rangle\;\langle |q|-\sqrt{2s}\rangle
s^{2-a}}\; .
$$
The second term is  bounded by $\langle |q|\rangle^{-1} \sim \langle
|q|-\sqrt{2|\alpha|}\rangle^{-1}$.
In the first term we consider two cases. If $|\wt\alpha|\leq \wt\beta:=
\lambda^2|\wt\alpha|^{1/2} +\eta$, then
$$
\int_0^2 \frac{\sqrt{s}\,\rd s}{
\;\big[(\wt\a - s)^2 + (\lambda^2|\wt\alpha|^{1/2} +\eta)^2\big]^{1-a/2}}
\leq \Big( \int_{|\wt\alpha -s|\leq 2\wt\beta} + 
\int_{|\wt\alpha -s|\ge 2\wt\beta}\Big)
$$
$$
   \leq \wt\beta^{-(2-a)}\int_{|s|\leq 2\wt\beta} \sqrt{s}\rd s
  + \int_{|\wt\alpha -s|\ge \wt\beta} \frac{\rd s}{|\wt\alpha -s|^{3/2-a}}
  = O(\wt\beta^{a-1/2}) \leq O(\lambda^{-2(1-a)}) \; ,
$$
by using that in the second regime $s$ and $|s-\wt\alpha|$ are comparable
and that $\wt\beta \ge \eta\ge \lambda^3$
in the last step. This proves \eqref{eq:A} for $|\wt\alpha| \leq 1$.

Next we consider the regime $|\wt\alpha|\ge 1$, then
$\cI(\wt\alpha)\leq c_1 |\wt\alpha|^{-1/2}$ and we have
\begin{align}
   I\leq & \int
 \frac{{\bf 1}(|\wt\alpha -s|\ge 1/2)
 \rd s}{\langle s\rangle^{1/2}\langle |q|- \sqrt{2s}\rangle
 |\wt\alpha-s|^{2-a}} \nonumber\\
 & + \int \frac{{\bf 1}(|\wt\alpha -s|\le 1/2)
  \rd s}{ \langle s\rangle^{1/2}
  \langle |q|-\sqrt{2s}\rangle \;
  \big[(\wt\a - s)^2 + (\lambda^2|\wt\alpha|^{-1/2} +\eta)^2\big]^{1-a/2}} \; .
\nonumber
\end{align}
The first integral is bounded by $C_a \langle \alpha \rangle^{-1/2}
\langle |q|-\sqrt{2|\alpha|}\rangle^{-1}$,
by using that $\langle \alpha \rangle \sim \langle \wt\alpha \rangle$. 
The second integral is bounded by
$$
    \frac{1}{\langle 
    \wt\alpha \rangle^{1/2}\langle |q|- \sqrt{2\alpha} \rangle}
    \int_{-1/2}^{1/2}
    \frac{\rd s}{[ s^2 + (\lambda^2|\wt\alpha|^{-1/2} +\eta)^2\big]^{1-a/2} }
   \leq  \frac{C_a\lambda^{-2(1-a)}}{\langle \alpha \rangle^{a/2}
   \langle |q|- \sqrt{2\alpha} \rangle}
$$
and this completes the proof of \eqref{eq:A}. $\;\;\Box$

\bigskip

We now prove
the more accurate estimate \eqref{eq:ladderint}.
We have
$$
\frac{ \lambda^{2}  }{|\alpha- \ov \om(p)-i\eta|^{2}}
=
  \frac{ \lambda^{2}  }{\lambda^2  {\cal I} (e(p))+ \eta} \; \mbox{Im} \;
  \frac 1 {\alpha- e(p) - \lambda^2  {\cal R} (e(p))
-i(\lambda^2  {\cal I} (e(p))+ \eta)} \; .
$$
{F}rom the resolvent identity and with the notations $e= e(p)$,
$\wt \alpha = \alpha-\lambda^2  {\cal R} (\alpha)$,  the last term equals to
$(I)+(II)+(III)$ with
\begin{align}
 (I):= &  \frac{ \lambda^{2}  }{\lambda^2  {\cal I}(\wt\alpha)
+ \eta} \; \mbox{Im} \;  \frac 1 {\wt\alpha- e
-i(\lambda^2  {\cal I}(\wt\alpha) + \eta)}
\label{2res} \\
(II):= &
- \frac{ \lambda^{2}  }{\lambda^2  {\cal I}(\wt\alpha) + \eta} \;
\frac{\lambda^2( {\cal I}(e) -{\cal I}(\wt\a) )}{\lambda^2 {\cal I}(e) +\eta}
\mbox{Im} \frac 1 {\alpha- e - \lambda^2\Theta(e) -i\eta}
\nonumber\\
(III):= &
  - \frac{ \lambda^{2}  }{\lambda^2  {\cal I}(\wt\alpha)+ \eta} \;
   \mbox{Im} \Bigg[  \frac 1 {\wt \alpha- e
-i(\lambda^2  {\cal I}(\alpha) + \eta)}
    \frac{\lambda^2 (\Theta(\a)-\Theta(e))}{\a -e -\lambda^2\Theta(e) -i\eta}
\Bigg] \; .
\nonumber
\end{align}
Our goal is to estimate $\int |\wh B(p-q)|^2 \big[ (I)+ (II)+ (III)\big]\rd p$.

We recall two continuity
properties of $\Theta_\e(\a, r)$ from Lemma 3.1 of \cite{ESYI}:
\begin{align}
  | \Theta_\e(\alpha, r) - \Theta_e (\alpha, r')|&\leq 
 C\big| \; |r|-|r'|\;\big|
\label{lipr}
\\
|\Theta_\e (\alpha, r) - \Theta_{\e'}(\alpha', r)|&\leq 
 C (|\e-\e'| + |\alpha - \alpha'|) \e^{-1/2}
\label{zhold}
\end{align}
if $\e\ge \e' >0$.
In estimating the integral of term (I), 
 we first use  \eqref{lipr}
to change $q$ to
$\wt q$ with $e(\wt q)=\wt\alpha$, then
we use \eqref{zhold} with $\e'\to 0+0$ and
$\e= \lambda^2  {\cal I}(\wt\alpha) + \eta =\cO(\lambda^2)$
 to obtain
\begin{align}
    \int |\wh B(p-q)|^2  (I) \; \rd p
& = \frac{ \lambda^{2}  }{\lambda^2 
 {\cal I}(\wt\alpha) + \eta} \Big[  {\cal I}(\wt\alpha)
+ \cO(\lambda) + O( |\wt\a - e(q)|^{1/2})\Big]
\nonumber\\
& \leq 1 + \cO( \lambda^2\eta^{-1} [ \lambda + |\a - \om(q)|^{1/2}]) \; .
\nonumber
\end{align}
In the second term of (\ref{2res}) we use (\ref{eq:holder}) and we drop the
positive ${\cal I}(\wt\alpha)$
and ${\cal I}(e)$ terms
in the denominators
\begin{align}
  |(II)|\leq &  \; C\Big( \frac{\lambda^2}{\eta}\Big)^2
  \frac{ |\lambda^2 {\cal I}(e(p)) +\eta|\;
|\wt\a -e(p)|^{1/2}}{|\wt\a -e (p) + \lambda^2({\cal R}(\a)-{\cal R}(e(p)))|^2
   + |\lambda^2 {\cal I}(e(p)) +\eta|^2} \nonumber\\
   \leq &  \; C\lambda^2 \Big( \frac{\lambda^2}{\eta}\Big)^2
  \frac{  |\wt\a -e(p)|^{1/2}}{|\wt\a -e (p)
 + \lambda^2({\cal R}(\wt\a)-{\cal R}(e(p)))|^2
   + \eta^2} \; ,
\nonumber
\end{align}
where we used  that $\big[ \lambda^2( {\cal R}(\wt\a) - {\cal R}(\a))\big]^2
=\cO(\lambda^6)\ll \eta^2$ based upon  (\ref{eq:holder}).

To perform the $\rd p$ integration, we distinguish two regimes
depending on whether $|\wt\alpha - e(p)|$ is bigger or smaller than $K\lambda^4$
for a sufficiently large fixed $K$. When $|\wt\alpha - e(p)|\ge K\lambda^4$,
then $\lambda^2|{\cal R}(\wt\a)-{\cal R}(e(p))| < \frac{1}{2}|\wt\a -e (p)|$, hence
$$
  |(II)|\leq  C\lambda^2 \Big( \frac{\lambda^2}{\eta}\Big)^2 
\frac{\eta^{-1/2} }{  |\wt\a -e(p)| +\eta}\; ,
$$
and, by using a bound analogous to \eqref{eq:Alog},
the corresponding integral is bounded by
$$
     C\lambda^6\eta^{-5/2}
\int \frac{|\wh B(p-q)|^2 \; \rd p}{|\wt\a -e(p)| +\eta} \leq 
    \cO(\lambda^6\eta^{-5/2}|\log\eta|) \; .
$$

When $|\wt\alpha - e(p)|\le K\lambda^4$, then
we can trivially estimate
$|(II)| \leq C(\lambda^2\eta^{-1})^4$ and after the co-area formula,
the volume  factor is given by
$$
    \int_0^\infty {\bf 1}( |\wt\a -s|\le K\lambda^4) \sqrt{s} S(s)
 \rd s = O(\lambda^4) \; ,
$$
with
$$
     S(e):=\int_{ S^{d-1}}
   |\wh B(\sqrt{2e}(\phi_r-\phi))|^2\rd\phi 
$$
where $\phi_r\in S^{d-1}$ is fixed. Recalling
the properties of $S(e)$ from the proof of Lemma 3.2 in \cite{ESYI}, 
we see that the contribution to
the integral  $\int |\wh B(p-q)|^2 (II) \; \rd p$
is of order $\cO(  (\lambda^3\eta^{-1})^4)$.

Finally, the last term  in (\ref{2res}) is estimated as
$$
  |(III)| \leq C\lambda^2 \Big( \frac{\lambda^2}{\eta}\Big)  \frac{1}{ |\wt\a -e|+\eta}
   \;\frac{ |\a-e|^{1/2}} {| \a -e + \lambda^2 {\cal R}(e)| + \eta} \; .
$$
In the regime where $|\a - e(p)| \ge K\lambda^2$ (with some large $K$) we obtain
$$
   |(III)| \leq C\lambda^2 \Big( \frac{\lambda^2}{\eta}\Big)
   \;\frac{ |\a-e|^{1/2}} {(| \a -e| + \eta)^2} \leq C\lambda^2 \eta^{-1/2}
   \Big( \frac{\lambda^2}{\eta}\Big)  \frac{1}{ |\a -e|+\eta}
$$
and after integration we collect $\cO( \lambda^4\eta^{-3/2}|\log\eta|)$.
In the regime where $|\a - e(p)| \le K\lambda^2$ we have
 $|(III)|\leq \cO(\lambda^5\eta^{-3})$
and the volume factor is $\cO(\lambda^2)$, therefore the integral is $\cO(\lambda^7\eta^{-3})$.
Collecting the error 
terms we arrive at the proof of Lemma \ref{le:opt}. $\;\;\Box$

\subsection{Proof of Lemma \ref{lemma:opt1}}

We can assume that $f$ is a real function and  write $f(p)= 
 \langle p \rangle^{-2d} g(p)$ with $\| g\|_{2d,0}<C\| f\|_{4d, 1}$.
We can restrict the integration regime in \eqref{eq:opt1}
 to  $|p|\leq \lambda^{-1}$
since the contribution of the  outside regime is $O(\lambda^{2d})$
by a Schwarz inequality (to separate the two denominators) and
a trivial application of Lemma \ref{le:A1} with $a=0$.
This large momentum cutoff will be done with the insertion
of a function $\chi(\lambda \langle p \rangle)$ with a smooth,
compactly supported $\chi$, $\chi\equiv 1$ on $[-1,1]$.

We can also assume that $|\alpha - e(p-r)|\leq \lambda$,
$|\beta-e(p+r)|\leq \lambda$,  otherwise
at least one
of the denominator can be estimated by $O(\lambda^{-1})$
and the other one integrated out by (\ref{eq:logest}) to give
$O(\lambda|\log\lambda|)$.
Since $|\alpha - e(p-r)|\ge |\alpha - e(p)| - C (|p| + |r|)|r|
\ge |\alpha -e(p)| - O(\lambda^{1+\kappa/4})$, we obtain that
$|\alpha -e(p)|\leq 2\lambda$ and similarly $|\beta -e(p)|\leq 2\lambda$,
in particular $|\alpha -\beta|\leq 4\lambda$ and $|\gamma -e(p)|\leq 2\lambda$.

We replace the first denominator of \eqref{eq:opt1} by
$\alpha - e(p) +p\cdot r - \lambda^2 \bar \Theta(\gamma) - i\eta$.
The error term of this replacement, by the resolvent expansion, is bounded by
\be
     \int \frac{ \lambda^{9/2}\chi(\lambda \langle p \rangle) |f(p)| 
\rd p} { | \alpha - \bar\om(p-r)  - i\eta|
  \; | \alpha - e(p) +p\cdot r - \lambda^2 \bar \Theta(\gamma) - i\eta|
  \; |\beta - \om(p+r)  +i\eta|} \; ,
\label{ress}
\ee
where we have used the estimate
$$
   \alpha - e(p-r) -\lambda^2\ov{\theta}(p-r) - i\eta
  = \alpha - e(p) +p\cdot r - \lambda^2 \bar \Theta(\gamma) - i \eta
   + O(\lambda^{5/2})\; ,
$$
which follows from the
 above restrictions on
the integration domain  $|r|\leq \lambda^{2+\kappa/4}$  and 
the H\"older continuity \eqref{eq:holder}. 
To estimate the error term \eqref{ress}, we bound
 the $\beta$ denominator  trivially by $\eta^{-1}$
and use the Schwarz inequality to separate the remaining two denominators
$$
    \frac{1}{  | \alpha - \bar\om(p-r)  - i\eta|
  \; | \alpha - e(p) + \ldots|} \leq 
  \frac{1}{  | \alpha - \bar\om(p-r)  - i\eta|^2} +
  \frac{1}{  | \alpha - e(p) + \ldots|^2}  \; .
$$
The integral of the first term can be bounded by Lemma \ref{le:A1} with $a=0$;
for the second term we rewrite $e(p)+ p\cdot r = e(p+r) - \frac{1}{2}r^2$
and use \eqref{eq:3aint}
after a shift
in $\alpha$ and $p$.
We arrive at $\Omega = \Omega_0+O(\lambda^{1/2- 4\kappa})$  with
$$
  \Omega_0 = \int \frac{\lambda^2 \chi ( \lambda\langle p\rangle) f(p)\rd p}{
  \Big( \alpha - e(p) +p\cdot r - \lambda^2 \bar \Theta(\gamma) - i\eta \Big)
   \Big( \beta - e(p) -p\cdot r - \lambda^2  \Theta(\gamma) + i\eta \Big)} \; .
$$
To compute $\Omega_0$ we can choose a coordinate system where
 the vector $r$ points in the $n$-th direction:  $r= |r|(0,\ldots, 0,1)$.
We can write
\be
  \Omega_0 = \lambda^2\int_\bR
 \frac{\rd p_\Vert}{ \a-\beta + 2|r|\, p_\Vert - 
2i[\lambda^2 \cI(\gamma)+\eta]}
\label{Y0}
\ee
$$
\times
 \int_{\bR^{n-1}} \Big[ 
\frac{1}{ \beta- e(p)
 -p_\Vert|r| - \lambda^2  \Theta(\gamma) + i\eta}
 -  \frac{1}{ \alpha - e(p)
 +p_\Vert \, |r| - \lambda^2 \bar \Theta(\gamma) - i\eta }
\Big] \chi ( \lambda\langle p\rangle) f(p)\rd p_\perp \; .
$$

\begin{lemma}\label{lem:Y}
Let $F$ be a $C^1$-function on $\bR$ with $|F(Q)|\leq C\langle Q\rangle^{-2}$
 and let
$$
    Y(z):= \int_0^\infty \frac{F(Q)}{z-Q} \; \rd Q
$$
for any $z=\alpha + i\e$ with $0<\e\leq 1/2$. Then
\be
      |Y(z)-Y(z')| \leq  C|F(0)| \;|\log z- \log z'|
 + |z-z'| |\log \e| \| F \|_{2d,1}\; ,
\label{Ylip}
\ee
where $z'=\alpha' + i\e'$ and $\e\ge \e'>0$.
\end{lemma}
{\it Proof.} This lemma is essentially
Lemma 3.10 in \cite{EY}. For completeness we recall the proof.
Choose a branch of the complex logarithm on the upper 
half plane and use integration by parts:
$$
     |Y(z)-Y(z')| \leq |F(0)| |\log z- \log z'|
  + \Big| \int_0^\infty F'(Q) \big[ \log (z-Q)- \log (z'-Q)\big] \rd Q\Big|\; .
$$
The second term is estimated by
$$
    \int_{\Gamma(z,z')} \rd |\xi| \int_0^\infty \frac{|F'(Q)|}{|\xi-Q|} \rd Q\; ,
$$
where $\Gamma(z,z')$ is any path in the upper half plane that connects
$z$ and $z'$ and $\rd |\xi|$ is the arclength measure.
A simple exercise shows
$$
    \int_0^\infty   \frac{|F'(Q)|}{|\xi-Q|}\;  \rd Q \leq C \| F \|_{2d, 1}
  |\log (\mbox{Im} \; \xi)| \; .
$$
Choose a path from $z=\a + i\e$ to $\a'+i\e$ then to $\a'+i\e'$
along straight line segments. After integration we obtain
\eqref{Ylip}. $\;\;\Box$

\bigskip

We now  change 
the denominators in the square bracket in \eqref{Y0}
 to $\gamma -e(p) \pm i\eta$.
This requires a change of order $O(\lambda)$ in the denominators using
the estimates on  $|\alpha - \gamma|$  and $ p_\Vert|r|$.
With the help of Lemma \ref{lem:Y} such change
 yields an
error of order $\lambda^3|\log\lambda| \eta^{-1}\| f\|_{2d,1}$ in $\Omega_0$.
After these changes, we can remove the cutoff $\chi(\lambda\langle p\rangle)$
at a price of $O(\lambda^{2d})$ as before, 
 and we have
\be
  \Omega_0 = \lambda^2\int_\bR
 \frac{\rd p_\parallel}{ \a-\beta + 2|r|p_\Vert - 2i[\lambda^2 \cI(\gamma)+\eta]}
 \int_{\bR^{d-1}} \Big[ 
\frac{1}{ \gamma- e(p)+ i\eta}
 -  \frac{1}{ \gamma - e(p)  - i\eta }
\Big]  f(p)\rd p_\perp
\label{L0}
\ee
modulo negligible errors involving $\| f\|_{4d, 1}$.

The inner integral is evaluated as
$$
  I:=  i \, \mbox{Im} \int_0^\infty \frac{f^*(u, p_\Vert)
  u^{\frac{d-3}{2}}\rd u}{\gamma - \frac{1}{2}p_\Vert^2 
   - \frac{1}{2} u +i\eta} \qquad \mbox{with}
\quad f^*(u, p_\Vert):= 
\int_{S^{d-2}} f(u^{1/2}\theta, p_\Vert) \rd\theta \; .
$$
For  $p_\Vert$ in the range
$\gamma -\frac{1}{2}p_\Vert^2 \leq -\lambda$, we have
$$
    |I|\leq \frac{C\eta \| f\|_{4d,0}}{\langle p_\Vert\rangle^{2d}} 
\int_0^\infty \frac{ u^{\frac{d-3}{2}}\rd u}{|\lambda +
\frac{1}{2}u|^2 \;\langle u\rangle^{2d}} = O(\eta\lambda^{-1}) \; ,
$$
by using the decay 
of $f^*$ inherited from $f$. The contribution of this regime to
$\Omega_0$ is therefore of order $\lambda |\log \lambda|$, and 
hence negligible.
In the regime $|\gamma -\frac{1}{2}p_\Vert^2| \leq \lambda$ one can estimate
$|I|\leq C\langle p_\Vert\rangle^{-2d}$.
 The $\rd p_\Vert$-volume is at most $O(\lambda^{1/2})$,
so the contribution of this
regime to $\Omega_0$ is  at most of order $\lambda^{5/2}\eta^{-1}$, and
hence also negligible.

Finally, we can concentrate on the regime $\gamma -\frac{1}{2}p_\Vert^2 \ge \lambda$.
We can use the
estimate (for $\e >\e'>0$)
$$
      \mbox{Im} \int_{-\e}^\infty \frac{g(x)}{x+i\e'}\;\rd x 
= -\pi g(0)  + O(\e'/\e) +
      O(\e'|\log\e'|)\; 
$$
if $g\in C^1$  with a bounded derivative. 
We obtain, with $\e = \gamma -\frac{1}{2}p_\Vert^2$, $\e'=\eta$, that
$$
   I = - 2\pi i (2\gamma -p_\Vert^2)^{\frac{d-3}{2}}
  f^*( 2\gamma -p_\Vert^2, p_\Vert)
+ O(\lambda^{1+ 4\kappa})\; ,
$$
where the error is  integrable in $p_\Vert$. Therefore it is negligible
in $\Omega_0$. Substituting the main term into \eqref{L0}, we obtain
the main term in \eqref{eq:opt1}. $\;\;\Box$.

\section{General estimates on circle graphs}\label{sec:genestcir}
\setcounter{equation}{0}

We define four operations on a  partition given on the vertex set of a
circle graph on $N$ vertices and we estimate how the $E$-value of
the partition changes. Operation I was already defined
in Section 9 of \cite{ESYI}, here we repeat the definition
and the corresponding estimate for convenience.

\bigskip

{\bf Operation I: Breaking up lumps}

\medskip

Consider a Feynman graph on $N$ vertices (Section \ref{sec:feyndef}).
Given a partition of the set $\cV\setminus \{ 0, 0^*\}$, 
$\bP=\{P_\mu\; :\; \mu \in I(\bP)\}
\in \cP_\cV$, we define a new partition $\bP^*$
by breaking up one of the lumps into two smaller nonempty lumps.
 Let $P_\nu = P_{\nu'} \cup P_{\nu''}$ with
$P_{\nu'}\cap P_{\nu''}=\emptyset$ and $\bP^*=\{ P_{\nu'},  P_{\nu''},
 P_\mu  \;  : \; \mu\in I(\bP)\setminus \{ \nu\}\}$.
In particular $I(\bP^*) = I(\bP)\cup \{ \nu', \nu''\} \setminus \{ \nu\}$
and $m(\bP^*)= m(\bP)+1$.
The following estimate was proven in Lemma 9.5 of \cite{ESYI}.

\begin{lemma}\label{lemma:breakup}
With the notation above, we have
$$
     E_{(*)g}( \bP, \bu, \balpha) \leq 
\int_{|r|\leq N\zeta}
 \rd r \; E_{(*)g}
    (\bP^*, \bu^*(r,\nu),
     \balpha)\; ,
$$
where the new set of momenta $ \bu^*=\bu^*(r,\nu)$ is given
by $u^*_\mu:=u_\mu$, $\mu\in I(\bP)\setminus \{\nu\}$
and $u^*_{\nu'}= u_\nu-r$, $u^*_{\nu''} = r$. In our estimates
we will always have $N\leq 2K$ and then
$$
    \sup_\bu E_{(*)g}( \bP, \bu, \balpha) \leq \Lambda
    E_{(*)g}
    \sup_\bu (\bP^*,\bu,
     \balpha)\; 
$$
with $\Lambda:= [CK\zeta]^d = O(\lambda^{-2d\kappa -O(\delta)})$
(see \eqref{def:K} and \eqref{def:mu}).
\end{lemma}

\medskip

{\bf Operation II: Removing the lump of a single vertex}

\medskip

Let $v\in \cV \setminus\{0, 0^*\}$ be a vertex and let $\bP\in \cP_\cV$
such that $P_\sigma = \{ v\}$ for some $\sigma\in  I(\bP)$,
i.e. the single element set
$\{ v \}$ is a lump.
 Define $\cV^*:=\cV\setminus\{ v\}$, $\cL(\cV^*):=\cL(\cV)\cup \{ (v-1, v+1)\} \setminus \{
(v-1,v), (v, v+1)\}$, i.e. we simply remove the vertex $v$ from the circle
graph and connect the vertices $v-1, v+1$. Let $\bP^*\in \cP_{\cV^*}$,
$\bP^* :=\bP\setminus \{ \; \{ v \} \; \}$ be $\bP$
after  simply removing the lump $\{ v\} $. In particular, $I(\bP^*) =
 I(\bP)\setminus \{ \sigma\}$.

\begin{lemma}\label{lemma:remove}
With the notations above
\be
   \sup_\bu E_{(*)g}(\bP, \bu,\balpha) \leq
 C\lambda\eta^{-1}\sup_{\bu^*} E_{(*)g+1}
    (\bP^*, \bu^*, \balpha) \; .
\label{eq:nontr}
\ee

If both neighbors of $0^*$, $v\neq v'$, form single
lumps in $\bP$, then both of these lumps can be simultaneously
removed to obtain a partition $\bP^*: = \bP\setminus\{ \{ v\}, \{ v'\} \}$
with the estimate
\be
   \sup_\bu  E_{*g}(\bP, \bu,\balpha) \leq C \lambda^2 \;
 \sup_{\bu^*} E_{g+2}
    (\bP^*, \bu^*, \balpha) \; .
\label{eq:2tru}
\ee
\end{lemma}

{\it Proof.}
The factor $\lambda$ in the estimate \eqref{eq:nontr}
is due to the fact that
 each vertex (apart from $0$ and $0^*$) carries
a factor $\lambda$ and $|\cV^*|=|\cV|-1$.
Let $P_\nu$ be the  lump of the vertex $v-1$ right before to $v$ in 
the circular  ordering
and assume $v-1\neq 0, 0^*$ (otherwise we consider $v+1$ and the proof is
 slightly modified).
 We use the trivial bound
\be
\frac{1}{|\alpha_{e_{v-}} - \om(w_{e_{v-}})+i\eta|}\leq \eta^{-1}
\label{linfty}
\ee
(recall that $\mbox{Im}\; \om \leq 0$ from  Lemma 3.2 of \cite{ESYI})
and  the bound
\be
     |\wh B(w_{e_{v+}}-w_{e_{v-}})|\;  |\wh B( w_{e_{v-}}-w_{e_{(v-1)-}})|
 \leq \frac{C}{\langle w_{e_{v+}}-w_{e_{(v-1)-}} \rangle^{2d}}
\label{mod}
\ee
(uniformly in $w_{e_{v-}}$) 
to obtain the necessary decay between the two newly consecutive momenta.
The same bound holds if some of the $\wh B(\cdot)$ on the left hand side
is replaced with 
$\langle \cdot \rangle^{-2d}$ due to the set $\cG$.
Now we integrate
$w_{e_{v-}}$ to
 obtain a new delta function from
$$
   \int \rd\mu(w_{e_{v-}}) \delta\Big(  w_{e_{v+}} - w_{e_{v-}} - u_\sigma\Big)
   \delta\Big( w_{e_{v-}}+\sum_{e\in L_\pm(P_\nu)\; : \; e\neq e_{v-}}
    \pm w_e - u_\nu \Big)
$$
$$
    \leq\delta\Big( w_{e_{v+}}
   +\sum_{e\in L_\pm (P_\nu)\; : \; e\neq e_{v-}}
    \pm w_e -( u_\sigma +  u_\nu)\Big)
$$
and clearly
$$
w_{e_{v+}}
+ \sum_{e\in L_\pm(P_\nu)\; : \; e\neq e_{v-}}
    \pm w_e =\sum_{e\in L_\pm(P_\nu^*)} \pm w_e \; .
$$
The new auxiliary momentum associated to $P_\nu$ is $u_\nu+u_\sigma$
and $P_\sigma$ disappeared, so the sum of the auxiliary momenta
remain unchanged, and
 (\ref{sumu}) continues to hold. This proves \eqref{eq:nontr}.

For the proof of \eqref{eq:2tru},
 if $v$ and $v'$ are the vertices on both sides of $0^*$,
then they can be removed and their neighbours can be connected
directly to $0^*$ yielding a non-truncated value of a
graph with two vertices less. This gives a factor
$\lambda^2$. The appropriate redefinition of
the auxiliary momenta is straightforward.
$ \;\;\;\Box$
\bigskip

{\bf Operation III: Removing  half of a gate}

\bigskip\noindent
Let $v, v+1\in \cV \setminus\{0, 0^*\}$ two subsequent vertices
and let 
$\bP\in \cP_\cV$
such that $v\equiv v+1 \;(\mbox{mod} \; \bP)$.
In the main application
this will arise when $v, v+1$ are connected and form a gate.
 Define $\cV^*:=\cV\setminus\{ v+1\}$,
$\cL(\cV^*):=\cL(\cV)\cup \{ (v, v+2)\} \setminus \{
 (v, v+1), (v+1, v+2)\}$, i.e. we simply remove the vertex $ v+1$
 from the circle graph with the adjacent edges
and add a new edge between the vertices $v, v+2$.
Let $\bP^*\in \cP_{\cV^*}$ be identical to the partition $\bP$
except that $v+1$ is
simply removed from its lump.
In particular, $I(\bP)=I(\bP^*)$.

\begin{lemma}\label{lemma:removehalfgate}
With the notations above
$$
    E_{(*)g}(\bP, \bu,\balpha) \leq C\lambda |\log\eta| \; 
E_{(*)g+1}
    (\bP^*, \bu, \balpha) \; .
$$
\end{lemma}

{\it Proof.} Note that the momentum $w_{e_{v+}}$ of the edge between
 $v$ and $ v+1$
does not appear in the delta functions in the definition of
 $E_{(*)g}(\bP, \bu,\balpha)$ (see \eqref{def:E}).
Before integrating out this momentum in \eqref{def:E},
 we use the bound
$$
     |\wh B(w_{e_{v+}}-w_{e_{v-}})|\;  |\wh B( w_{e_{v-}}-w_{e_{(v-1)-}})|
$$
$$
 \leq \frac{C}{\langle w_{e_{v+}}-w_{e_{(v-1)-}} \rangle^{2d}}
 \Big[ \frac{1}{\langle w_{e_{v+}}-w_{e_{v-}} \rangle^{2d}}
+\frac{1}{\langle w_{e_{v-}}-w_{e_{(v-1)-}} \rangle^{2d}}\Big]
$$
to ensure the decay between the momenta
$w_{e_{v-}}$ and  
$w_{e_{(v+2)-}}$, that are consecutive in the new graph.
The same bound holds if some of the $\wh B(\cdot)$ is
already replaced with $\langle \cdot \rangle^{-2d}$.
The integration of $w_{e_{v-}}$ yields 
$C|\log\eta|$ by using
\eqref{eq:logest}.
   $\;\;\Box$

\bigskip

{\bf Operation IV: Removing an isolated gate}

\bigskip\noindent
Let $v, v+1\in \cV \setminus\{0, 0^*\}$ be two subsequent vertices
and let  a partition  $\bP\in \cP_\cV$
such that $v\equiv v+1 \; (\mbox{mod} \; \bP)$.
 Define $\cV^*:=\cV\setminus\{ v,v+1\}$,
$\cL(\cV^*):=\cL(\cV)\cup \{ (v-1, v+2)\} \setminus \{
(v-1,v), (v, v+1), (v+1, v+2)\}$, i.e. we simply remove the gate.
Let $\bP^*\in \cP_{\cV^*}$ be $\bP$
after  removing the lump $\{ v, v+1\} $.
 Combining Operations III and II, we immediately obtain:

\begin{lemma}\label{lemma:removegate}
With the notations above
$$
   \sup_\bu E_{(*)g}(\bP, \bu,\balpha)
 \leq C\lambda^2\eta^{-1}|\log\eta| \;
  \sup_{\bu^*} E_{(*)g+2}
    (\bP^*, \bu^*, \balpha) \; . \qquad  \Box
$$
\end{lemma}
Note that this bound is not optimal. 
The removal of a gate
affects the value of the graph
only a  by constant factor, but the corresponding
estimate is more complicated and we do not aim at optimizing
the value of $\kappa$.

\end{document}